\documentclass{jfm}
\usepackage{graphicx}
\usepackage{epstopdf, epsfig}
\usepackage{listings}
\usepackage[normalem]{ulem}
\usepackage[sort, numbers]{natbib}
\usepackage{booktabs}
\usepackage[dvipsnames]{xcolor}
\usepackage{appendix}
\usepackage{tabularx}
\usepackage{comment}

\usepackage{textcomp}

\def\imagetop#1{\vtop{\null\hbox{#1}}}

\lstdefinestyle{base}{
  language=C,
  emptylines=1,
  breaklines=true,
  basicstyle=\ttfamily\color{black},
  moredelim=**[is][\color{red}]{@}{@},
  moredelim=[is][\sout]{@@}{@@},
}

\shorttitle{Nudging the turbulent flow around a square cylinder}
\shortauthor{
M. Zauner, V. Mons, O. Marquet \& B. Leclaire
}

\title{Nudging-based data assimilation of the turbulent flow around a square cylinder}

\author{M. Zauner\aff{1},
  V. Mons\aff{1},
  O. Marquet\aff{1},
 \and B. Leclaire\aff{1}}

\affiliation{\aff{1}DAAA,
ONERA, Universit\'{e} Paris Saclay, F-92190 Meudon, France}

\begin{document}

\maketitle

\begin{abstract}
We investigate the prediction of the turbulent flow around a canonical square cylinder at $Re=22000$ solving the unsteady Reynolds-averaged Navier-Stokes (URANS) equations. The limitations of URANS modelling are overcome through the application of a data assimilation technique involving a feedback control term, allowing to drive the URANS predictions towards reference data. Using regularly-spaced temporally-resolved reference data that are extracted from DNS, we assess the abilities of the data assimilation methodology in improving the unsteady URANS predictions. While improving the low-frequency oscillations of the wake flow, we also demonstrate the ability to recover the high-frequency dynamics of Kelvin-Helmholtz instabilities. The influence of spatial resolution of reference data is systematically investigated. The data resolution which is required to improve the prediction of flow structures of interest is related to their wavelength. Using a spacing of the order of one cylinder length between data points, we already observe synchronisation of the low-frequency vortex shedding that leads to a significant decrease in temporal and spectral errors as computed by spectral proper orthogonal decomposition. Furthermore, improved accuracy in terms of mean flow prediction of URANS is achieved. Using spacing of the order of the wavelength of the Kelvin-Helmholtz vortices, we even recover those shear layer structures which are absent in the baseline simulation.
\end{abstract}

\begin{keywords}
\end{keywords}

\section{Introduction}



Fine characterization of turbulent flows is usually performed via 
experiments, scale-resolving numerical approaches, or simulation methods relying on turbulence models. Experiments reproduce the entire complexity of flow physics, but the spatial and temporal resolutions of experimental data are generally limited, which prevents from a full characterization of the flow of interest. Direct numerical simulations (DNS) may provide such detailed results with high spatio-temporal resolution for the full flow field. But 
those scale-resolving simulations require massive computational power, that may even be excessive 
when investigating high Reynolds number flows of interest for the aeronautics, civil engineering or automotive industry. 
Lower-fidelity steady or unsteady methods based on Reynolds-averaged Navier-Stokes (RANS) equations are popular low-cost alternative approches.
 However, the accuracy of such computations may largely be affected by the fidelity of the closure model for the Reynolds stress tensor.


In order to overcome the respective limitations of the above mentioned approaches, data assimilation \citep{Lewis2006_cambridge,Hayase2015_fdr} appears as a valuable tool. Data assimilation consists in merging experimental, or reference data with computationally efficient numerical models. On one hand, it allows to infer or compensate for uncertainties in input and/or model parameters in simulations. On the other hand, it provides an augmented and full flow estimation from the possibly sparse reference data. Popular data assimilation techniques include adjoint-based \citep{LeDimet1986_tellus,Papadakis2008_siam,Franceschini2020_prf,Li2020_jfm} and (ensemble) Kalman filter-based methodologies \citep{Evensen2009_springer,Suzuki2012_jfm,Meldi2017_jcp,DaSilva2020_jfm}. Despite their significant differences in terms of formulation and implementation, these two classes of approaches may require a similar computation cost, which is typically of the order of between $10$ and $100$ simulations \citep{Mons2016_jcp}.

As a cost-efficient alternative, we here investigate the application of the nudging technique \citep{Hoke1976_mwr,Lakshmivarahan2013}, which may also be referred to as state observer in the framework of control theory, or more specifically measurement-integrated simulation \citep{Hayase2015_jfcmv} in the context of Computational Fluid Dynamics (CFD). Nudging consists of adding a feedback term to the flow governing equations that acts at specified measurement locations and is proportional to the difference between reference data and numerical prediction. Incompressible flows have generally been considered in previous studies discussed below, and, depending on the considered type of observation (namely velocity or pressure measurements), the feedback term may act on the momentum equations \citep{Imagawa2010} and/or on the pressure Poisson equation \citep{Neeteson2020_mst}. The intensity of this feedback term is tuned through a single scalar, which thus allows to adjust to which extent the simulation is driven towards the reference data. Nudging may be considered as a simplification of the Kalman filter. Indeed, Kalman filtering also amounts to introduce a feedback term in the governing equations that is proportional to the discrepancies between observations and prediction, but this feedback term is there weighted through a matrix gain instead of a single scalar. Nudging may also be categorized as a special case of proportional–integral–derivative (PID) controller that only includes the proportional contribution. The benefits of considering temporal integral and derivative of the discrepancies between observations and prediction have been recently investigated in \citep{Neeteson2020_mst,Saredi2021_cf}. In contrast to previously mentioned data assimilation approaches such as adjoint- and Kalman filter-based techniques, the supplementary computational cost that is associated to nudging is essentially negligible compared to that of a baseline simulation.

In the context of CFD, the application of nudging was first performed in a purely numerical context by \cite{Hayase1997,Imagawa2010} considering the turbulent flow in a square pipe. Based on punctual observations of the velocity field of a reference DNS, nudging was employed to drive a second simulation towards the reference one. Nudging was able to compensate for uncertainties in the initial condition \citep{Imagawa2010} or numerical errors due to the use of a coarser grid in the second simulation \citep{Hayase1997}, and thus successfully reproduced the reference solution. \cite{Neeteson2020_mst} also considered nudging and more general PID controllers to drive a coarse-grid simulation based on synthetic pressure data of the flow past a square cylinder at $Re=100$. In \cite{DiLeoni2020}, nudging was applied to homogeneous turbulence, still relying on DNS. In particular, nudging was used to recover a reference forced isotropic turbulent flow, assuming that the forcing of the reference flow was unknown. Various types of observations of the reference flow were considered among Eulerian observations, as in \cite{Hayase1997,Imagawa2010,Neeteson2020_mst}, Lagrangian data, or in spectral space. The effect of the observation density on the ability of nudging in recovering unmeasured scales was also investigated. While the reference data were generated with the same numerical methods as those employed for data assimilation in the above studies, nudging in conjunction with a turbulence model was investigated by \cite{Buzzicotti2020_pof} in the context of large eddy simulations of isotropic turbulence. Filtered reference data were generated by DNS, and nudging was performed assimilating the full reference data at all resolved scales, and was thus only useful to correct model errors.

Aside from these purely numerical studies, \cite{Nisugi2004,Yamagata2008} demonstrated the ability of nudging to deal with various types of experimental data of the flow past a square cylinder at $Re=1200$, such as wall pressure and particle image velocimetry (PIV) measurements, based on the laminar Navier-Stokes equations. \cite{Suzuki2009a_ef,Suzuki2009b_ef} performed the assimilation of particle tracking velocimetry (PTV) measurements of flows past airfoils at low Reynolds numbers. Their proposed methodology slightly differs from that in the above studies in the fact that the measurements were rearranged before the data assimilation procedure in order to be directly comparable with a flow state estimated by DNS. To efficiently handle experimental data of higher Reynolds number flows, \cite{Nakao2009} considered relying on unsteady RANS (URANS) modelling to perform nudging. Recently, steady nudging in conjunction with RANS was performed by \cite{Saredi2021_cf} based on three-dimensional PIV measurements of the mean flow around a wall-mounted bluff obstacle. Nudging proved to be effective in correcting deficiencies in the RANS prediction, such as the overestimation of the extent of the recirculation region past the obstacle. However, while these studies confirmed the feasibility of applying nudging to experimental data and ambitious flow configurations, the use of only experimental data precluded a quantitative assessment of the improvement in the estimation of the considered high Reynolds number flows outside of the measurement locations.

In the present contribution, we aim to assess in detail the full potential of the nudging approach for unsteady flow estimation in an aerodynamic context, based on URANS modelling and synthetic velocity data. The unsteady turbulent flow around a square cylinder at $Re=22000$ is chosen as flow configuration \citep{Lyn1995_jfm,Trias2015} because it exhibits several flow phenomena occurring at various time and spatial scales that are of interest for aeronautical application. They include low-frequency quasi-periodic vortex shedding (VS) past the square cylinder, along with broad-band high-frequency Kelvin-Helmholtz (KH) fluctuations in the shear layers at the top and bottom sides of the cylinder. URANS simulations aim at capturing the fluctuation of the statistically-averaged flow, and are most accurate when the fluctuation frequency, such that of vortex shedding, is much lower than that of the turbulence itself, which is modelled. For the present flow configuration, \cite{Iaccarino2003_ijhff} showed that, while being inevitably more expensive than RANS simulations, URANS simulations significantly improve the prediction of the time-independent mean flow. On the other hand, they do not necessarily capture the small-scale high-frequency fluctuations associated to Kelvin-Helmholtz vortices and, more generally, their accuracy remains determined by the adequacy of the closure model for the Reynolds stress tensor \citep{Bosch1998_ijnmf,Palkin2016_ftc}, here the Spalart-Allmaras model \citep{Spalart1994}. The objective of the paper is thus to assess the potential of the nudging method in unsteady flow estimation, not only for the large-scale low-frequency fluctuations that are solved by standard URANS simulations, but also for the small-scale high-frequency fluctuations that are not captured by these same simulations. Synthetic observations for the considered two-dimensional URANS simulations will be generated through a spanwise average of a three-dimensional DNS of the present flow configuration. Nudging will then be employed to enhance the URANS predictions and reconstructing the full reference flow from sparse velocity observations, whose spatial density will be systematically varied. A crucial aspect of the present study will be to correctly reproduce flow structures evolving coherently in space and time. This will be assessed through the spectral proper orthogonal decomposition \citep{Towne2018}.

The paper is organised as follows. Standard numerical simulations of the turbulent flow around a square cylinder are described in \S \ref{sec:standard}. After introducing the numerical methods used for DNS and URANS simulations in \S \ref{sec:DNS} and \S\ref{sec:URANS}, respectively, we describe methods to analyse dynamic as well as mean properties of flow fields in \S \ref{sec:errors}. DNS and standard URANS results are discussed in \S \ref{sec:DNSvsURANS}. The potential of nudged URANS simulations is then discussed in \S \ref{sec:unsteadynudging}. After a description of the nudging method and used data-sets of measurements in \S \ref{sec:nudging_methodology}, results of nudged simulations using a dense set of measurements are compared to standard URANS and DNS data in \S \ref{sec:results_concept}. Then, the performance of nudging with respect to low- and high-frequency flow phenomena is discussed in sections \S \ref{sec:results_low_frequency} and \S \ref{sec:results_high_frequency}, respectively. Implications for experimental studies are discussed in \S \ref{sec:criterion}. Finally, concluding remarks are drawn in \S\ref{sec:conclusions}.

\section{Standard URANS simulations compared to DNS} \label{sec:standard}

We investigate the turbulent flow developing around a square cylinder of length $D$ facing an upstream uniform flow of velocity $U_{\infty}$. The Reynolds number based on these reference length and velocity scales is $Re= U_{\infty} D / \nu = 2.2 \cdot 10^4$. 
Three-dimensional direct numerical simulation (DNS) presented in \S \ref{sec:DNS} captures the full range of flow scales, unlike standard two-dimensional Unsteady Reynolds-Averaged Navier-Stokes (URANS) equations introduced in \S \ref{sec:URANS} that capture only the low-frequency dynamics of statistically averaged flow structures. A turbulence model then accounts for the effect of incoherent structures so that the overall computational costs are significantly lower for URANS computations compared to DNS.
In order to assess the quality of URANS predictions, instantaneous DNS snapshots are averaged in the spanwise direction and provide reference data.
After describing respectively temporal and modal diagnosis tools in \S\ref{sec:errors}, detailed comparison between DNS and URANS simulations is provided in \S \ref{sec:DNSvsURANS}.

\subsection{Direct numerical simulations}\label{sec:DNS}

Three-dimensional DNS of the turbulent flow around the square cylinder are performed with the ONERA's compressible solver FastS \citep{Mary2002,Dandois2018}. The three-dimensional computational domain is formed by a two-dimensional circle of radius $50 D$ in the $\boldsymbol{x}=(x,y)$ plane, centered around the square cylinder, which is extruded in the spanwise homogeneous direction $z$ with length $L_z=4 D$. The two-dimensional mesh consists of approximately $2.6 \cdot 10^5$ grids points and {$N_z=960$} planes uniformly discretise the spanwise direction. Periodic boundary conditions are applied in the spanwise direction. Homogeneous freestream conditions are enforced at the outer boundaries, where only the horizontal velocity component (parallel to upper and lower cylinder surfaces) is non-zero and equal to $U_{\infty}$. The simulation is run close to incompressible flow regime at $M=0.1$ enforcing no-slip conditions at the cylinder surface. 
In the following, all quantities are nondimensionalised based on the reference length $D$ and free stream parameters $U_{\infty}$, $\rho_{\infty}$. 
The DNS time step is $\Delta t_{DNS} = 3.3188 \cdot 10^{-4}$, which corresponds to a maximal CFL number of approximately $0.7$.
More details about this DNS solver are provided in \cite{Dandois2018,Mary2002,Franceschini2020_prf}. 

The unsteady three-dimensional velocity fields $\boldsymbol{u}$ so obtained can be decomposed as  
\begin{eqnarray}
\boldsymbol{u}(\boldsymbol{x},z,t) = \bar{\boldsymbol{u}}_{r}(\boldsymbol{x},t) + \boldsymbol{u}^{'}(\boldsymbol{x},z,t)
\end{eqnarray}
where $\boldsymbol{\bar{u}}_{r}$ is the statistically-averaged two-dimensional velocity field of interest while $\boldsymbol{u}^{'}$ refers to the associated three-dimensional deviation, not used in the following. Then, it is assumed that $\boldsymbol{\bar{u}}_{r}$ can be extracted from DNS velocity fields $\boldsymbol{u}$ by averaging instantaneous snapshots in the spanwise direction according to
\begin{eqnarray}\label{eq:reference_u_r}
\bar{\boldsymbol{u}}_{r}(\boldsymbol{x},t) =  \frac{1}{L_z} \int_{0}^{L_z} \boldsymbol{u}(\boldsymbol{x},z,t) \, dz   \sim \frac{1}{N_z} \sum_{k=1}^{N_z}  
 \boldsymbol{u}(\boldsymbol{x},z=k \Delta z,t) \end{eqnarray}
where $\Delta z \sim 0.004$ is the uniform discretization step in the spanwise direction. This spatial average well approximates the statistical average if the spanwise extent $L_z$ of the computational domain and the number of planes $N_z$ are large enough. This two-dimensional field extracted from DNS will be used as reference solution (hence the subscript $r$), not only to assess the accuracy of the standard URANS simulation methods in \S\ref{sec:DNSvsURANS}, but also to improve their predictive capability thanks to the nudging method in \S \ref{sec:unsteadynudging}. To complete the description of this reference data set, it is obtained by down-sampling the DNS snapshots every $63$ time steps. The time interval $\Delta t_{r}$ between snapshots in this reference data set is thus $\Delta t_{r} = 63 \times \Delta t_{DNS} = 0.0209$ convective time units. 

\subsection{Unsteady Reynolds-averaged Navier-Stokes simulations}\label{sec:URANS}

\begin{figure}
    \begin{tabular}{ll}
(a) & (b)  \\
\includegraphics[height=0.35\columnwidth]{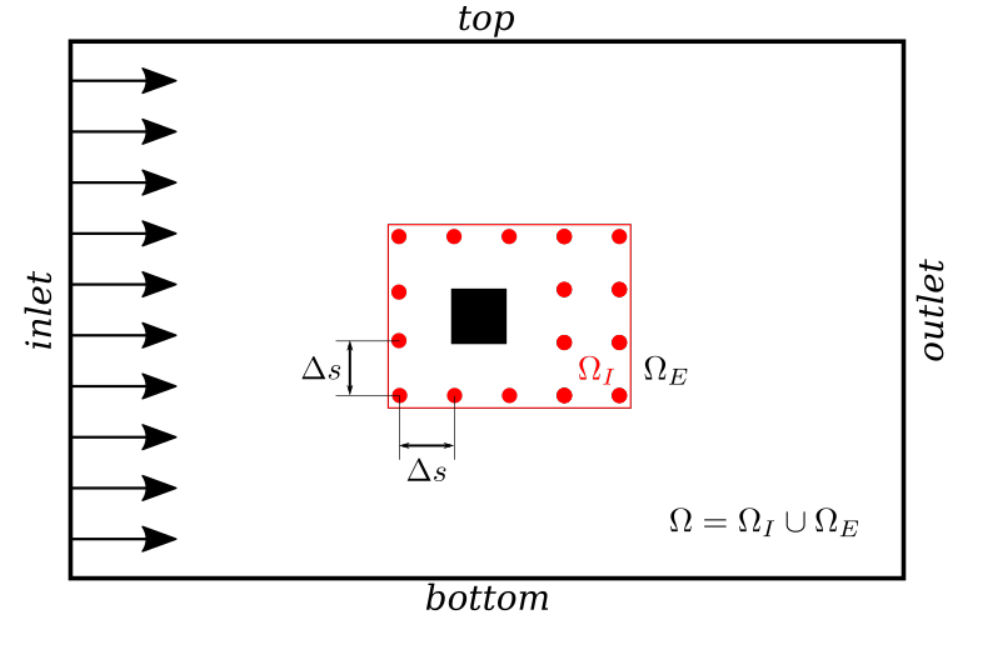} & \includegraphics[height=0.35\columnwidth]{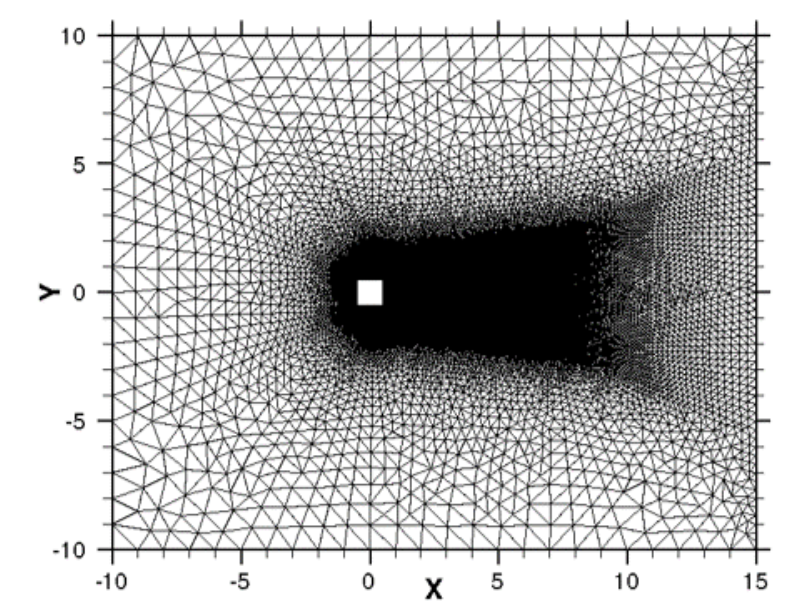}
    \end{tabular}
\caption{(a) Sketch of the computational domain $\Omega$ for URANS simulations, where red dots denote positions $(\mathbf{x}_{k})_{k=1 \cdots M}$ of velocity measurements in the internal domain $\Omega_{I}$ (bounded by the red frame). These so-called nudging points are uniformly separated in both directions by a spacing-distance $\Delta s$. The domain that is external to the measurement area is denoted as $\Omega_{E}$. (b) Typical triangular mesh used for the standard and nudged URANS simulations around a square cylinder.
}
\label{fig:scketch-nudging}
\end{figure}

Rather than extracting the temporal variation of the spatially-averaged flow from DNS simulations, one may rather estimate the latter through the two-dimensional URANS equations which are supplemented with a turbulence model to take into account the effect of the non-resolved component. The two-dimensional resolved velocity ($\bar{\boldsymbol{u}}$) and pressure ($\bar{p}$) fields then satisfy the following (incompressible) flow equations
\begin{eqnarray}\label{eqn:URANS} 
   \frac{\partial \bar{\boldsymbol{u}}}{\partial t} + (\bar{\boldsymbol{u}} \cdot \boldsymbol{\nabla}) \bar{\boldsymbol{u}} + \boldsymbol{\nabla}\bar{p} - \boldsymbol{\nabla}\cdot \left[ 2\left( Re^{-1}+ \nu_t(\tilde{\nu}) \right)  \boldsymbol{\nabla}_{\mathrm{s}} \bar{\boldsymbol{u}} \right] &=& \boldsymbol{0} \;, \nonumber \\
    \boldsymbol{\nabla}\cdot \bar{\boldsymbol{u}} &=& 0,  
 \end{eqnarray}
where $\boldsymbol{\nabla}_{\mathrm{s}} \bar{\boldsymbol{u}}$ refers to the symmetric part of the velocity gradient and $\nu_t$ denotes the turbulent eddy-viscosity. Here, the latter is determined using the Spalart-Allmaras model \citep{Spalart1994}, where the following governing equation for the intermediate eddy-viscosity-like variable $\tilde{\nu}$ is solved
\begin{eqnarray}\label{eqn:SA} 
   \frac{\partial \tilde{\nu}}{\partial t} + (\bar{\boldsymbol{u}} \cdot \boldsymbol{\nabla}) \tilde{\nu}  &=& s(\tilde{\nu},\bar{\boldsymbol{u}}).
 \end{eqnarray}
The source term $s$ accounts for the production, dissipation and diffusion terms for $\tilde{\nu}$. Details about the present  implementation of the Spalart-Allmaras model are provided in \cite{Franceschini2020_prf}.

The URANS equations (\ref{eqn:URANS})-(\ref{eqn:SA}) are discretised in space using the finite-element method implemented in the software FreeFEM++ \citep{Hecht2012_jnm}. Second-order polynomial elements $(P_2)$ are employed for the velocity field, while first-order elements $(P_1)$ are considered for the pressure and eddy-viscosity variables. For the relatively high Reynolds number flow of the present contribution ($Re=22000$), streamline-upwind Petrov-Galerkin (SUPG) \citep{Brooks1982} and grad-div \citep{Olshanskii2009} stabilisations are implemented. The rectangular computational domain is sketched in figure \ref{fig:scketch-nudging}(a) with an inlet located at $x=-10$, an outlet at $x=15$ and top and bottom planes at $y=10$ and $y=-10$, respectively. Remaining details of figure \ref{fig:scketch-nudging}(a) will be discussed later when introducing the nudging approach. The unstructured mesh, shown in figure \ref{fig:scketch-nudging}(b), consists of triangles with a total number of $46228$ nodes.  At the inlet, the boundary condition $(\bar{u},\bar{v},\tilde{\nu})=(1,0,0)$ is imposed. At the cylinder surface, the no-slip boundary condition $\bar{\boldsymbol{u}}=\boldsymbol{0}$ is enforced in conjunction with $\tilde{\nu} = 0$. The conditions $\frac{\partial \tilde{\nu}} {\partial x}= 0$ and $2(Re^{-1}+\nu_t)\boldsymbol{\nabla}_{\mathrm{s}}\bar{\boldsymbol{u}} \cdot \boldsymbol{n} - \bar{p} \boldsymbol{n} = \boldsymbol{0}$ are used at the outlet, where $\boldsymbol{n}$ denotes the normal vector. At the top and bottom boundaries, symmetry conditions are imposed according to $(\frac{\partial \bar{u}}{\partial y}, \bar{v},\frac{\partial \tilde{\nu}}{\partial y})=\boldsymbol{0}$.

Time integration is performed in a fully implicit way based on a second-order accurate finite-difference approximation of the time derivative. A quasi-Newton method is used to solve the resulting nonlinear problem at each iteration. The time step of URANS simulation is denoted by $\Delta t$. Results of the standard simulations presented in \S \ref{sec:DNSvsURANS} are obtained with $\Delta t = \Delta t_{r}$, i.e. matching the time interval of the reference data set. 
For nudged URANS simulations in \S \ref{sec:unsteadynudging}, the value of $\Delta t$ will be equal to the time interval of data sets (not necessarily the reference data set) inserted into the simulations through the feedback term. By varying the sampling of these data sets, we will thus modify the time step $\Delta t$. 

\subsection{Temporal and spectral errors}\label{sec:errors}

We introduce in this paragraph various mathematical definitions for assessing errors between the reference data set and the output of the standard or nudged URANS simulations. Note that the latter have not yet been introduced, but we will use the same notation for denoting the corresponding variables in both simulations. In the following, both standard and nudged URANS solutions may be referred to as the estimated flow, while the field $\bar{\boldsymbol{u}}_r$ which is extracted from DNS in (\ref{eq:reference_u_r}) is referred to as the reference flow, as mentioned above. Definitions of the instantaneous and time-averaged errors are introduced in this section before considering the spectral error based on a SPOD analysis of the data.\\

The instantaneous error field $e(\boldsymbol{x},t)$ between the estimated velocity field $\bar{\boldsymbol{u}}$ and the reference one $\bar{\boldsymbol{u}}_{r}$ is defined as
\begin{eqnarray}\label{eqn:instantaneous-error}
    e(\boldsymbol{x},t)=\left({(\bar{u}_{r} -\bar{u})^2+( \bar{v}_{r} -\bar{v})^2}\right)^{\frac{1}{2}}.
\end{eqnarray}
The instantaneous and time-averaged global errors are then respectively defined as
\begin{eqnarray}\label{eqn:total-error}
    E(t)=\left(\int_{\Omega} e(\boldsymbol{x},t)^2 \, d\boldsymbol{x}\right)^{\frac{1}{2}}, \quad \langle E \rangle = \frac{1}{T_{e}} \int_{t_i}^{t_i+T_{e}} E(t) \, dt ,
\end{eqnarray}
where $\Omega$ denotes the entire  computational domain, $t_i>0$ is an arbitrary time that is chosen to avoid transient effects (occurring in $[0,t_i]$) and $T_{e}$ refers to the considered time window for averaging the instantaneous global error. Note that the time average is denoted $\langle \cdot \rangle$, and thus $\langle \bar{\boldsymbol{u}}\rangle$ is the time-averaged velocity field of the estimated flow. Its error compared to the time-averaged reference velocity $\langle \bar{\boldsymbol{u}}_{r}\rangle$ is then defined as 
\begin{eqnarray}\label{eqn:total-meanerror}
  e_{\langle \bar{\boldsymbol{u}} \rangle }  (\boldsymbol{x}) = \left(\left(  \langle \bar{u}_r \rangle -\langle \bar{u} \rangle \right)^2 + \left( \langle \bar{v}_r \rangle - \langle \bar{v} \rangle \right)^2\right)^{\frac{1}{2}}. 
\end{eqnarray}

To better understand the unsteady effect, a spectral analysis of the data is performed which relies on the spectral proper orthogonal decomposition (SPOD) described in \cite{Towne2018}. Taking a temporal series of velocity snapshots as inputs, the SPOD analysis provides a set of SPOD modes $\boldsymbol{\Phi}=(\Phi_{u},\Phi_{v})^{T}$ that each oscillate at a single non-dimensional frequency (Strouhal number) $St=f D/U_{\infty}$. To each frequency corresponds a set of normalized SPOD modes ($\int_{\Omega} \boldsymbol{\Phi}^{*} \boldsymbol{\Phi} d\Omega = 1$, where the asterisk denotes the Hermitian transpose) ranked by their kinetic energy $\lambda$, that are the eigenvalues of the cross-spectral density tensor which is estimated from snapshots. In the following, we will focus on the dominant SPOD modes, namely which are associated to the largest eigenvalue, for frequencies of interest. Denoting hereinafter the SPOD modes of the reference and URANS data set as 
$[\lambda_{r},\boldsymbol{\Phi}_{r}]$ and 
$[\lambda,\boldsymbol{\Phi}]$ respectively, a spectral error field may be defined as 
\begin{eqnarray}\label{eqn:modal-error}
   \Phi_e=\left(\left( \sqrt{\Phi_{u,r}^* \Phi_{u,r}} - \sqrt{\Phi_{u}^* \Phi_{u}}\right)^2 + \left( \sqrt{\Phi_{v,r}^* \Phi_{v,r}} - \sqrt{\Phi_{v}^* \Phi_{v}}\right)^2\right)^{\frac{1}{2}}.
\end{eqnarray}
In the following, the dynamical content of standard and nudged URANS simulations will thus be assessed and compared to the reference solution through their frequency content, the kinetic energy of the dominant SPOD modes, and the shape of the latter through (\ref{eqn:modal-error})


\subsection{Comparisons between standard URANS and DNS}\label{sec:DNSvsURANS}




\begin{figure}
\centering
\vspace{0.25cm}
\centering
\begin{tabular}[t]{ll}
(a) & (b)     \\
\imagetop{\includegraphics[width=0.45\columnwidth]{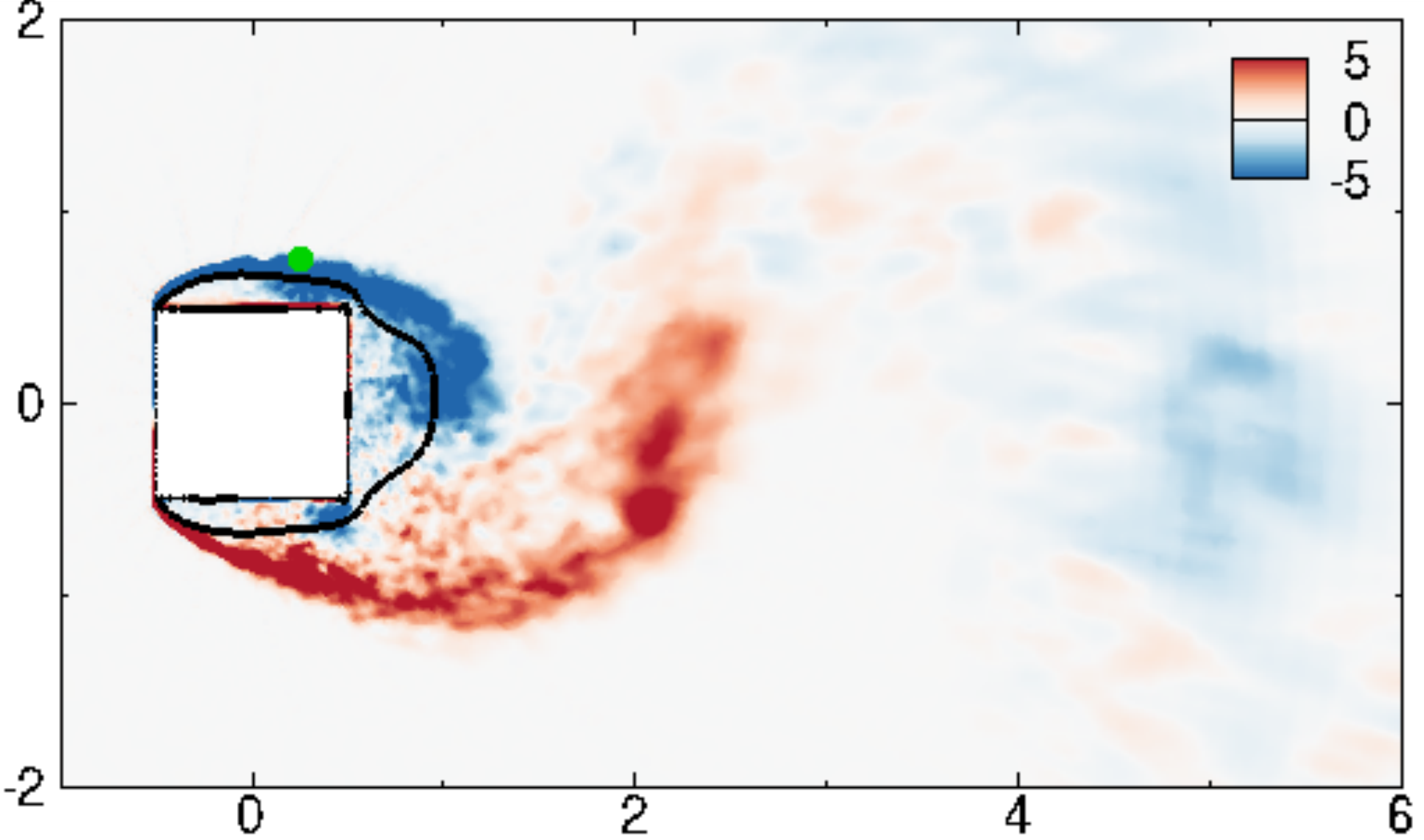}} & \imagetop{\includegraphics[width=0.34\columnwidth]{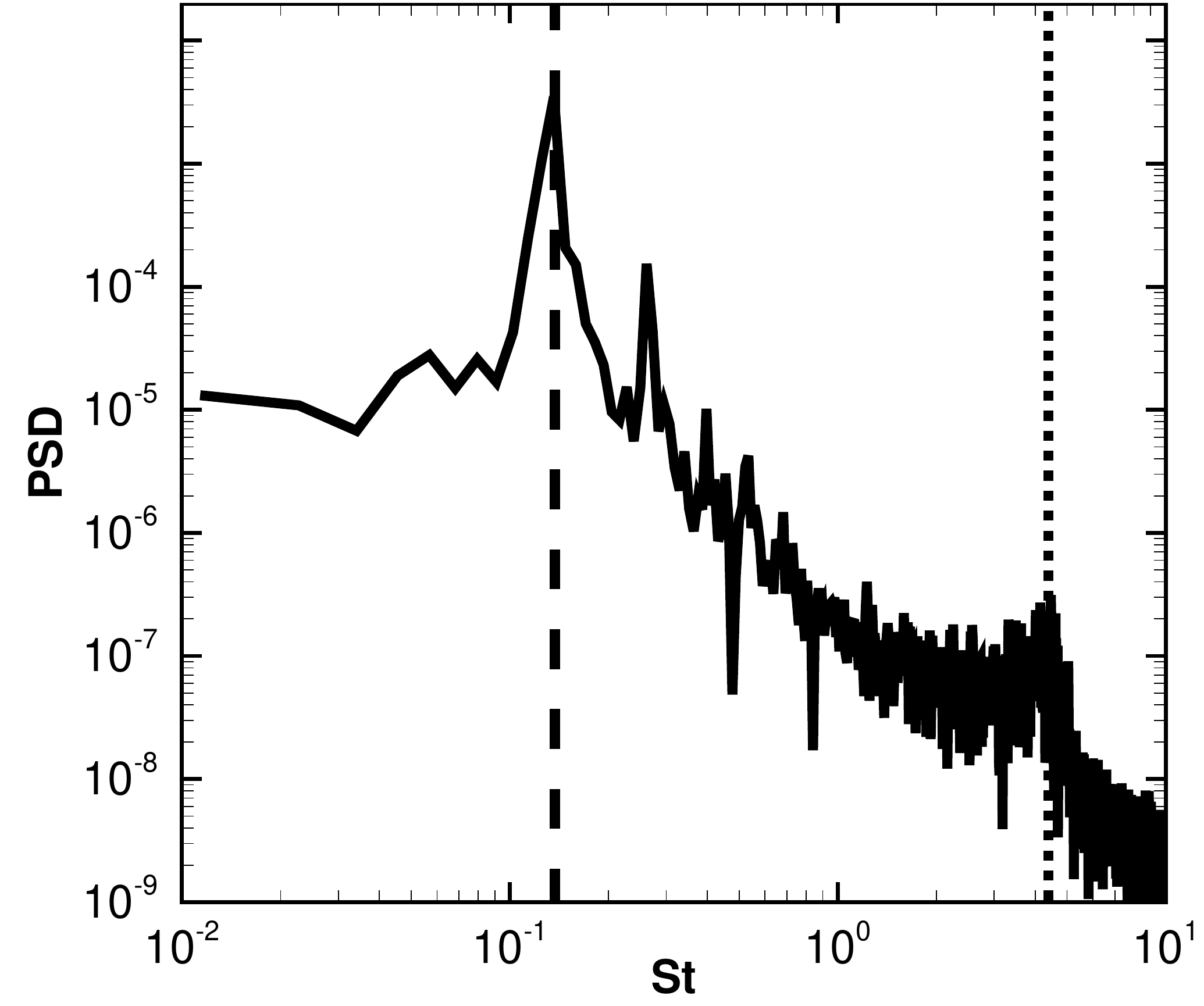}} \\
(c) & (d)   \\
\imagetop{\includegraphics[width=0.45\columnwidth]{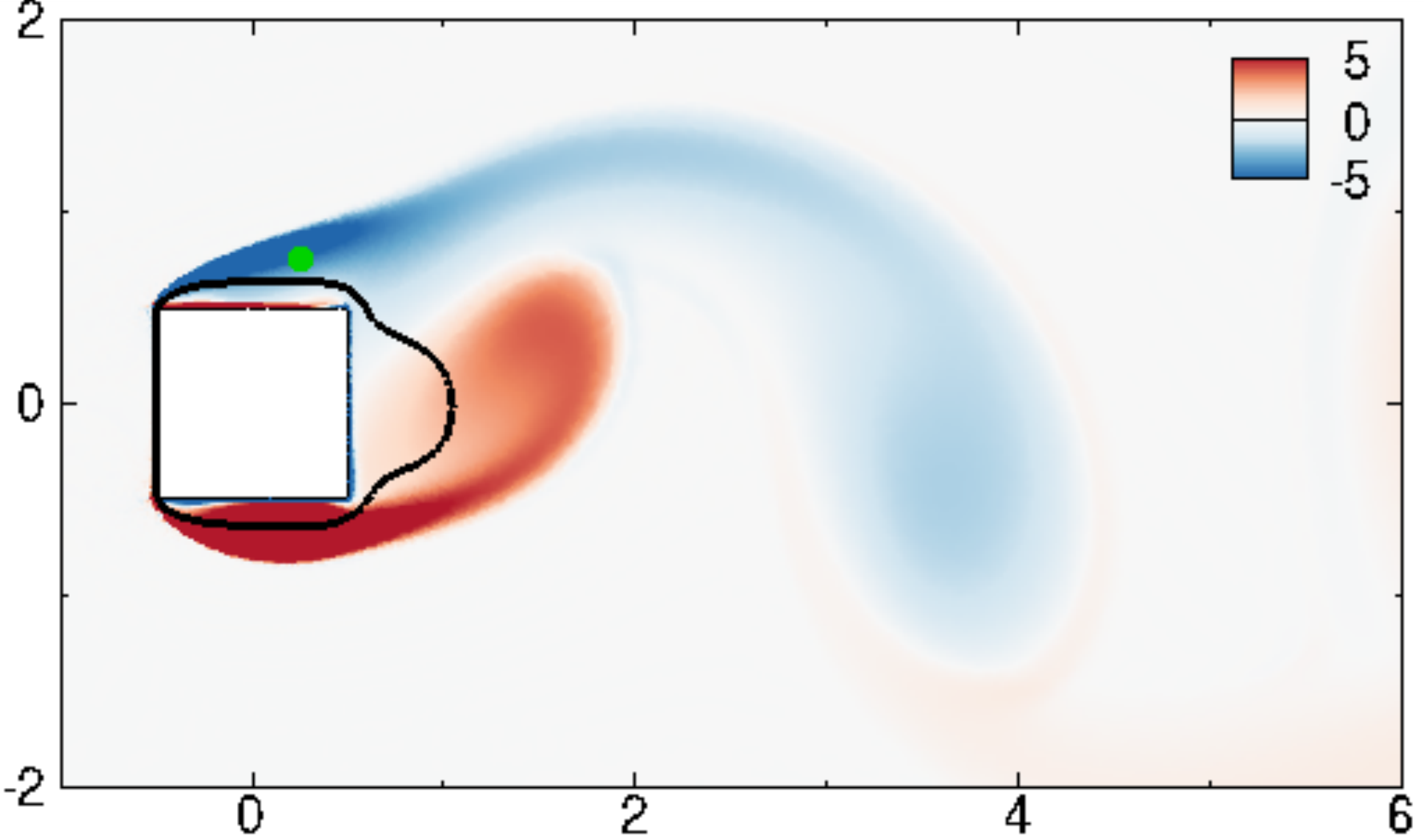}} & \imagetop{\includegraphics[width=0.34\columnwidth]{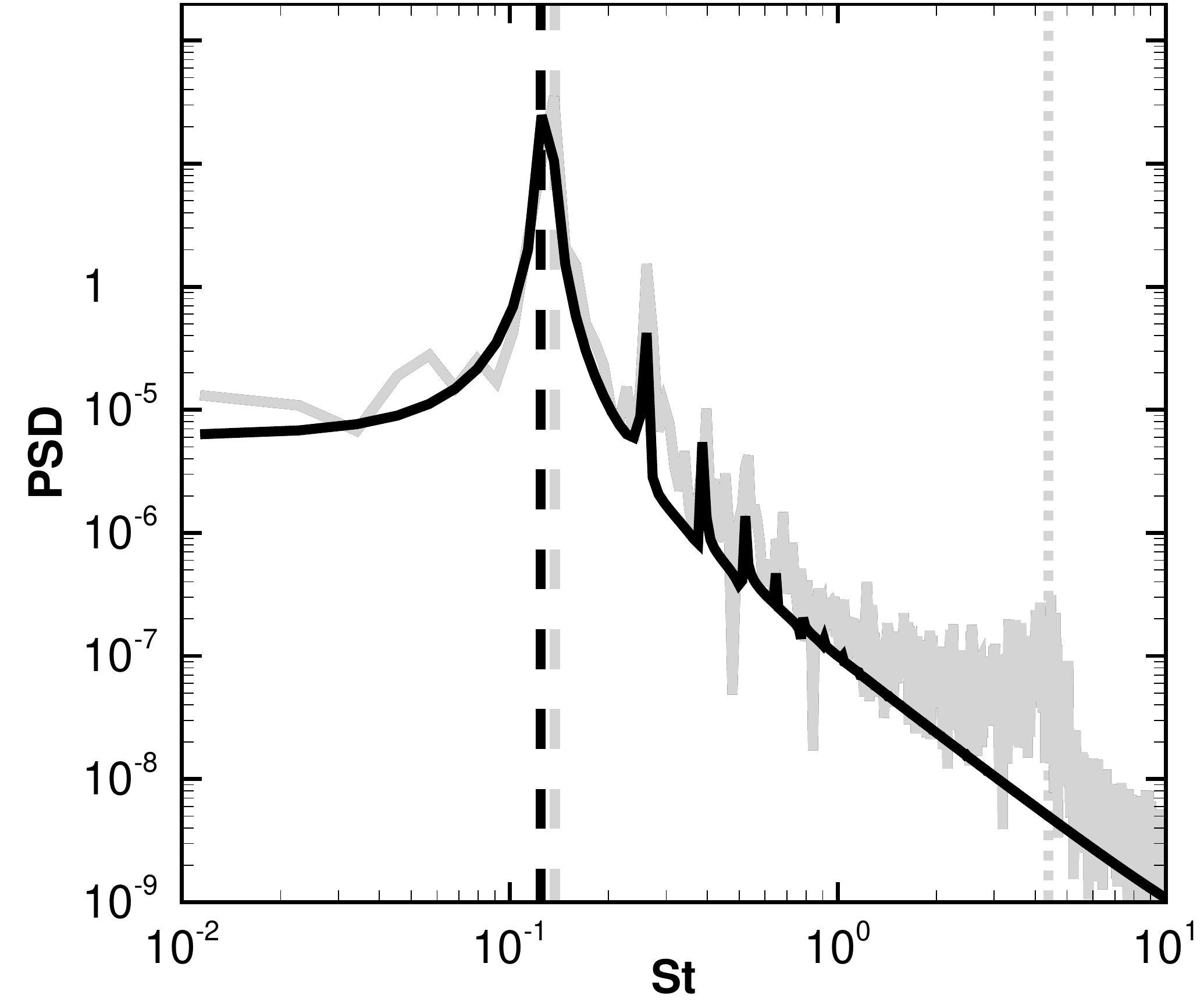}} 
\end{tabular}
  \caption{Results for (a-b) reference (DNS) and (c-d) estimated (standard URANS) flows. (a,c): Instantaneous spanwise vorticity field $\bar{\omega}_z$ at $t=50$, where black iso-curves denote $\left< \overline{u} \right> =0$. (b,d): Fourier spectrum of the streamwise velocity at the green monitor point in (a,c) (full black line). For the sake of comparison, the DNS spectrum is duplicated in (d) with the grey curve. Vertical dashed and dotted lines indicate the low-frequency peak associated with large-scale vortex-shedding ($St_{VS}=0.137$ for DNS while $St_{VS}=0.126$ for URANS) and the high-frequency bump linked to Kelvin-Helmholtz instabilities ($St_{KH}=4.384$ for DNS), respectively.
 }
\label{fig:reference}
\end{figure}


A preliminary URANS simulation is performed so as to obtain a periodic evolution of the aerodynamic coefficients. It allows to determine an instant which minimises the discrepancy between the lift coefficients computed from URANS and DNS simulations. The flow field at that instant is then used as initial condition $t=0$ for the URANS simulation that is discussed below. By doing so, the DNS and URANS solutions are initially phased in time and we can thus properly compare the temporal evolution of the error between the two.

Figure \ref{fig:reference} provides a first comparison of the flow extracted from DNS [figures \ref{fig:reference}(a-b)] and estimated by the URANS equations [figures \ref{fig:reference}(c-d)]. The instantaneous vorticity contours are shown in figures \ref{fig:reference}(a) and (c) for $t=50$.
In both cases, we observe that large-scale clockwise (blue) and counter-clockwise (red) vortices are shed in the wake of the square cylinder. Small-scale structures are also visible in the reference flow field shown in figure \ref{fig:reference}(a). 
In the upper and lower shear layers, emerging from the leading edge corners of the cylinder, they correspond to two-dimensional roll-up structures associated with Kelvin-Helmholtz instabilities. In the wake, they interact with the large-scale vortices of the von-Kármán vortex street.
Such small-scale structures are clearly absent in the estimated vorticity field that is shown in figure \ref{fig:reference}(c).
Large-scale and small-scales structures are associated with different parts of the Fourier spectra displayed in figures \ref{fig:reference}(b,d), which are obtained from the streamwise velocity $u$ monitored at a grid point located in the upper shear layer, at $(x,y)=(0.25,0.75)$ [green dot in figures \ref{fig:reference}(a,c)]. The large-scale periodic vortex shedding is associated to several peaks, one at the fundamental frequency $St_{VS}=U_{\infty} f_{VS} / D$ (indicated in both figures by the vertical dashed lines) and the others at multiples of the fundamental frequency, i.e. $k St_{VS}$ with $k\ge 2$. The fundamental frequency is $St_{VS} = 0.137$ for the DNS simulation and $St_{VS}= 0.126$ for the standard URANS simulation. The small-scale structures that are observed in the shear layers in the reference snapshot (figure \ref{fig:reference}(a)) correspond to a high-frequency broadband bump which is centered around $St_{KH}\approx4.384$, as indicated by the vertical dotted line in figure \ref{fig:reference}(b). This high-frequency bump is absent in the spectrum of the URANS signal, displayed with the black curve in figure \ref{fig:reference}(d). This confirms that the URANS simulation does not capture the emission of Kelvin-Helmholtz vortices in the shear layers.

\begin{figure}
\vspace{0.25cm}
\centering
\begin{tabular}[t]{lll}
(a) & (b) & (c) \\
\includegraphics[width=0.3\columnwidth]{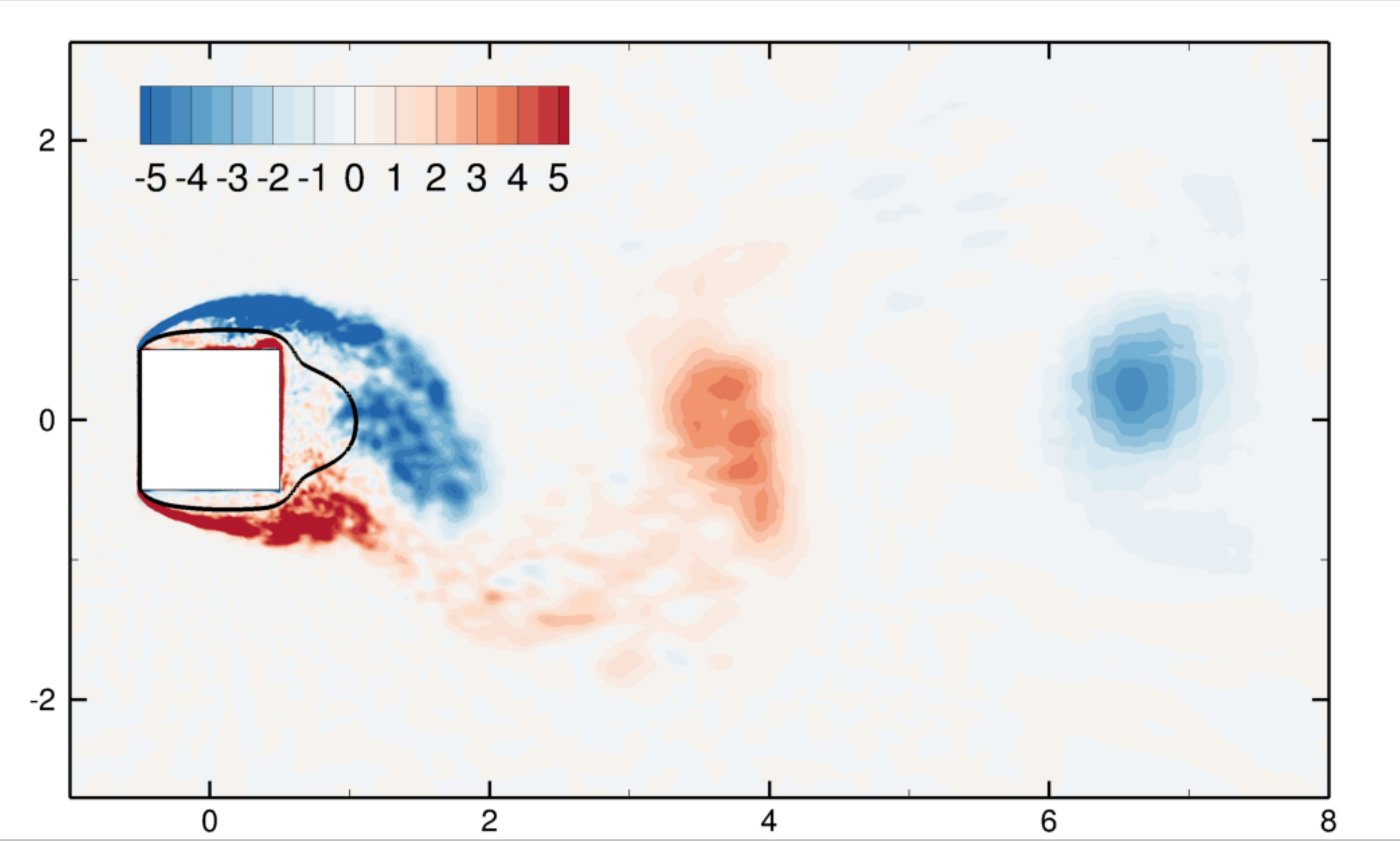} & \includegraphics[width=0.3\columnwidth]{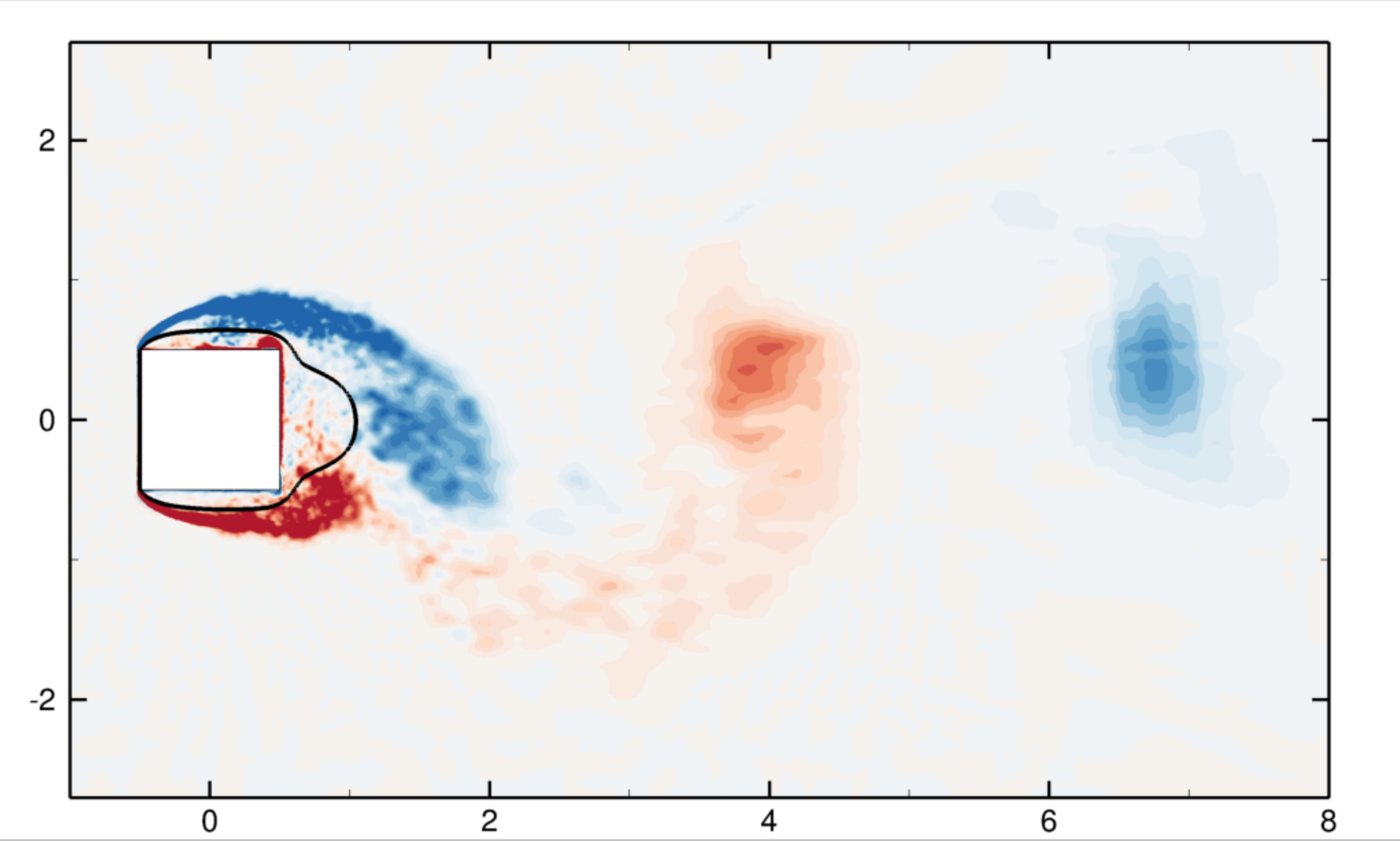} & \includegraphics[width=0.3\columnwidth]{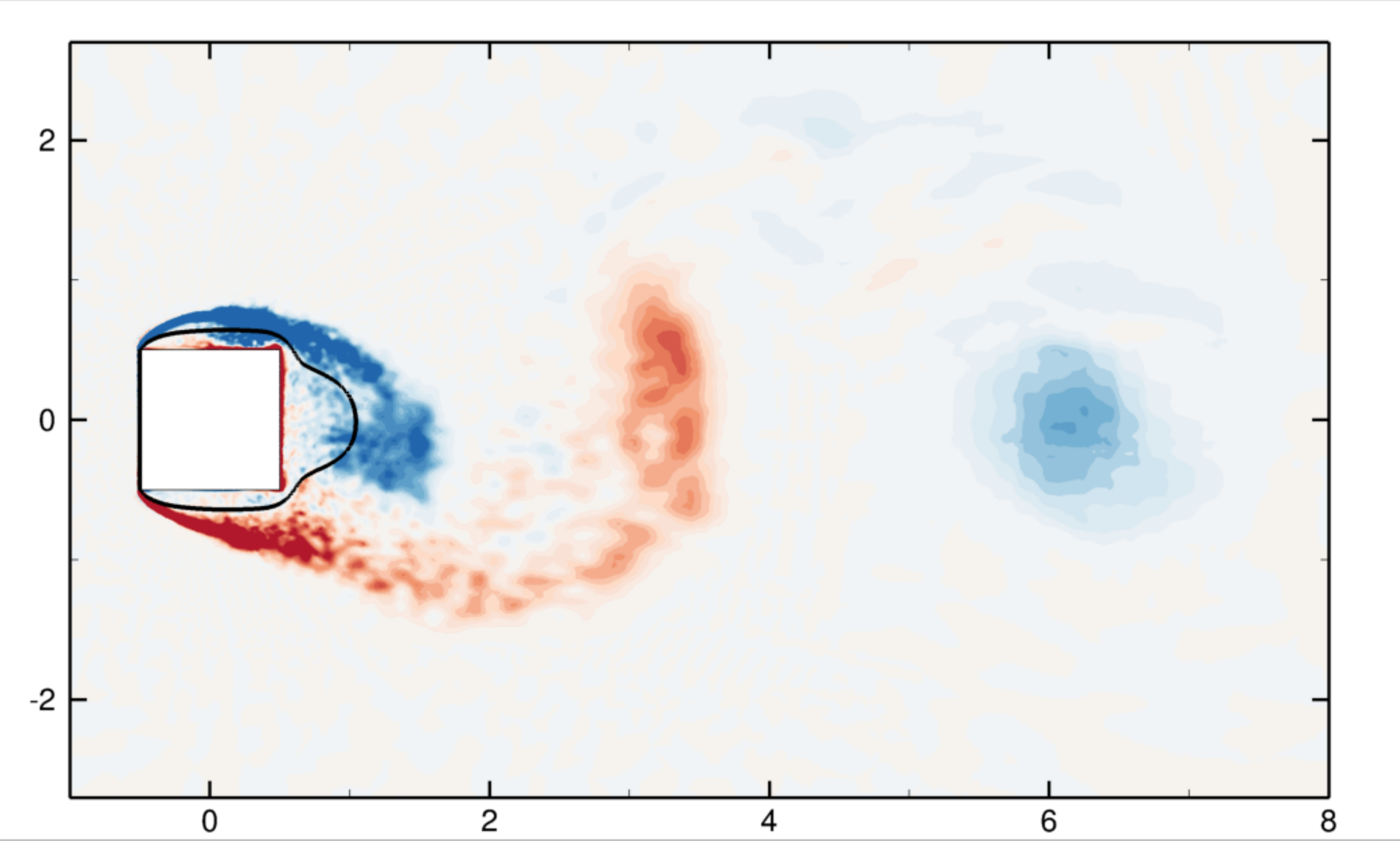} \\
(d) & (e) & (f) \\
\includegraphics[width=0.3\columnwidth]{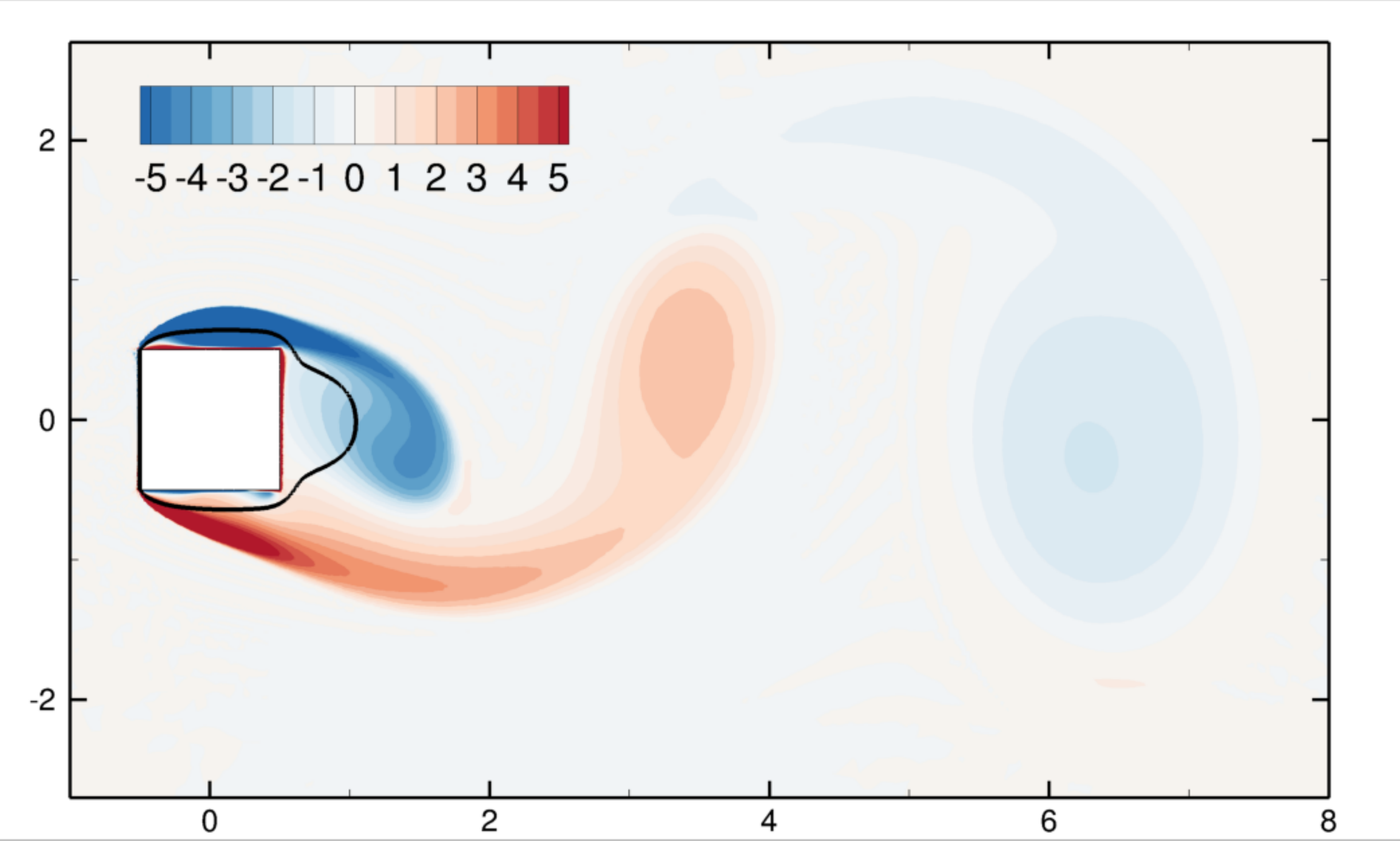} & \includegraphics[width=0.3\columnwidth]{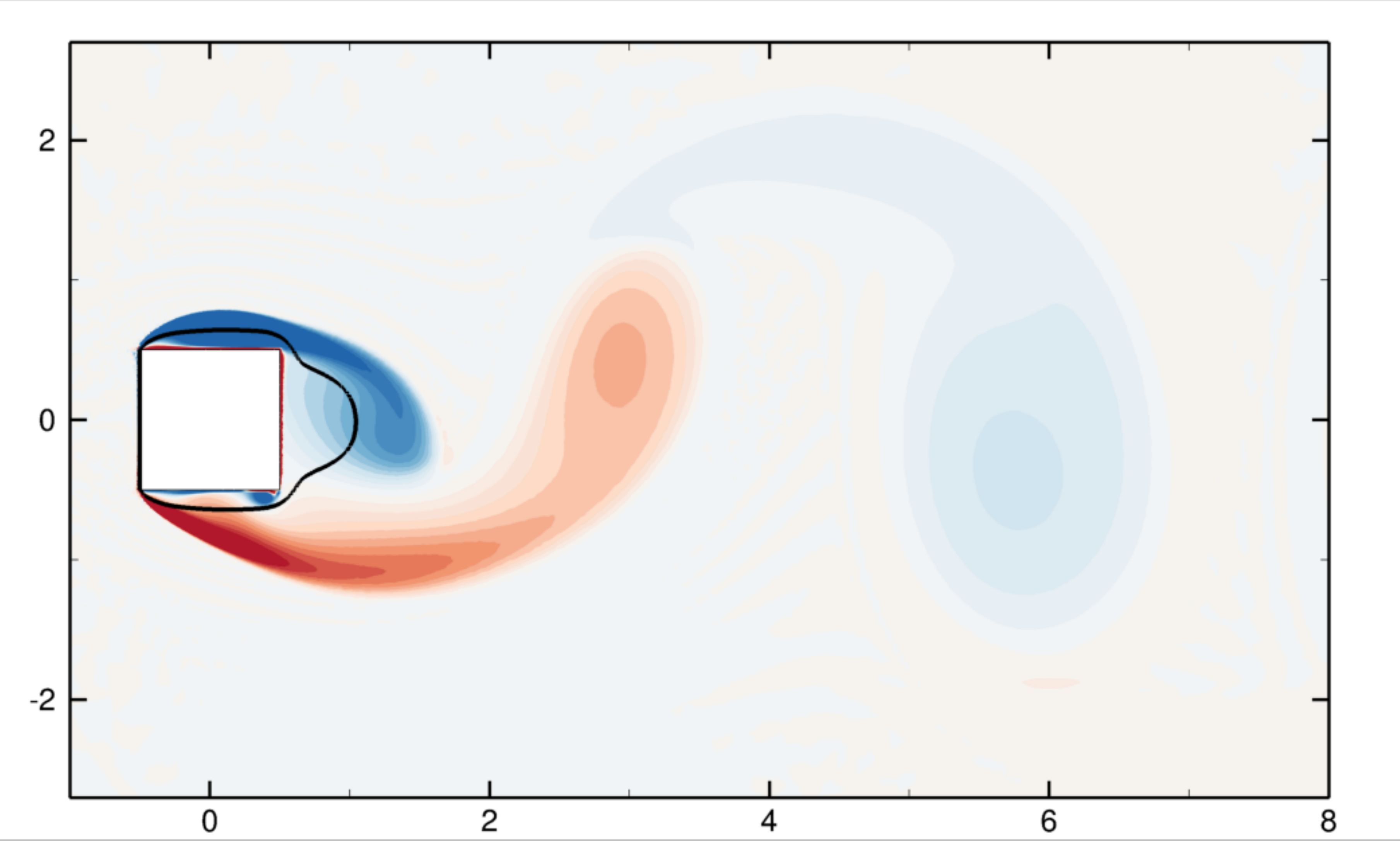} & \includegraphics[width=0.3\columnwidth]{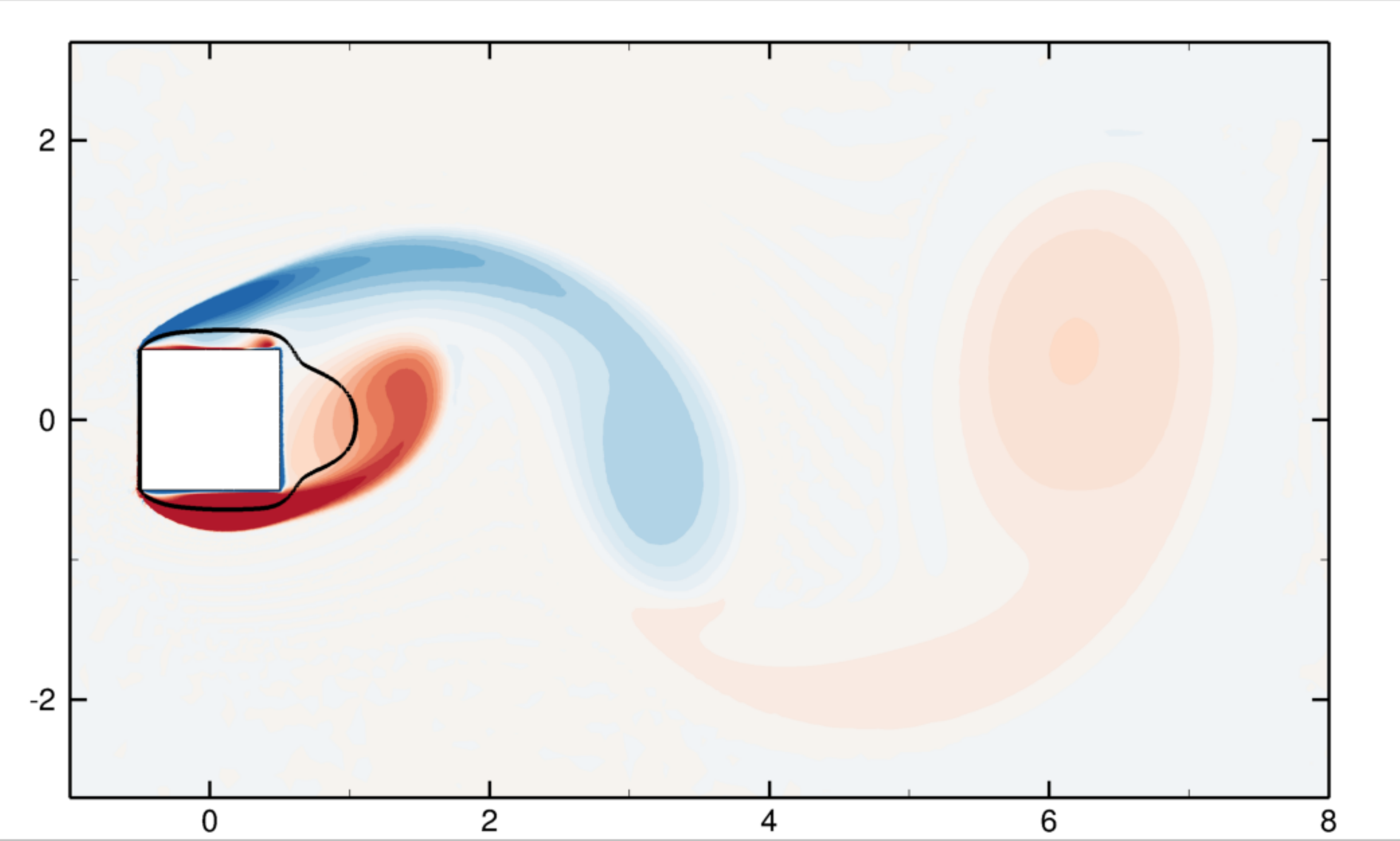} \\
(g) & (h) & (i) \\
\includegraphics[width=0.3\columnwidth]{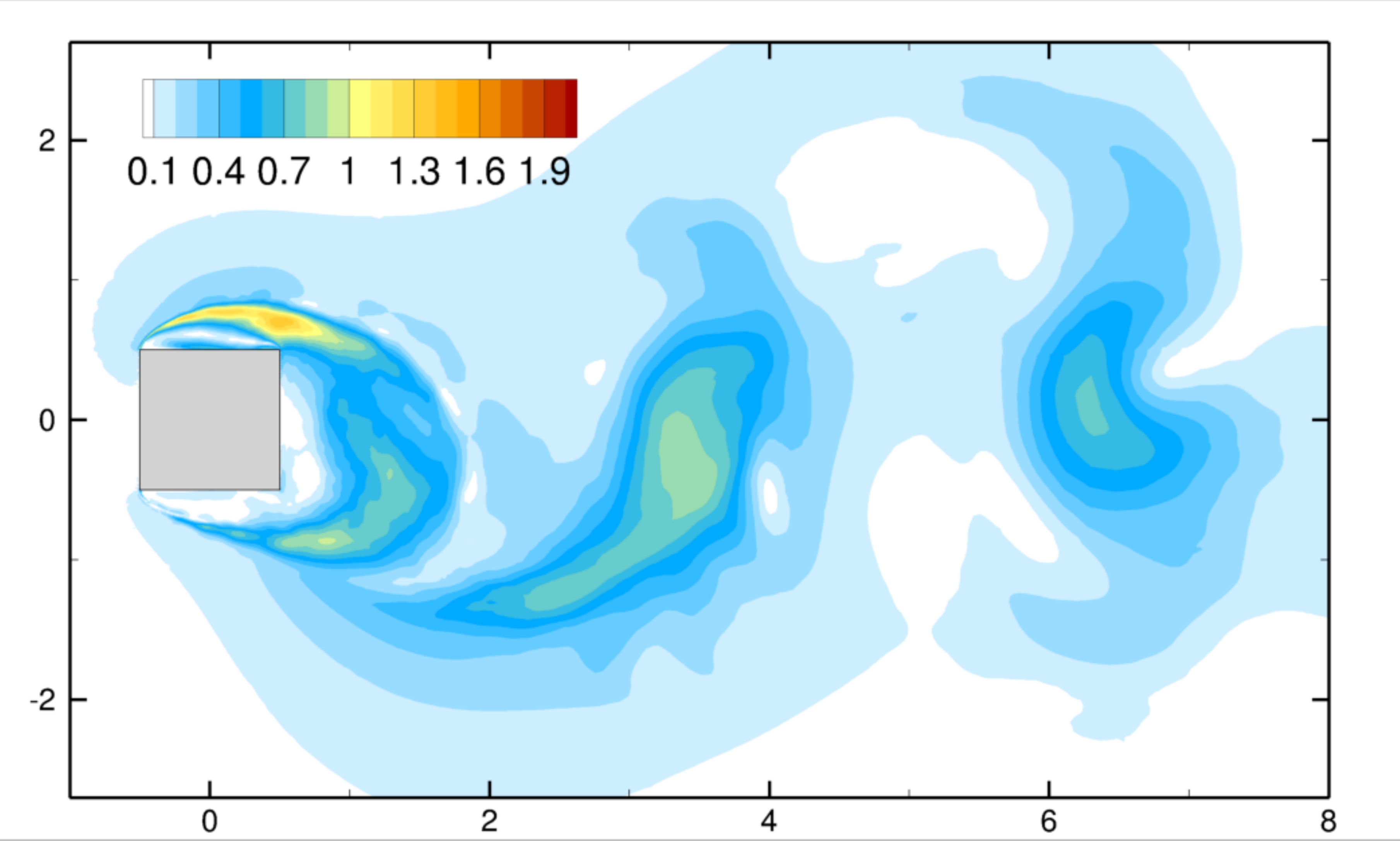} & \includegraphics[width=0.3\columnwidth]{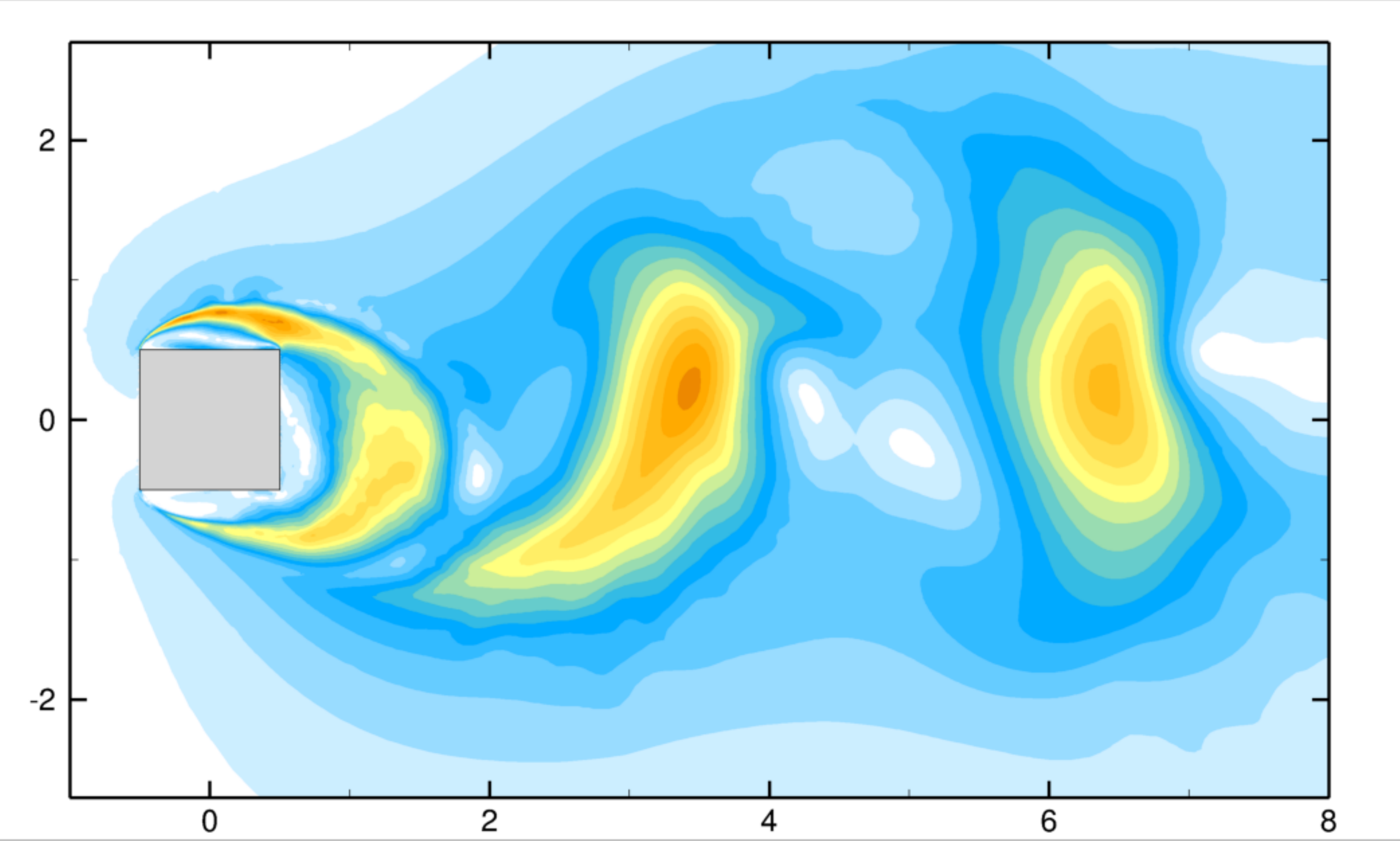} & \includegraphics[width=0.3\columnwidth]{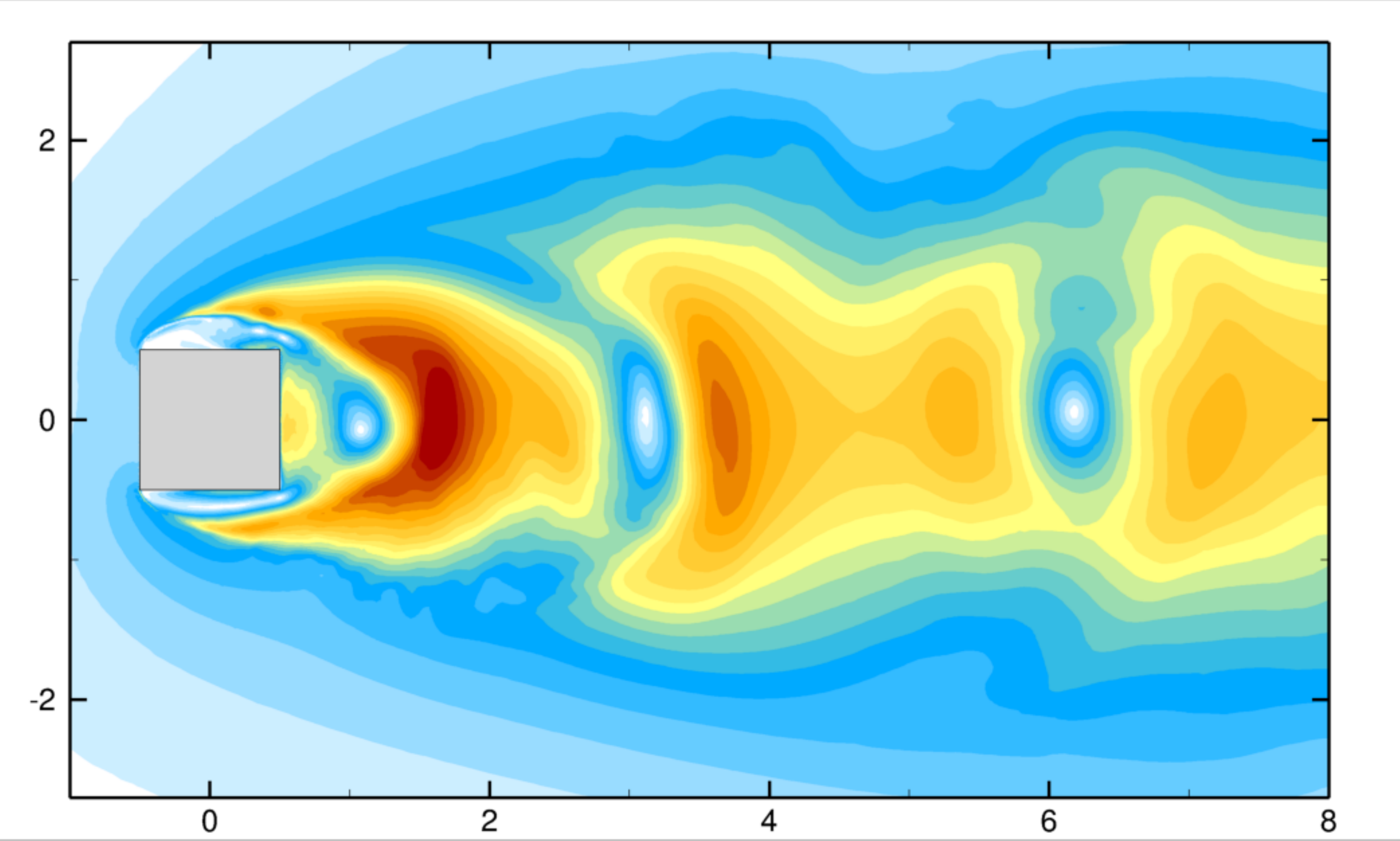} 
\end{tabular}
\begin{tabular}[t]{l}
(j) \\
\includegraphics[width=0.9\columnwidth]{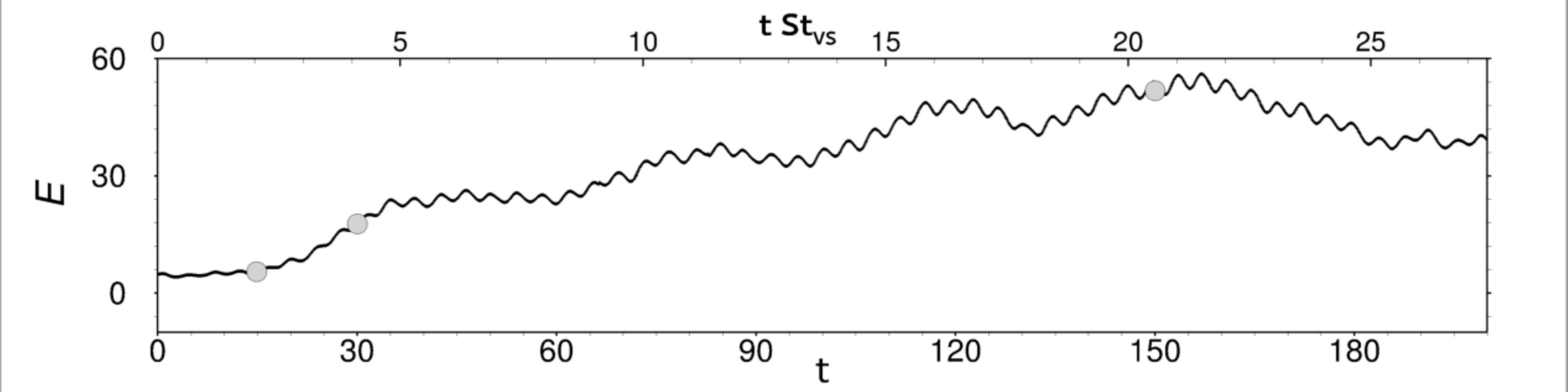}
\end{tabular}
 \caption{
Instantaneous vorticity fields of (a-c) the reference flow $\boldsymbol{\bar{u}}_{r}$ (DNS) and (d-f) estimated flow $\boldsymbol{\bar{u}}$ (standard URANS) at times (a,d) $t=15$, (b,e) $t=30$ and (c,f) $t=150$. (g-i) Instantaneous error fields $e(\boldsymbol{x},t)$ between the reference and estimated flow (definition in eq. \ref{eqn:instantaneous-error}) at the corresponding instants. (j) Temporal evolution of the global error $E(t)$ (see eq. \ref{eqn:total-error}) where the bottom axis reports the (nondimensional) time while the top axis reports the number of low-frequency cycles determined as $t/\tau_{VS}=t \ St_{VS}$ with $St_{VS} = 0.137$. The grey circles indicates time $t=15$, $30$ and $150$.}
\label{fig:unsynchronisation}
\end{figure}

Although the URANS simulation well approximates the periodic vortex-shedding phenomenon, the difference between Strouhal numbers noticed above is responsible for desynchronisation of the wake flow as illustrated in figure \ref{fig:unsynchronisation}. 
The instantaneous vorticity fields displayed in figures \ref{fig:unsynchronisation}(a-c) and (d-f) for DNS and URANS simulations, respectively, seem in phase after $2$ vortex-shedding periods (a,d), but, after $4$ shedding cycles (b,e), large vortex structures observed in URANS snapshots clearly lag behind corresponding structures in DNS snapshots. After $19.5$ periods (c,f), the two solutions are out-of-phase by a half-wavelength.
This desynchronisation of the large-scale structures results in large values of the instantaneous local error in the wake of the cylinder, as illustrated by figures \ref{fig:unsynchronisation}(g-i). The temporal evolution of the instantaneous global error $E(t)$ in (\ref{eqn:total-error}) is reported in figure \ref{fig:unsynchronisation}(j) from $t=0$ to $t=200$, which corresponds to a time window of $26$ vortex-shedding periods. The error $E(t)$ increases during the first $20$ periods ($0 \le  t \le 150)$, in agreement with the previous observations, before slightly decreasing. This effect is still related to the lack of synchronisation between reference and estimated flows. 


\begin{figure}
\vspace{0.25cm}
\begin{tabular}[t]{@{}l@{}l@{}l}
(a)  & (b) & (c)  \\
\imagetop{\includegraphics[width=0.33\columnwidth]{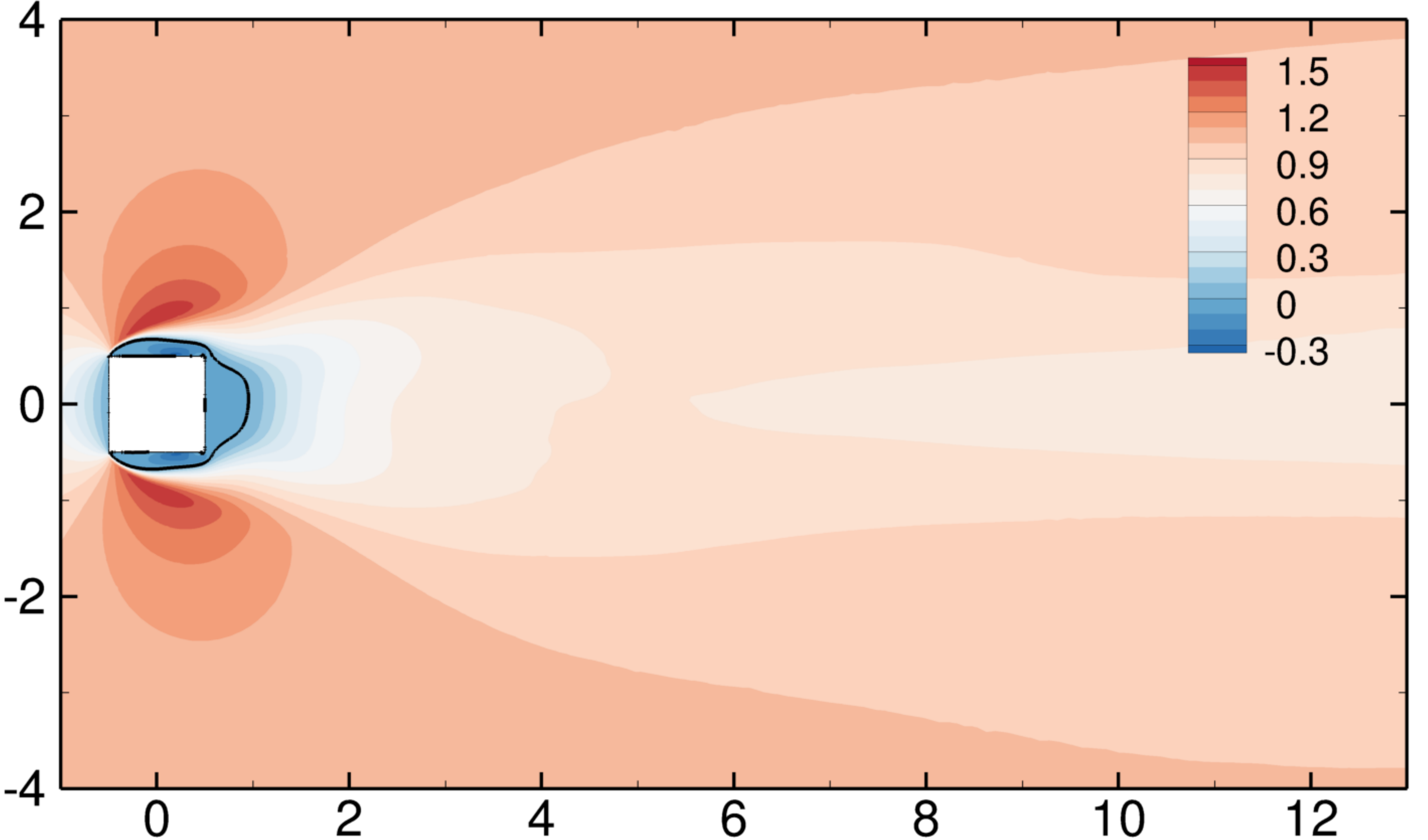}} & \imagetop{\includegraphics[width=0.33\columnwidth]{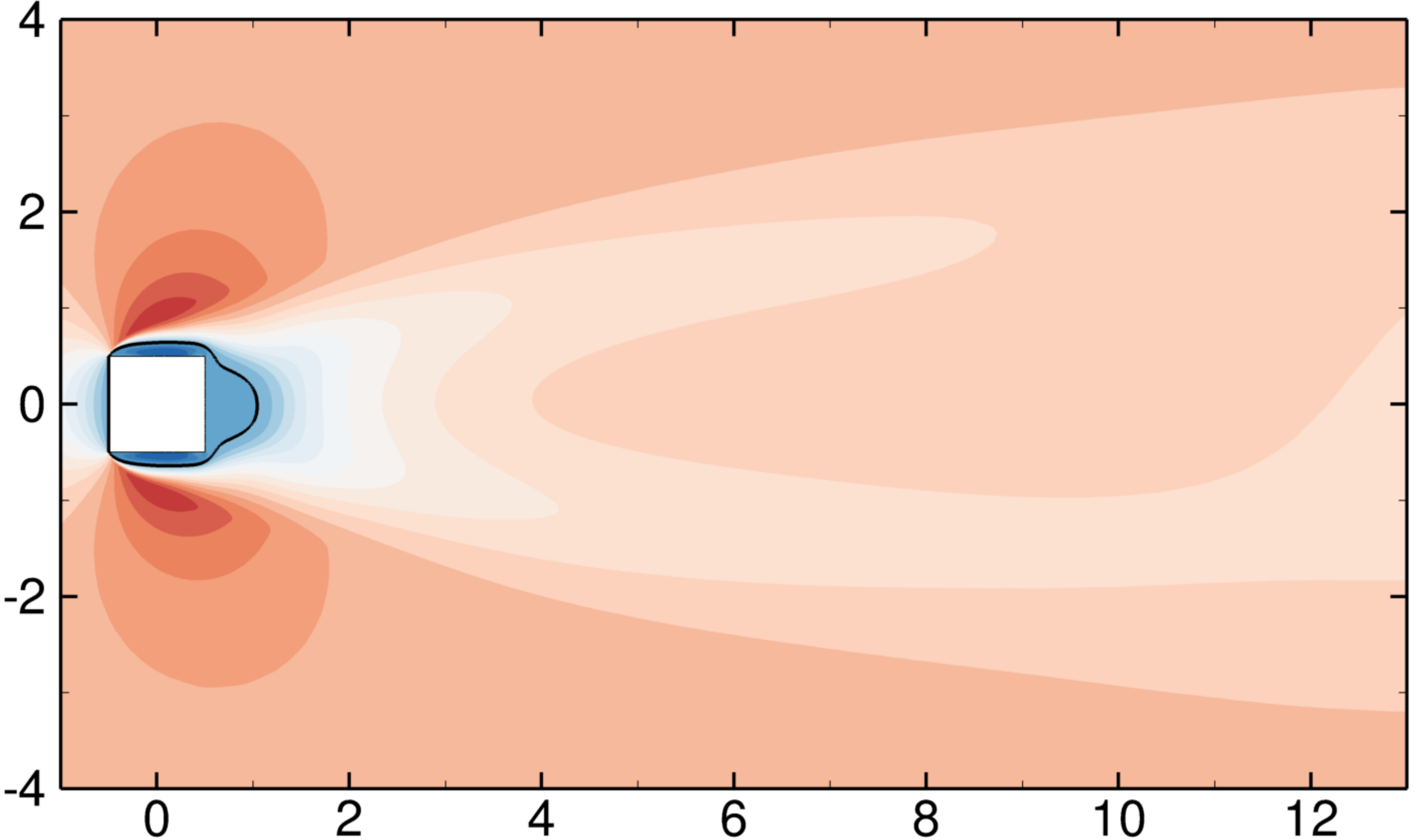}} & \imagetop{\includegraphics[width=0.33\columnwidth]{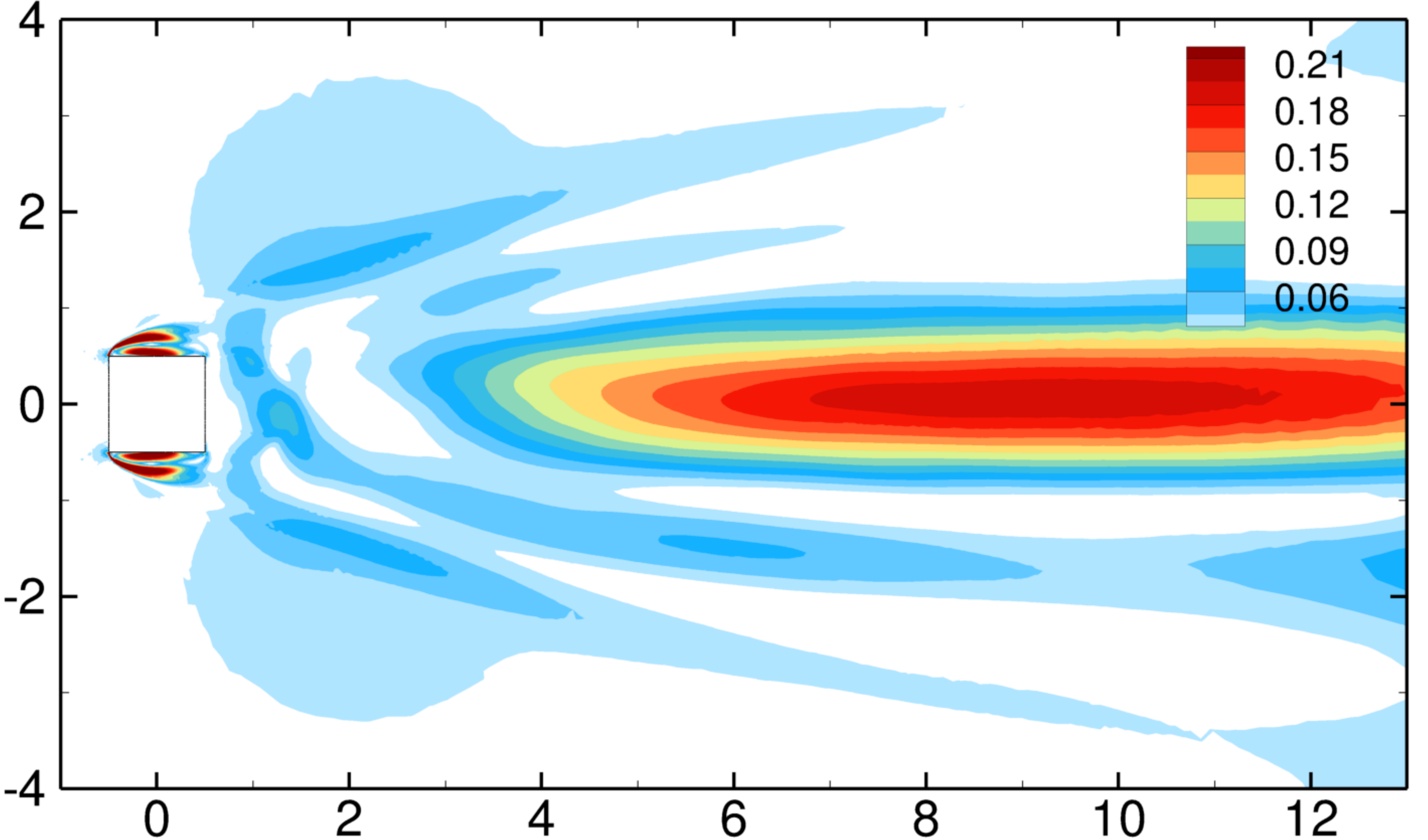}}
\end{tabular}
  \caption{(a-b) Streamwise component of the time-averaged (a) reference flow $\langle \boldsymbol{\bar{u}}_{r} \rangle$ (DNS) and (b) estimated flow $\langle \boldsymbol{\bar{u}} \rangle$ (standard URANS). (c) Associated discrepancy field $e_{\langle \bar{\boldsymbol{u}} \rangle}$ as defined in (\ref{eqn:total-meanerror}). In (a-b), black curves correspond to contours of zero streamwise time-averaged velocity, which delineate the recirculation region.}
\label{fig:MeanField}
\end{figure}

The statistical and dynamic discrepancies between estimated and reference flows are further examined in the following, and we first consider the time average of these flows. The streamwise velocity of the time-averaged reference (DNS) and estimated (URANS) flows are depicted in figures \ref{fig:MeanField}(a) and \ref{fig:MeanField}(b), respectively, while the associated discrepancy field $e_{\langle \bar{\boldsymbol{u}} \rangle}(\boldsymbol{x})$ in (\ref{eqn:total-meanerror}) is displayed in figure \ref{fig:MeanField}(c). Interestingly, the mean flows agree qualitatively well, and the error field $e_{\langle \bar{\boldsymbol{u}} \rangle}(\boldsymbol{x})$ reaches overall significantly lower values than the discrepancy field $e(\boldsymbol{x},t)$ for the instantaneous flow in figure \ref{fig:unsynchronisation}(c). 
Largest errors are observed in the shear layers that emerge from the cylinder upstream corners. In the near wake, the recirculation region, which is delineated by black curves, is slightly larger for the estimated flow ($L=0.55$) compared to that of the reference flow ($L=0.45$). Substantial errors are also visible in the far wake where streamwise velocity is overestimated by URANS.

The SPOD decomposition of the reference and estimated flows are both performed by considering as input $8420$ snapshots that are sampled at $\Delta t_{r}=0.021$ in a time interval of around $176$ convective times. Using the Matlab implementation provided by \cite{Towne2018}, this series is typically divided in $3$ overlapping bins, each bin containing about $4200$ instantaneous snapshots, and overlapping by $50\%$ of their size. The time interval of a bin is thus around $88$ convective time units.
\begin{figure}
\vspace{0.25cm}
\centering
\begin{tabular}[t]{lcl}
(a) & & (b) \\
\imagetop{\includegraphics[width=0.45\columnwidth]{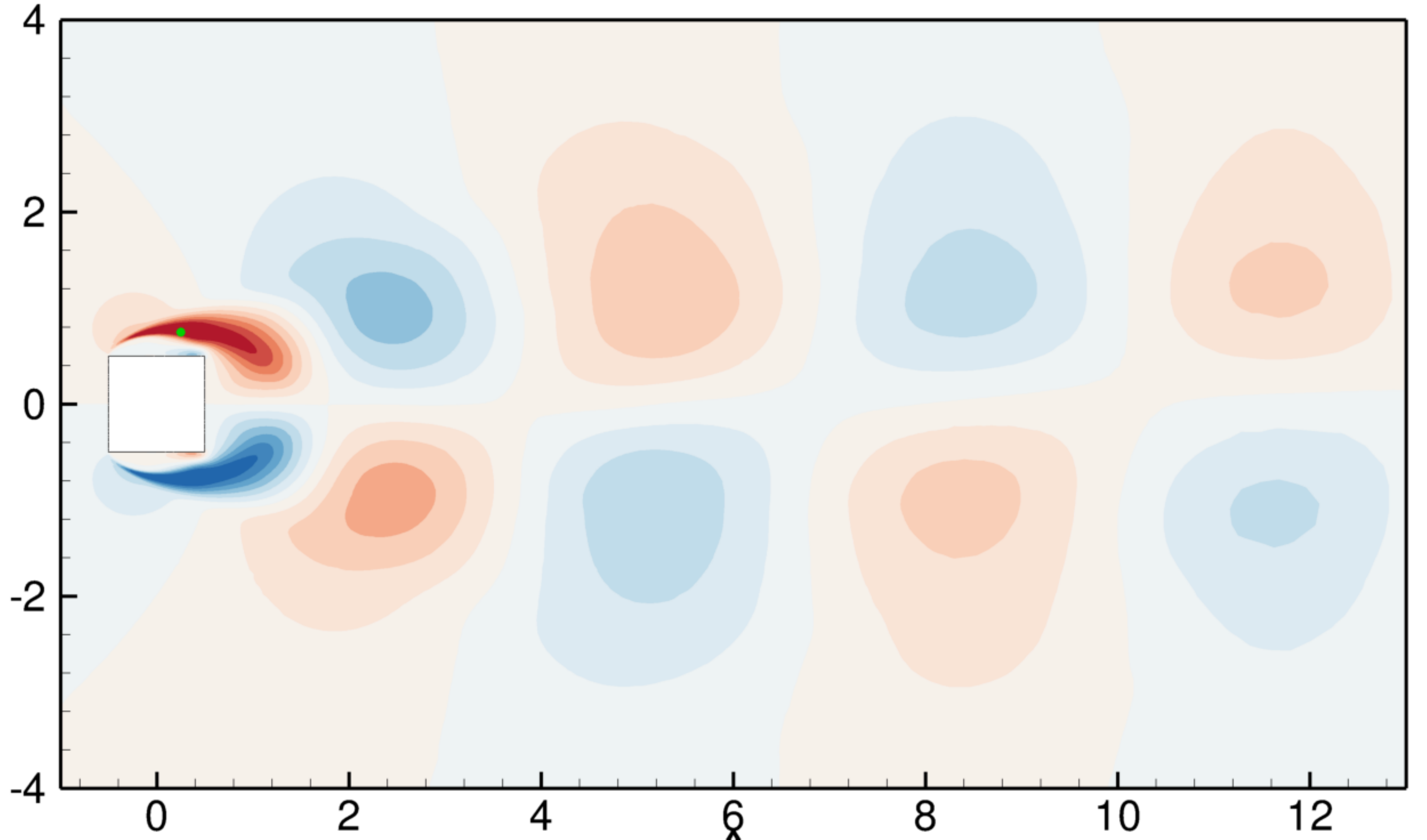}} &  & \imagetop{\includegraphics[width=0.45\columnwidth]{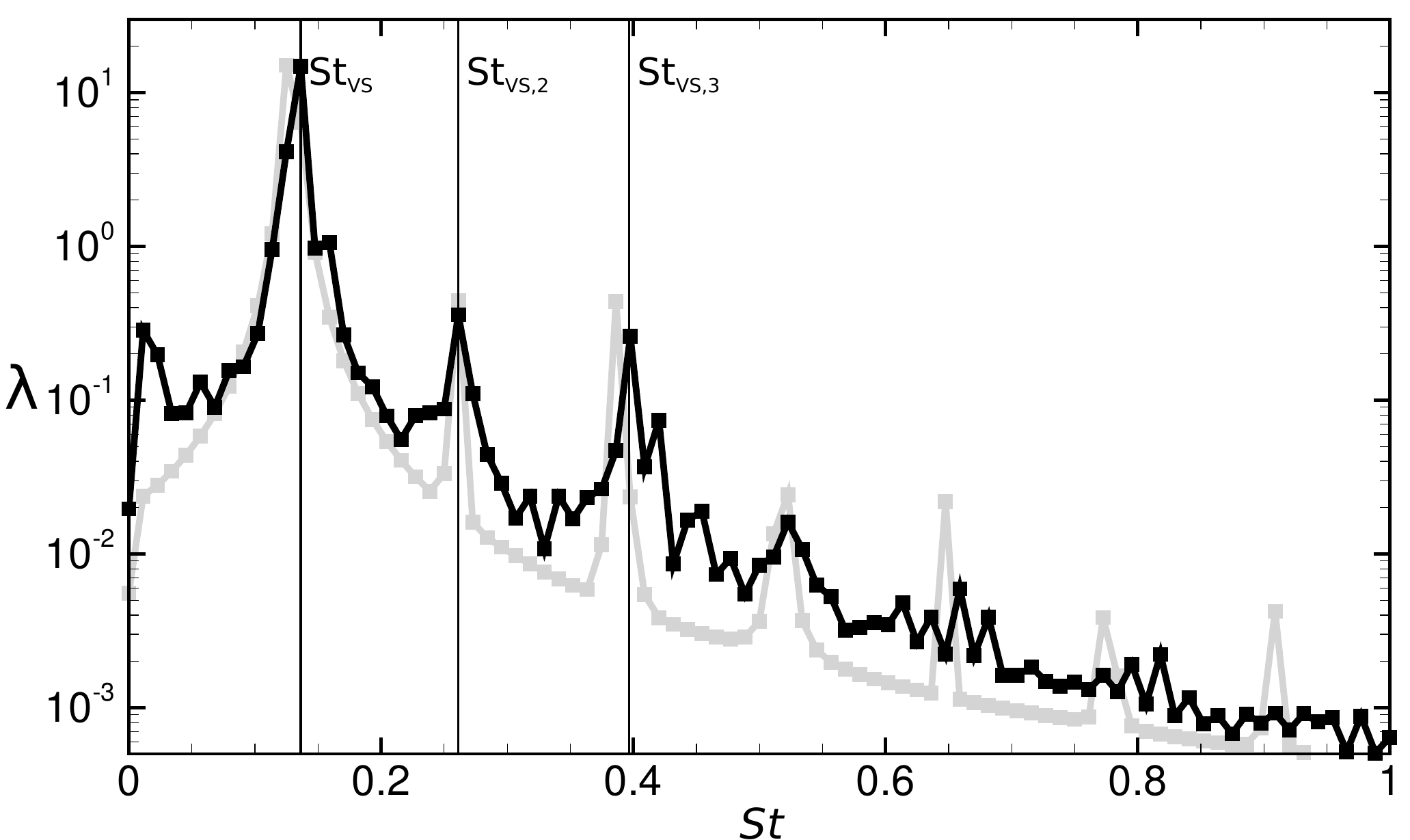}} \\
(c) & & (d) \\
\imagetop{\includegraphics[width=0.45\columnwidth]{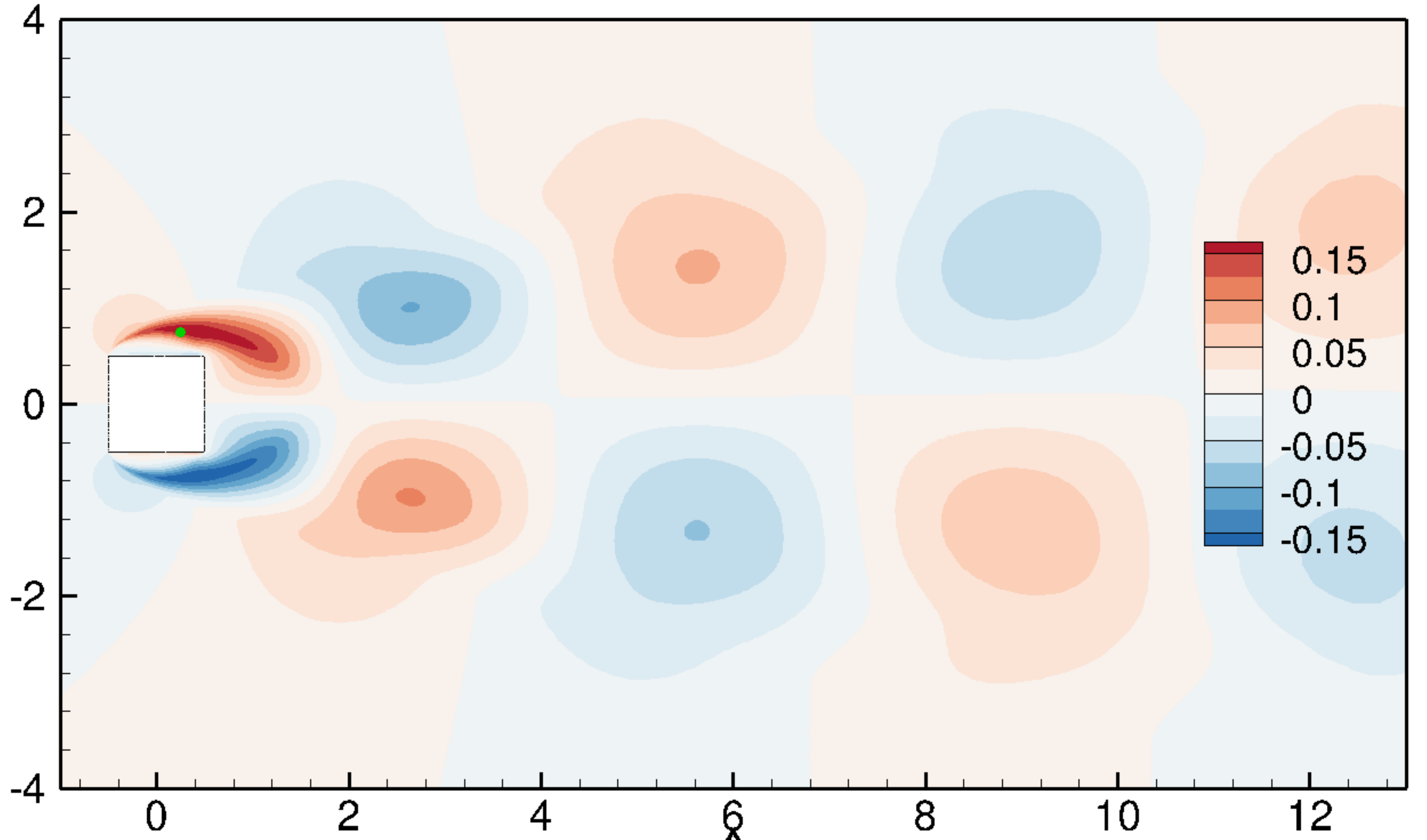}} &  & \imagetop{\includegraphics[width=0.45\columnwidth]{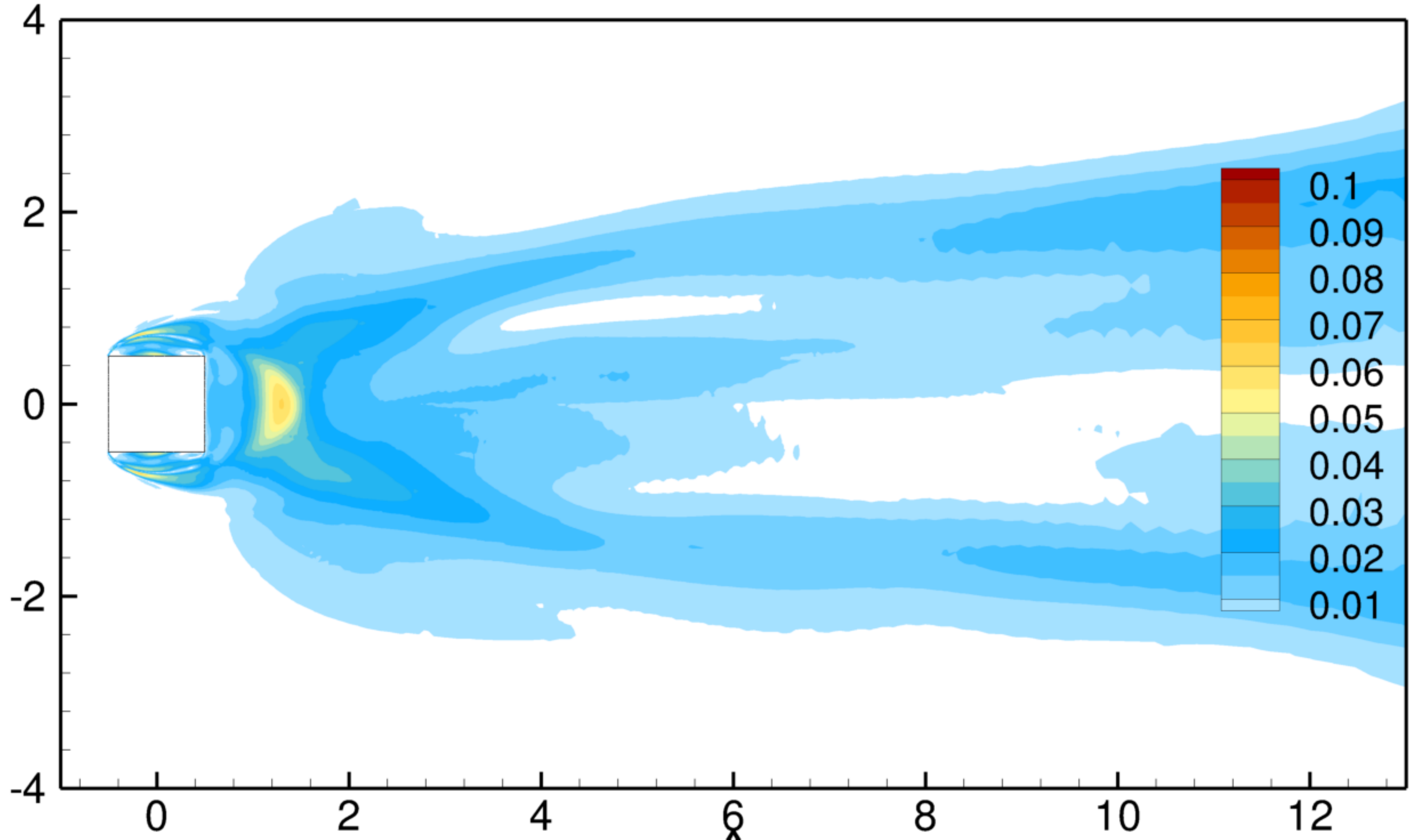}} 
\end{tabular}
  \caption{SPOD analysis of the vortex-shedding phenomenon in the reference (DNS) and estimated (standard URANS) flows: streamwise velocity contours of the imaginary part of the dominant (a) DNS and (c) URANS SPOD modes for the fundamental frequency $St_{VS}$.
  Modes are phased so that each of them reaches its maximum value at the green monitor point.
  (b) Spectra showing the largest eigenvalue $\lambda$ as a function of the frequency $St$ in the low-frequency range $0 \le St \le 1$ for DNS (black curve) and URANS (grey curve). The vertical lines indicate low-frequency peaks at $St_{VS}=0.137$ and its harmonics for DNS results. 
  The modal discrepancy field $\Phi_e$ in (\ref{eqn:modal-error}) between the modes in (a) and (c) is reported in (d).}
\label{fig:SPOD_Spec_DNS_VS}
\end{figure}

\begin{figure}
\vspace{0.25cm}
\centering
\begin{tabular}[t]{lcl}
(a) & & (b) \\
\imagetop{\includegraphics[width=0.42\columnwidth]{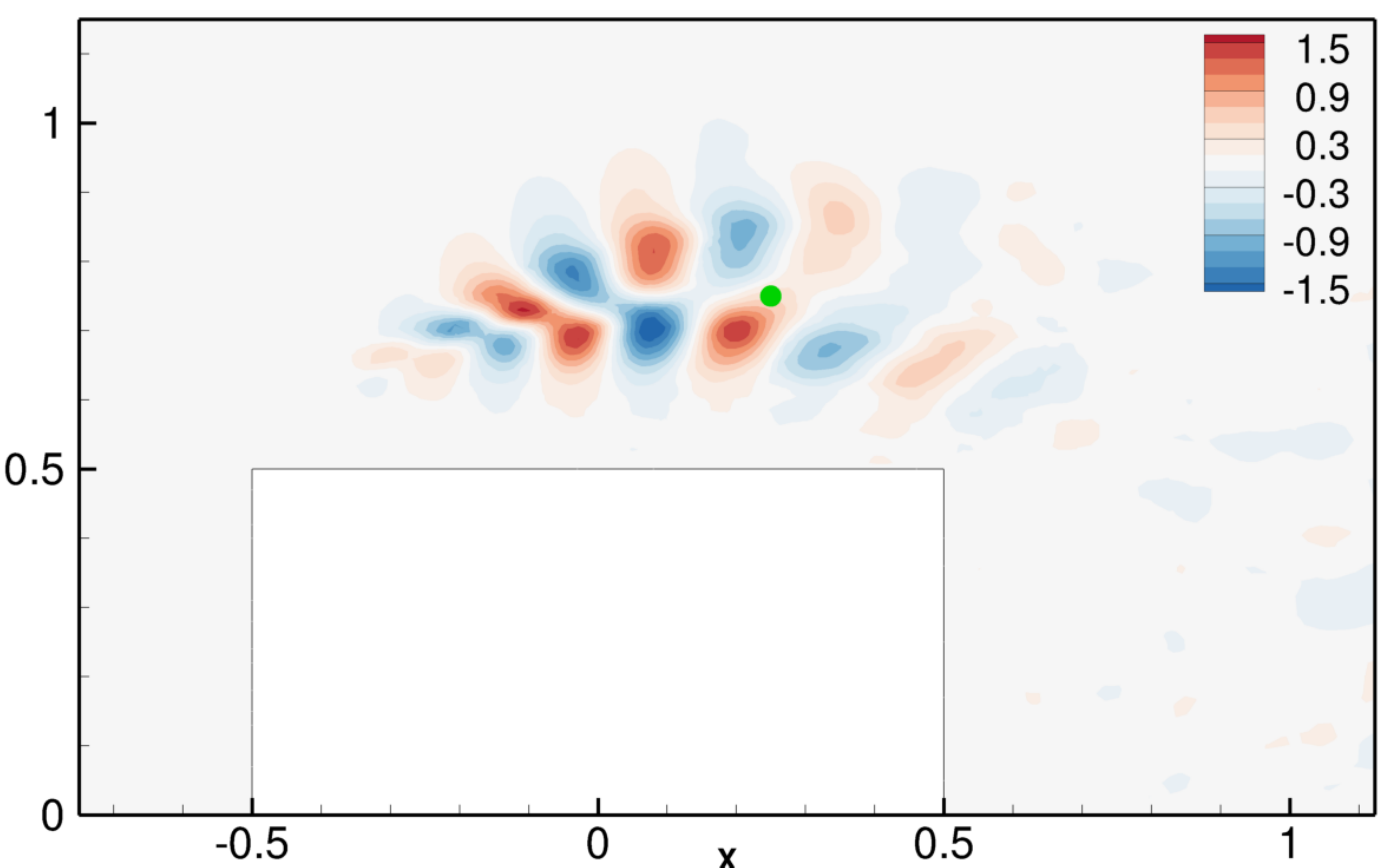}} &  & \imagetop{\includegraphics[width=0.44\columnwidth]{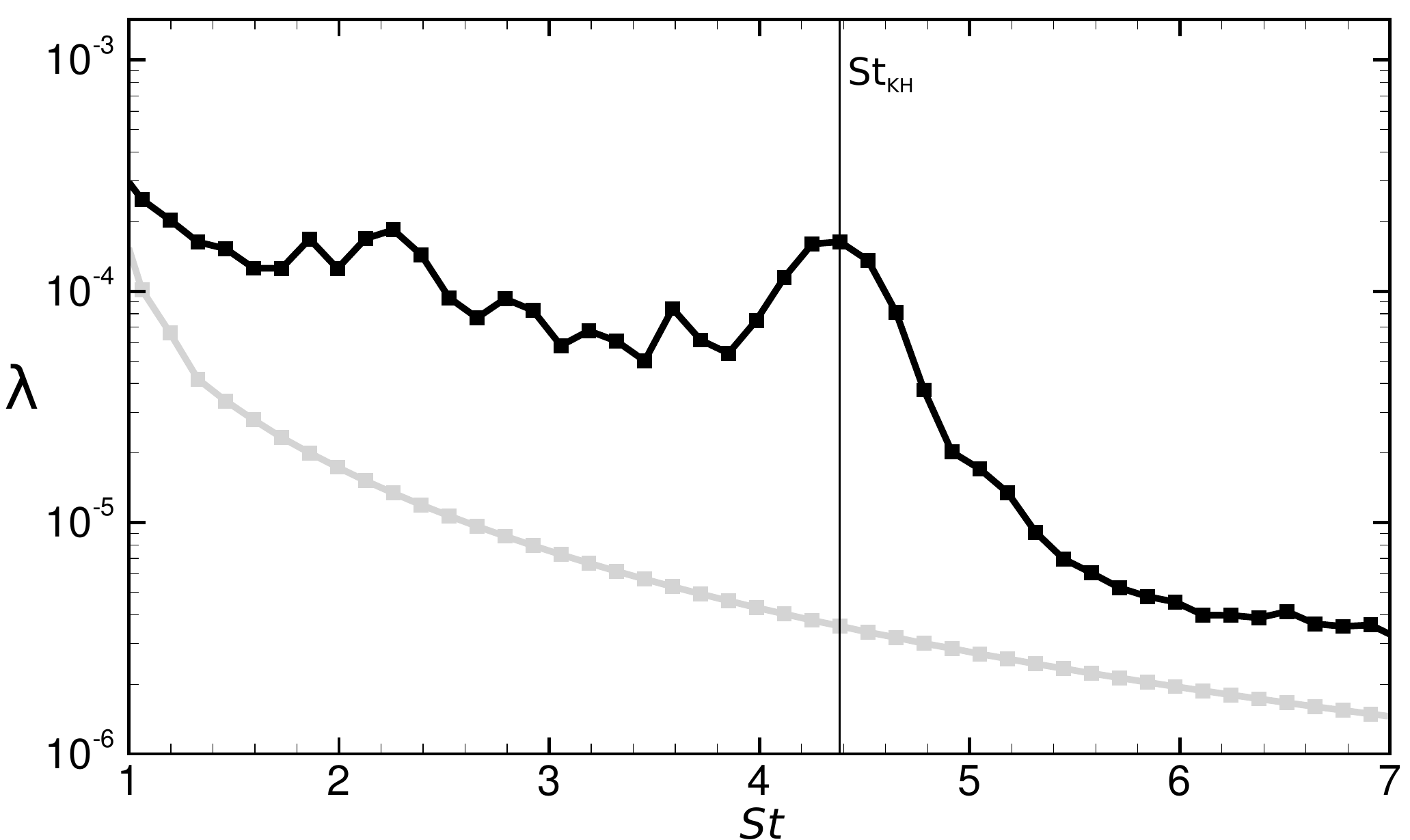}}
\end{tabular}
  \caption{
  (a) Contours of streamwise velocity fluctuations are shown for the real part of the dominant SPOD mode of DNS that is associated to Kelvin-Helmholtz instabilities ($St_{KH}=4.384$, only captured with DNS).
  (b) Spectra showing the largest eigenvalue $\lambda$ as a function of the frequency $St$ in the high-frequency range $1 \le St \le 7$. Black and light grey curves correspond to DNS and URANS results, respectively.
}
\label{fig:SPOD_Spec_DNS_KH}
\end{figure}

The SPOD analysis of the vortex-shedding phenomenon as predicted by DNS and standard URANS is first illustrated in figure \ref{fig:SPOD_Spec_DNS_VS}.
Spectra of the most-energetic SPOD modes for a low-frequency range ($0 \le St \le 1$) as functions of Strouhal number $St$ are reported in figure \ref{fig:SPOD_Spec_DNS_VS}(b).
DNS and URANS results are denoted by black and light grey dots, respectively. As in the Fourier analysis of figure \ref{fig:reference}, several peaks are observed at the fundamental frequency $St_{VS}=0.137$ of the vortex-shedding phenomenon and at frequencies corresponding to second and third harmonics at $St_{VS,2}=0.261$ and $St_{VS,2}=0.398$, respectively.  
Frequencies of URANS modes (grey peaks) are slightly lower, peaking at $St_{VS}=0.126$, $St_{VS,2}=0.261$, and $St_{VS,2}=0.386$.
Streamwise velocity contours of the imaginary part of the dominant DNS and URANS SPOD modes for the fundamental frequency $St_{VS}$ are displayed in figures \ref{fig:SPOD_Spec_DNS_VS}(a) and \ref{fig:SPOD_Spec_DNS_VS}(c), respectively. Both modes are phased so that they reach their maximum amplitude at the monitor point indicated by the green dot, which is the same as in figure \ref{fig:reference}. Their spatial structure is typical of modes corresponding to the alternate shedding of vortices in the wake, with a streamwise wavelength equal to $\sim 4$ cylinder lengths. Furthermore, we show in figure  \ref{fig:SPOD_Spec_DNS_VS}(d) the so-called modal error field $\Phi_e(\boldsymbol{x})$ which is defined in equation \ref{eqn:modal-error}. Dominant errors are observed in boundary and shear layers, as well as in the near-wake region, which may be related to the slightly longer recirculation region predicted by URANS.  

Figure \ref{fig:SPOD_Spec_DNS_KH}(b) shows the spectrum of the most dominant SPOD modes corresponding to high-frequency fluctuations. A broadband peak centered around $St_{KH}=4.384$ is clearly visible in the DNS spectrum (black) but not in the URANS one (grey). In order to prevent modes from being corrupted by numerical oscillations in regions of decreased grid resolution, we limit the domain that is considered for SPOD targeting Kelvin-Helmholtz phenomena to $-1.5<x<1.5$ and $-1.5<y<1.5$. Furthermore, we consider $1440$ snapshots at a sampling rate of $\Delta t_{r}=0.021$ in a time interval of around $30$ convective times using $6$ overlapping bins, each containing about $33$ Kelvin-Helmholtz cycles.
In order to better visualise convective small-scale instabilities originating from the front edge of the cylinder, a magnification of the upper-side shear layer in figure \ref{fig:SPOD_Spec_DNS_KH}(a) shows the real part of streamwise velocity contours for the dominant SPOD mode from DNS results at $St_{KH}=4.384$. Typical wavelengths of these structures are around $0.25$, i.e. a quarter of the cylinder's length. Similar spatial structures are also observed on the lower side of the cylinder, but of much weaker magnitude. As expected when considering the spatial symmetry of the time-averaged flow, a second SPOD mode with spatial structures predominant on the lower-side shear layer is also obtained but associated to a slightly lower kinetic energy.




The results of the SPOD analysis are summarized in table \ref{tab:DNSvsURANS}, which confirms a rather good agreement between DNS and standard URANS in terms of the prediction of low-frequency vortex-shedding structures. Interestingly, a small discrepancy in the estimation of the fundamental frequency $St_{VS}$ (about $10\%$) leads to large temporal errors due to desynchronisation effects. It may thus be anticipated that adding sparse information to the low-frequency content of the estimated flow may significantly reduce the corresponding temporal error. In the next section, we will use the nudging approach to improve the prediction of the large-scale structures oscillating at low frequency, but also to recover the small-scales Kelvin-Helmholtz structures, which are not captured with standard URANS simulations. 

\begin{table}
  \begin{center}
\def~{\hphantom{0}}
  \begin{tabular}{r|cc|cc|cc|cc|cc}
& $\Delta C_L$ & $\overline{C_D}$ & $St_{VS}$ & $\lambda_{VS}$ & $St_{VS,2}$ & $\lambda_{VS,2}$ & $St_{VS,3}$ & $\lambda_{VS,3}$ & $St_{KH}$ & $\lambda_{KH}$ \\ 
   \midrule
DNS & $4.683$ & $2.091$ & $0.137$ & $14.847$ & $0.261$ & $0.359$ & $0.398$ & $0.261$ & $4.384$ & $0.000164$ \\ 
standard URANS & $4.198$ & $2.017$ & $0.126$ & $15.082$ & $0.261$ & $0.442$ & $0.386$ & $0.440$ & $-$ & $-$  
     \end{tabular}
  \caption{Time-averaged and unsteady characteristics of the flow obtained with DNS and standard URANS simulations. The first two columns report the maximal lift variation and time-averaged drag coefficients, respectively. The remaining columns report the dominant eigenvalue $\lambda$ for various frequencies $St$ from the SPOD analysis, considering the fundamental harmnonic for the von-Kármán vortex shedding ($VS$), its second (${VS,2}$) and third (${VS,3}$) harmonics, and the dominant frequency for the broadband emission of Kelvin-Helmholtz vortices ($KH$).}
  \label{tab:DNSvsURANS}
  \end{center}
\end{table}


\section{Nudging-based data-assimilation} \label{sec:unsteadynudging}

Now that the shortcomings of the URANS prediction compared to the DNS have been characterised, we will introduce a data assimilation technique with the objective of alleviating them. The proposed nudging approach, which amounts to incorporating pointwise measurements extracted from the reference DNS into the URANS simulations, is introduced in \S\ref{sec:nudging_methodology} and is then applied in
\S\ref{sec:results_concept}-\S\ref{sec:results_high_frequency} considering various scenarios in terms of temporal and spatial resolutions of measurement data. 
First, results for a baseline configuration using dense measurement data for nudging are presented in \S\ref{sec:results_concept}.
As a further step, the influence of reduced resolution of measurement data is assessed in \S\ref{sec:results_low_frequency} and \S\ref{sec:results_high_frequency} with respect to low- and high-frequency phenomena, respectively.

\subsection{Data assimilation methodology}\label{sec:nudging_methodology}

\subsubsection{Nudged equations and measurements}\label{sec:nudged_URANS_equation}

Nudging allows a dynamic adjustment of the estimated flow through the addition of a feedback term in the URANS momentum equations in (\ref{eqn:URANS}), which is proportional to the discrepancies between flow measurements denoted $\boldsymbol{m}$ and the estimated velocity field. 
The intensity of this feedback term is adjusted through a parameter called $\alpha$. The estimated velocity field, which is now denoted $\bar{\boldsymbol{u}}_{\alpha}$, satisfies the so-called nudged equations
\begin{equation}\label{eqn:nudged-urans}
    \frac{\partial \bar{\boldsymbol{u}}_{\alpha}}{\partial t} + (\bar{\boldsymbol{u}}_{\alpha} \cdot \boldsymbol{\nabla}) \bar{\boldsymbol{u}}_{\alpha} + \boldsymbol{\nabla}\bar{p}_{\alpha} - \boldsymbol{\nabla}\cdot \left[ 2\left( Re^{-1}+\nu_t(\tilde{\nu}_{\alpha}) \right)  \boldsymbol{\nabla}_{\mathrm{s}} \bar{\boldsymbol{u}}_{\alpha} \right] = \alpha \, \mathcal{H}^{\dagger} [ \mathcal{H}(\bar{\boldsymbol{u}}_{\alpha})-\boldsymbol{m}],
\end{equation}
which are supplemented by the divergence-free condition in (\ref{eqn:URANS}) and the Sparlart-Allmaras model (\ref{eqn:SA}) to determine the eddy-viscosity  field
$\nu_t(\tilde{\nu}_{\alpha})$ in the above equation. The feedback term on the right-hand-side involves the measurement operator $\mathcal{H}$, which acts on the estimated velocity field $\bar{\boldsymbol{u}}_{\alpha}$ and returns a vector $\mathcal{H}(\bar{\boldsymbol{u}}_{\alpha})$ in the measurement space. Further details about the measurement operator and its adjoint $\mathcal{H}^{\dagger}$ are provided below.
Setting $\alpha=0$, one recovers the standard URANS equations, while for $\alpha>0$, the feedback term in the momentum equations drives the estimated flow $\bar{\boldsymbol{u}}_{\alpha}$ towards measurements $\boldsymbol{m}$. 
The rationale behind the choice of an appropriate value for $\alpha$ is detailed in appendix \ref{sec:alphastudy}, where minor sensitivity of results is shown as soon as $\alpha>1$. The value $\alpha=100$ is thus chosen in the following nudged URANS simulations.\\

In the present study, the measurement operator $\mathcal{H}$ is defined so as to extract the two components of a velocity field at selected spatial locations, thus mimicking the Particle Image Velocimetry (PIV) approach but without taking into account any measurement error that may occur in real experiments. These $M$ `nudging points', which are denoted by $\boldsymbol{x}_{k}$ with ${k = 1\cdots M}$, are sketched with red dots in figure \ref{fig:scketch-nudging}(a). They are chosen to be equidistantly distributed (in both directions) by a spatial distance $\Delta s$ inside a nudging region $\Omega_I$ which is delimited by the red frame in this figure. Note that the complementary region is denoted as $\Omega_{E}$. In the following, the nudging region is fixed to $\Omega_I=\left[-1.5,3.5 \right] \times \left[-1.5,1.5\right]$ and we will investigate the effect of the spatial sampling $\Delta s$. 

The measurements $\boldsymbol{m}$ are here extracted from the reference velocity field $\bar{\boldsymbol{u}}_{r}$ from DNS results (see \S\ref{sec:DNS}). More precisely, $\boldsymbol{m}$ is defined as
\begin{eqnarray}\label{eqn:measurements}
    \boldsymbol{m}(t) = \mathcal{H}(\bar{\boldsymbol{u}}_{r}(t)) = 
    \left[ \bar{\boldsymbol{u}}_{r}(\boldsymbol{x}_{1},t) , \cdots ,  \bar{\boldsymbol{u}}_{r}(\boldsymbol{x}_{M},t) \right]^{T} = \left[ \boldsymbol{m}_{1}(t), \cdots ,  \boldsymbol{m}_{M}(t) \right]^{T} 
\end{eqnarray} 
where $\boldsymbol{m}_{k}$ is a vector that contains the two velocity components at measurement point $\boldsymbol{x}_{k}$. The size of the data vector $\boldsymbol{m}$ is therefore $2M$. At an instant $t$, the right-hand side term in (\ref{eqn:nudged-urans}) is obtained from the discrepancy between the estimated flow and this data vector, i.e. 
$[ \mathcal{H}(\bar{\boldsymbol{u}}_{\alpha})- \boldsymbol{m}(t)] = \left[ \bar{\boldsymbol{u}}_{\alpha}(\boldsymbol{x}_{1}) - \boldsymbol{m}_{1} , \cdots ,  \bar{\boldsymbol{u}}_{\alpha}(\boldsymbol{x}_{M}) - \boldsymbol{m}_{M} \right]^{T}$, 
and the adjoint measurement operator $\mathcal{H}^{\dagger}$ then allows to transform this discrepancy vector of size $2M$ into a field (in the modelled flow space) of pointwise forcings at the same locations $\boldsymbol{x}_k$. \\

The temporal and spatial methods that are used to discretise the nudging equation (\ref{eqn:nudged-urans}) are similar to those described in \S \ref{sec:URANS} for the standard URANS simulations. Note that the feedback term is discretized as an implicit term so as to improve the numerical stability of the algorithm, thus allowing to choose the time step $\Delta t$ arbitrarily large. For all simulations considered hereinafter, it is assumed that the measurement vector $\boldsymbol{m}(t_k)$ is available at every instant $t_k=k \Delta t$ of the simulation. In other words, we do not make any difference  between the time step $\Delta t$ of the simulation and the time interval between measurements. In the following, $\Delta t$ is therefore used to characterize the temporal sampling of a data set that can be written as
\begin{eqnarray}
\left[ \boldsymbol{m}(t_1), \cdots , \boldsymbol{m}(T_m) \right] 
\end{eqnarray}
where $N_{m}$ is the number of measurement vectors and $T_{m}=t_{N_{m}}=N_{m} \Delta t$ is the measurement time (or equivalently here the time window of the simulations). In the next paragraph, it will be detailed how the spatial sampling $\Delta s$, the temporal sampling $\Delta t$ and the measurement time $T_{m}$ are chosen to define several data sets that are used to assess the nudging method in its ability to improve the fidelity of the estimated flow at different spatial and temporal flow scales.


\subsubsection{Measurement data-sets}\label{sec:data_sets}

\begin{table}
\centering
\begin{tabular}{r|ccc|ccc|cc}  
Data-set group & $St$ & $\tau$ &  $\lambda_{x}$ & $\Delta t$ & $N_{m}$ & $T_{m}=N_{m} \Delta t$ & $\tau/\Delta t$ & $T_{m}/\tau$  \\
\hline
VS (\S \ref{sec:results_low_frequency}) & $0.137$ & $7.335$ & $4.00$ & $0.209$ & $908$ & $189.85$  & $35$  & $26$ \\
\hline
KH (\S \ref{sec:results_high_frequency}) & $4.384$  & $0.228$ & $0.25$ &$0.021$ & $1441$ & $30.13$    & $11$  & $132$  \\
\hline
VS+KH (\S \ref{sec:results_concept}) & $0.137$ & $7.335$ & $4.00$& $0.021$ & $8421$ & $176.1$  & $349$  & $24$ \\
& $4.384$ & $0.228$ & $0.25$ & \multicolumn{3}{c|}{}    & $11$  & $772$ \\
\hline
\end{tabular}
    \caption{
    Description of three groups of measurement data-sets  which are named from the targeted flow phenomenon by the nudging approach: the von-Kármán Vortex Street (VS), the Kelvin-Helmholtz vortices (KH) and both phenomena (VS+KH). Their frequency $St$ and period $\tau=1/St$ as captured by DNS are recalled in the second and third columns, respectively. These data-set groups are characterized by the time interval between consecutive measurements $\Delta t$ (fifth column), the number of measurement snapshots $N_m$ (sixth column) and the total measurement time $T_m$ (seventh column). The last two columns report the number of snapshots per cycle (of the target flow phenomenon) and the total number of cycles.}
    \label{tab:datasetgroup}
  \end{table}
\begin{table}
\centering
   \begin{tabular}{r|c|c|c|c|c|c|c|c|c|c|c|}
        $\Delta s$ & $0.03125$ & $0.0625$ & $0.125$ & $0.25$ & $0.5$ & $0.75$ & $1$ & $1.25$ &  $1.5$ & $1.66$ & $2$ \\
        \midrule
        Data-set group & (a) & (b) & (c) & (d) & (e) & (f)  & (g) & (h) & (i) & (j) & (k)\\
      \midrule
     VS (\S \ref{sec:results_low_frequency}) & $128$ & $64$ & $32$ & $16$ & $8$ & $5.33$ & $4$ &  $3.2$ & $2.66$ & $2.41$ & $2$\\
      \midrule
      KH (\S \ref{sec:results_high_frequency})  & $8$ & $4$ & $2$ & $1$ & $\times$ & $\times$ & $\times$ & $\times$ & $\times$ & $\times$ & $\times$ \\
      \midrule
      VS+KH (\S \ref{sec:results_concept}) & $(128;8)$ & $\times$  & $\times$ & $\times$  & $\times$  & $\times$ & $\times$ & $\times$ & $\times$ & $\times$ & $\times$ \\
      \midrule
   \end{tabular}
  \caption{Test matrix indicating, for the three data-set groups VS, KH and VS+KH in table \ref{tab:datasetgroup}, the number of measurement points $N_{s} = \lambda_{x}/\Delta s$ that sample the wavelength $\lambda_{x}$ of the corresponding spatial structures (indicated in the fourth colmun of table \ref{tab:datasetgroup}) when varying the spatial sampling $\Delta s$. For instance, the data-set VS(a) corresponds to $(\lambda_{x}=4.00,\Delta s=0.03125)$, thus $N_s=128$.}
  \label{tab:arrays}
\end{table}
Three groups of data-sets are defined in table     \ref{tab:datasetgroup} to specifically investigate the ability of nudging to improve the prediction of low-frequency vortex shedding [group VS], to reproduce high-frequency Kelvin-Helmholtz vortices [group KH], and to capture both phenomena [group VS+KH]. The frequency $St$, period $\tau=1/St$ and typical wavelength $\lambda_{x}$ of the structures that are associated to each flow phenomenon are first recalled in the first four columns of this table. The time interval $\Delta t$ between consecutive measurements and the total time $T_{m}$ of the simulations are then chosen to properly capture one of the two phenomena (VS or KH) or both of them (VS+KH). The ratios $\tau/\Delta t$ and $T_m/\tau$, which are indicated in the last two columns, correspond to the number of measurement snapshots per cycle and to the total number of cycles, respectively. They confirm that the temporal resolution and time window of the data assimilation experiments are sufficiently large to capture the physics of interest.

It may be noted that data-sets of group VS+KH combine the temporal resolution of group KH ($\Delta t= 0.021$) and the time window of group VS ($T_{m}=211.5$), which makes them sufficiently well resolved for describing the high-frequency shedding of Kelvin-Helmholtz vortices ($12$ snapshots per cycle), while also capturing a significant number ($29$) of the low-frequency von-Kármán vortex shedding cycles. Data-sets in this group (VS+KH) are therefore characterized by a large number of measurement snapshots ($N_{m} \sim 10^4$), which makes them less convenient for parametric studies. They will mainly be used in \S \ref{sec:results_concept} to demonstrate the ability of the nudging method to reconstruct both phenomena based on measurements that are dense in both time and space. 

The influence of the spatial resolution $\Delta s$ on the performances of the data assimilation procedure is mainly investigated using the data-set groups VS and KH in \S\ref{sec:results_low_frequency} and \S\ref{sec:results_high_frequency}, respectively. Table \ref{tab:arrays} summarizes the values of spatial sampling $\Delta s$ that are tested for each data-set group. More precisely, it indicates the number of measurement points $N_s=\lambda_{x}/\Delta_{s}$ that sample the characteristic wavelength $\lambda_{x}$ of the corresponding flow structure. The data-sets of group VS cover a wide range of spatial resolutions, from the highest resolution with $128$ points per wavelength when $\Delta s=0.03125$, to the lowest spatial resolution with $4$ points per wavelength for $\Delta s =1$. The spatial resolution covered by the data-sets of group KH is more restricted, from $8$ points to only $1$ point per wavelength of Kelvin-Helmholtz vortices.

\subsection{Results with the data-set of highest spatio-temporal resolution (data-set VS+KH(a))} \label{sec:results_concept}

\begin{figure}
\vspace{0.25cm}
\centering
\begin{tabular}[t]{lcl}
(a) & & (b)   \\
\imagetop{\includegraphics[width=0.45\columnwidth]{NewFigures/URANS_T50.pdf}} & & \imagetop{\includegraphics[width=0.33\columnwidth]{NewFigures/Spectrum_URANS.pdf}} \\
& \\
(c) & & (d)   \\
\imagetop{\includegraphics[width=0.45\columnwidth]{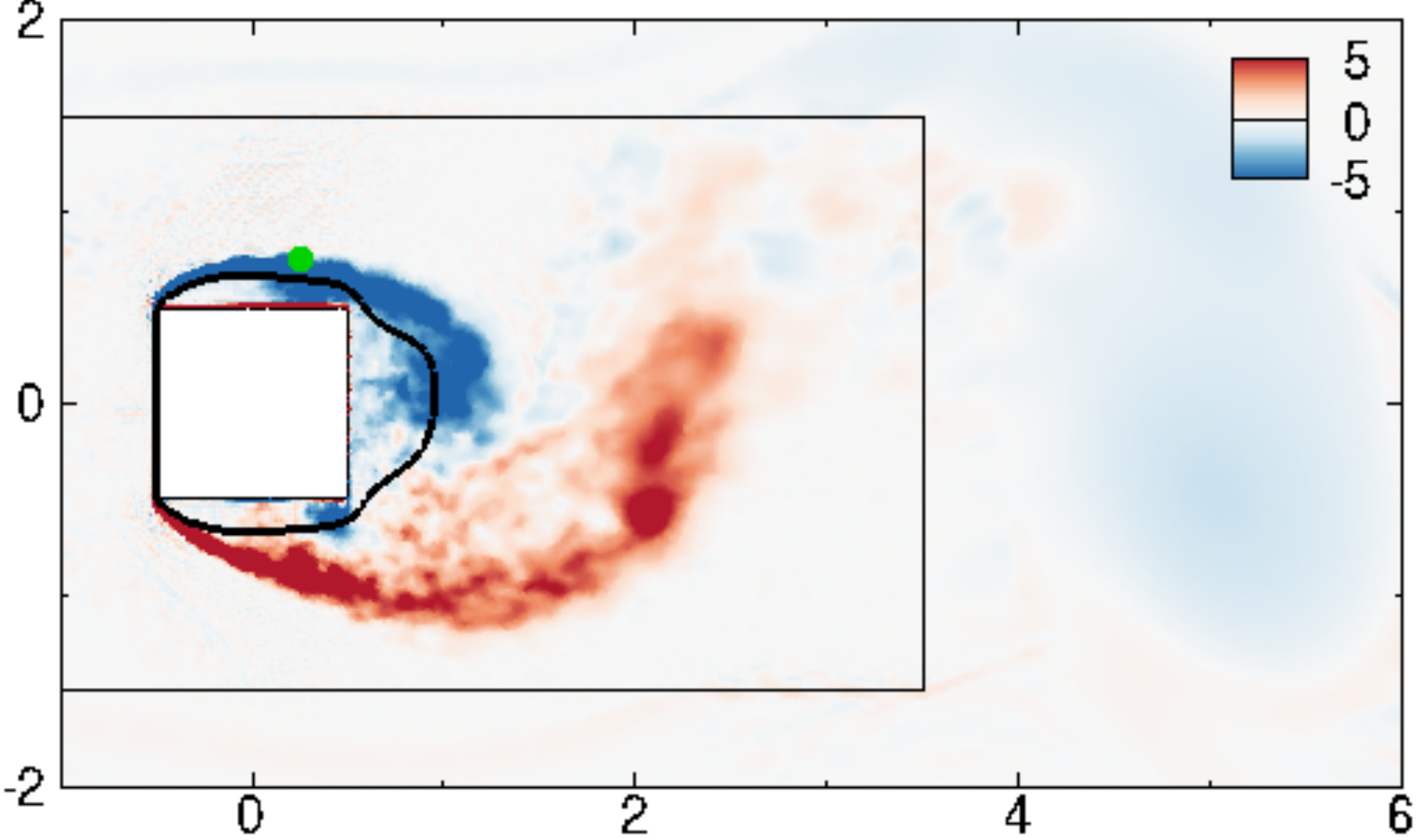}} & & \imagetop{\includegraphics[width=0.33\columnwidth]{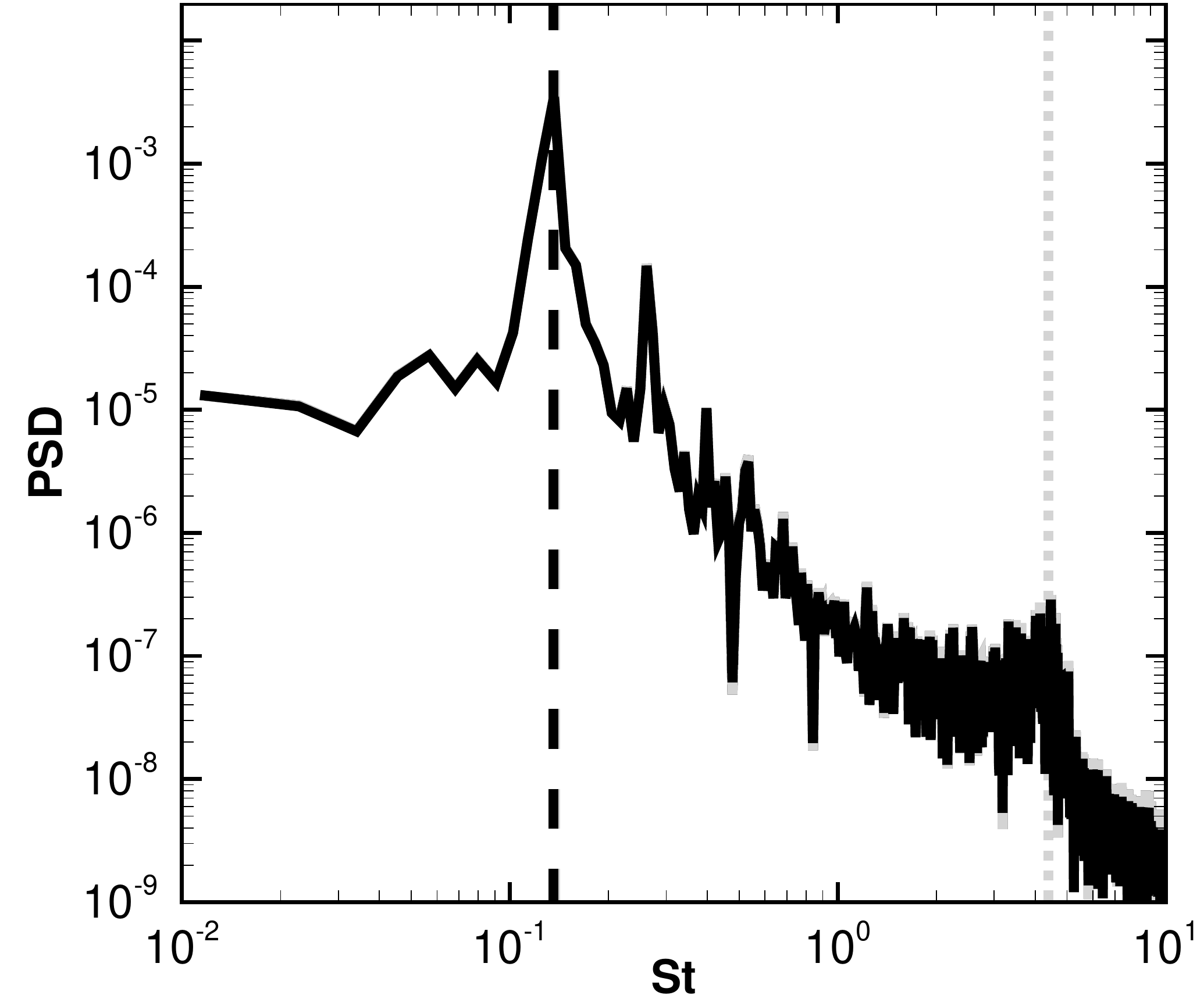}} \\
\end{tabular}
  \caption{
  Results for (a-b) standard and (c-d) nudged URANS with the data-set VS+KH(a). (a,c): Instantaneous spanwise vorticity field $\bar{\omega}_z$ at $t=50$, where black iso-curves denote $\left< \overline{u} \right> =0$. The extent of the nudging region is delineated by black lines in (c). (b,d): Fourier spectrum of the streamwise velocity at the green monitor point in (a,c) (full black line). For the sake of comparison, the DNS spectrum is also reported (full grey line, well-hidden in (d)). Vertical dashed and dotted lines indicate the low-frequency peak associated with large-scale vortex-shedding and the high-frequency bump linked to Kelvin-Helmholtz instabilities, respectively.
  }
\label{fig:standardURANS}
\end{figure}

We first investigate results of the nudging method using the data set VS+KH(a), which corresponds to the highest temporal and spatial resolutions investigated here, as detailed in tables \ref{tab:datasetgroup} and \ref{tab:arrays}. Results from the nudged URANS simulation are first examined and compared with standard URANS through the instantaneous vorticity field at $t=50$, similarly as in figure \ref{fig:reference} which reported DNS and standard URANS results. Small-scale structures, which are missing in the URANS prediction without nudging (figure \ref{fig:standardURANS}(a)), are clearly recovered with nudging (figure \ref{fig:standardURANS}(c)), and the thus estimated flow looks very similar to the reference one (figure \ref{fig:reference}(a)), especially in the nudging region. More downstream ($x>3.5$), we may already note that these small scale structures are missing, as further discussed in the following.

Figures \ref{fig:standardURANS}(b) and (d) display the Fourier spectrum of the streamwise velocity at $(x,y)=(0.25,0.75)$ for the estimated flow without and with nudging, respectively. As mentioned in \S\ref{sec:DNSvsURANS}, the high-frequency content ($1<St<10$) corresponding to Kelvin-Helmholtz fluctuations in the reference flow (grey curve in figure \ref{fig:standardURANS}(b)) is not captured by the estimated flow without nudging (black curve in figure \ref{fig:standardURANS}(b)). On the other hand, the Fourier spectrum corresponding to the URANS prediction with nudging (black curve in figure \ref{fig:standardURANS}(d)) is perfectly superimposed to the reference spectrum, for all frequencies. In particular, the low-frequency peak corresponding to vortex-shedding (dashed vertical line) is reached at $St_{VS}=0.137$ with nudging, instead of $St_{VS}=0.126$ without nudging, which is in prefect agreement with the reference value. 

\begin{figure}
\vspace{0.25cm}
\centering
\begin{tabular}[t]{lll}
(a) & (b) & (c) \\
\includegraphics[width=0.32\columnwidth]{NewFigures/Fig3a1.pdf} & \includegraphics[width=0.32\columnwidth]{NewFigures/Fig3a2.pdf} & \includegraphics[width=0.32\columnwidth]{NewFigures/Fig3a3.pdf} \\
(d) & (e) & (f) \\
\includegraphics[width=0.32\columnwidth]{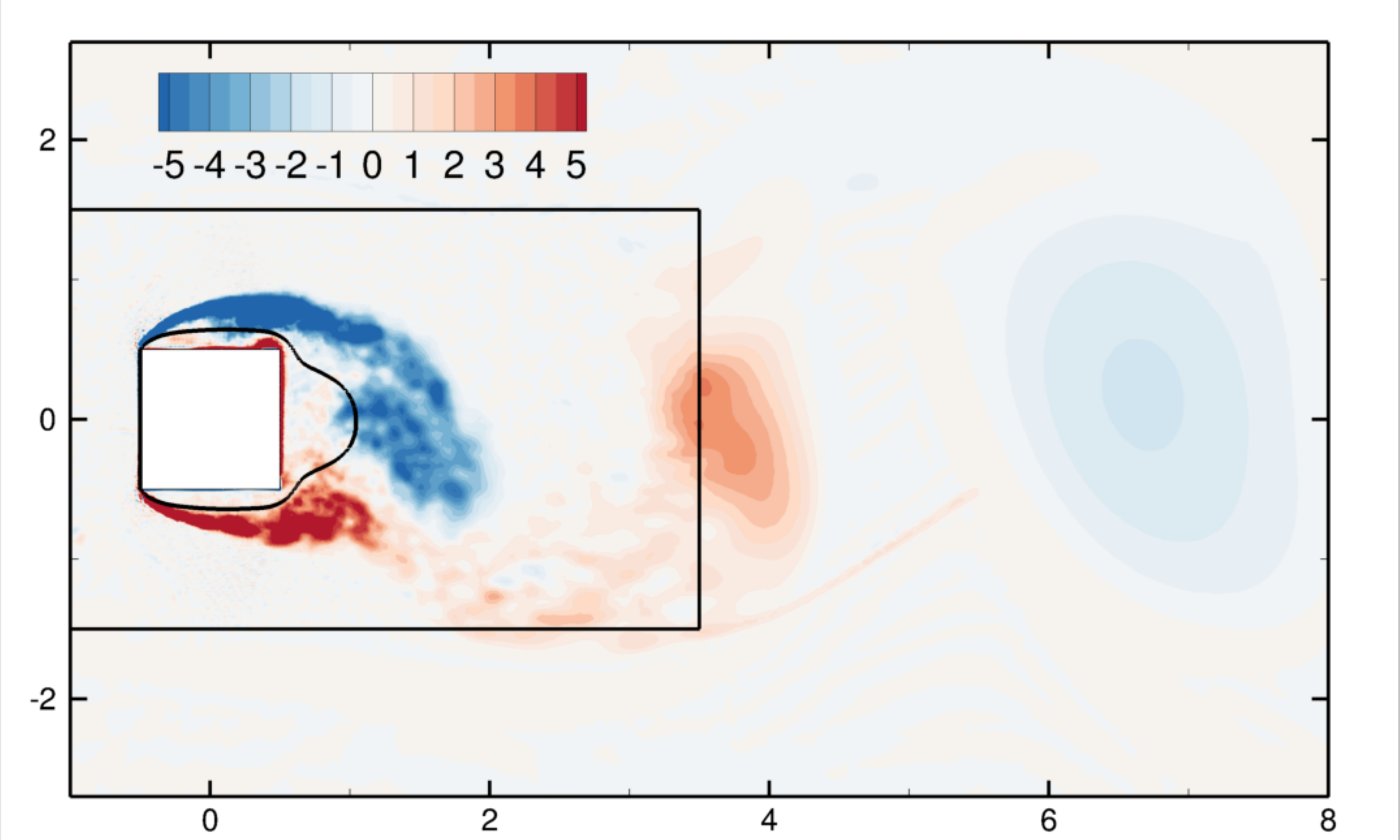} & \includegraphics[width=0.32\columnwidth]{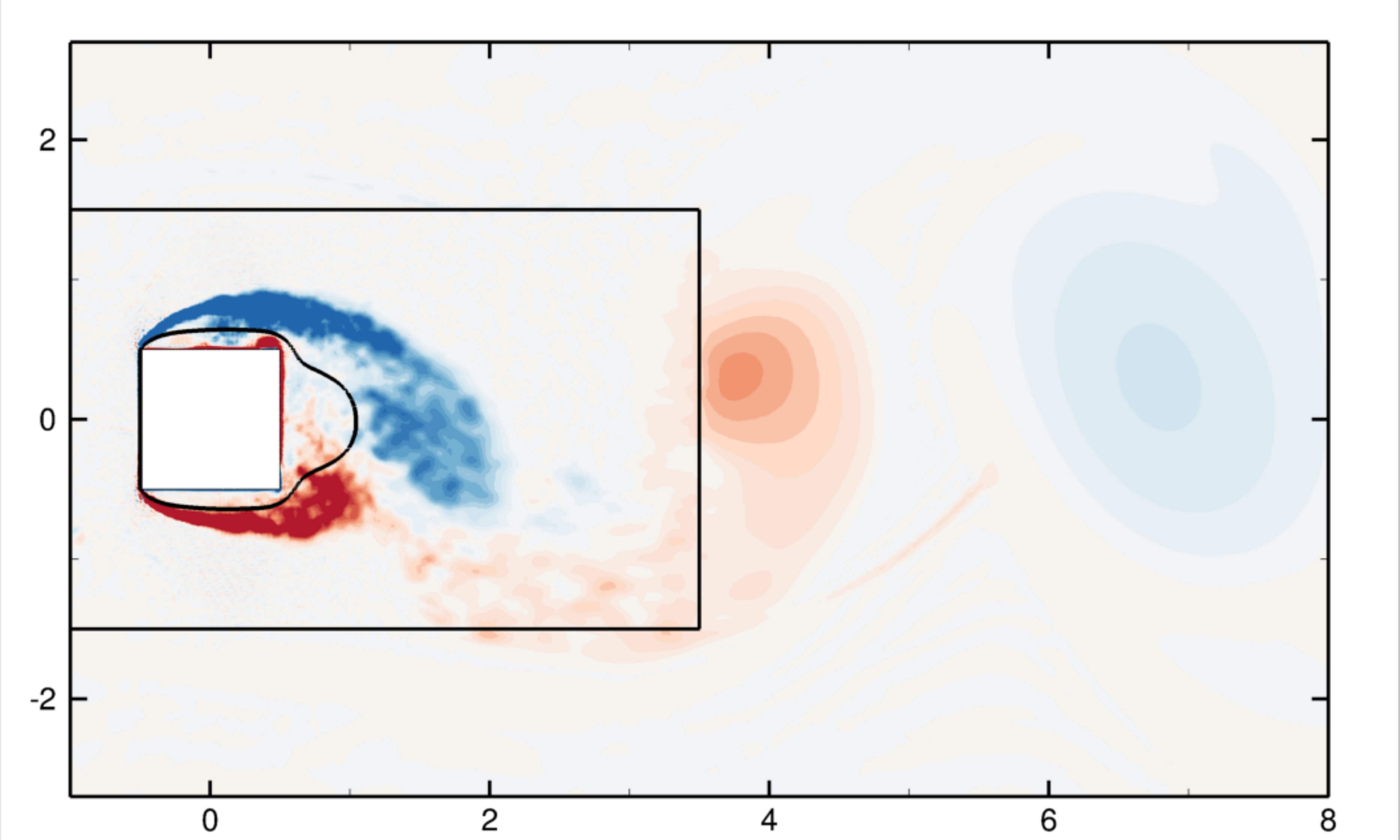} & \includegraphics[width=0.32\columnwidth]{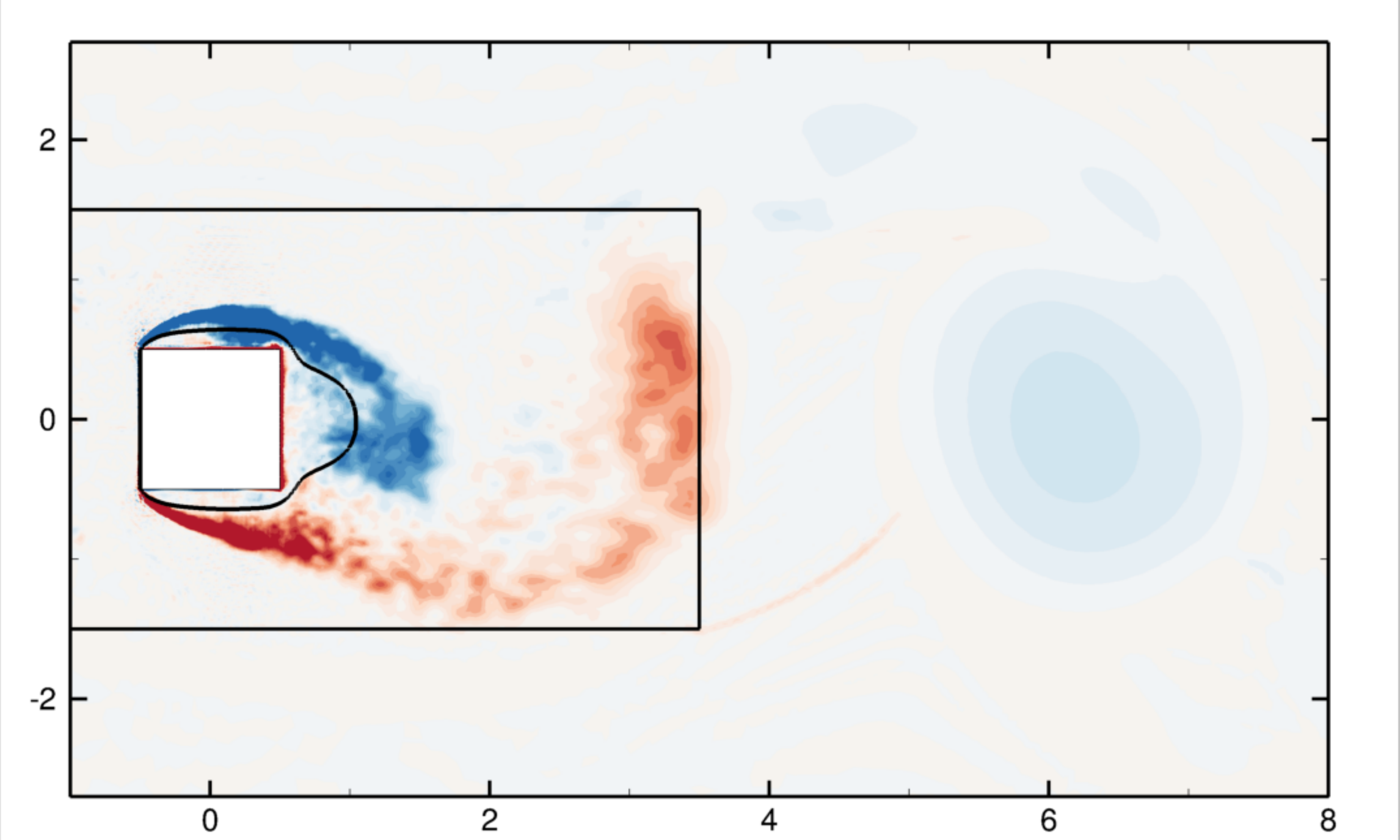} \\
(g) & (h) & (i) \\
\includegraphics[width=0.32\columnwidth]{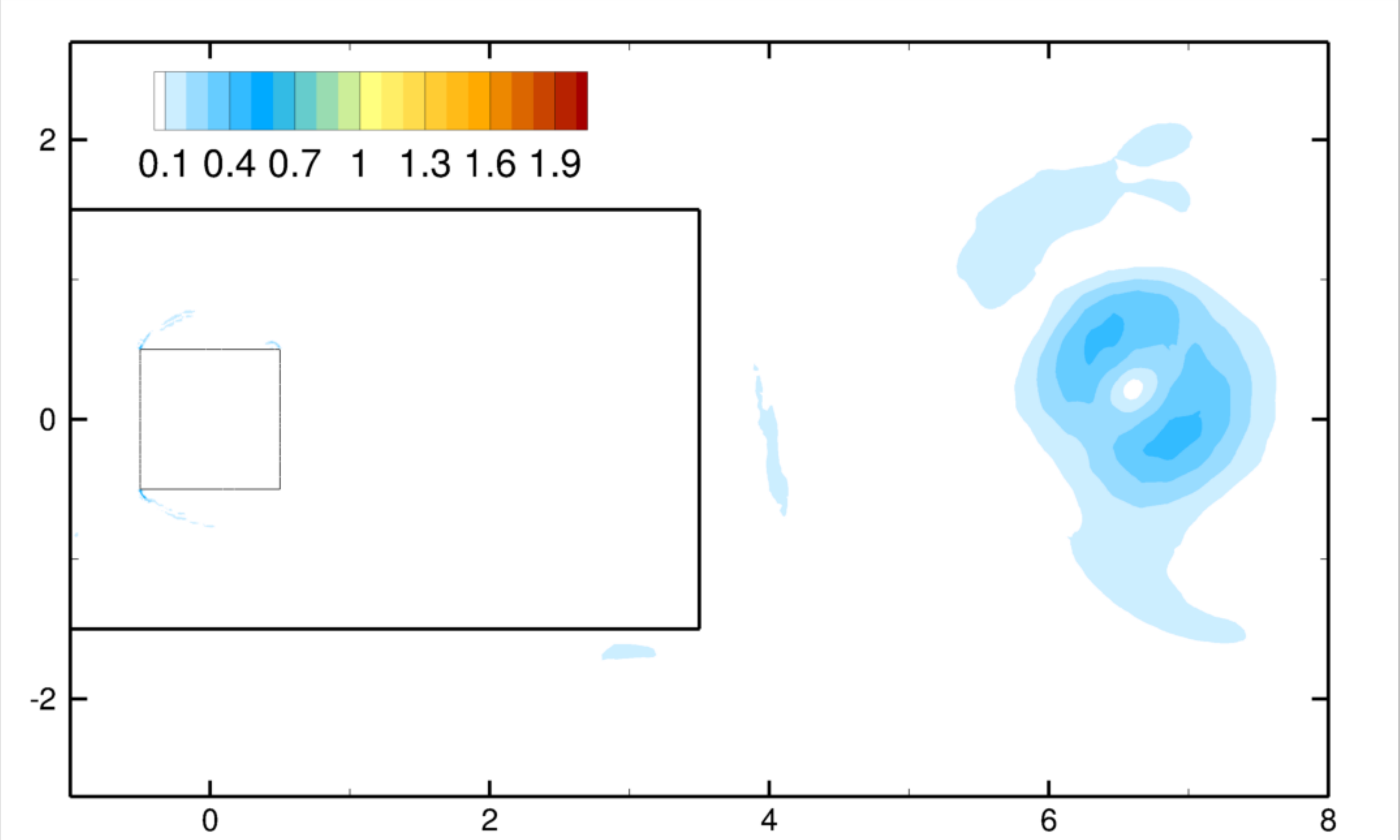} & \includegraphics[width=0.32\columnwidth]{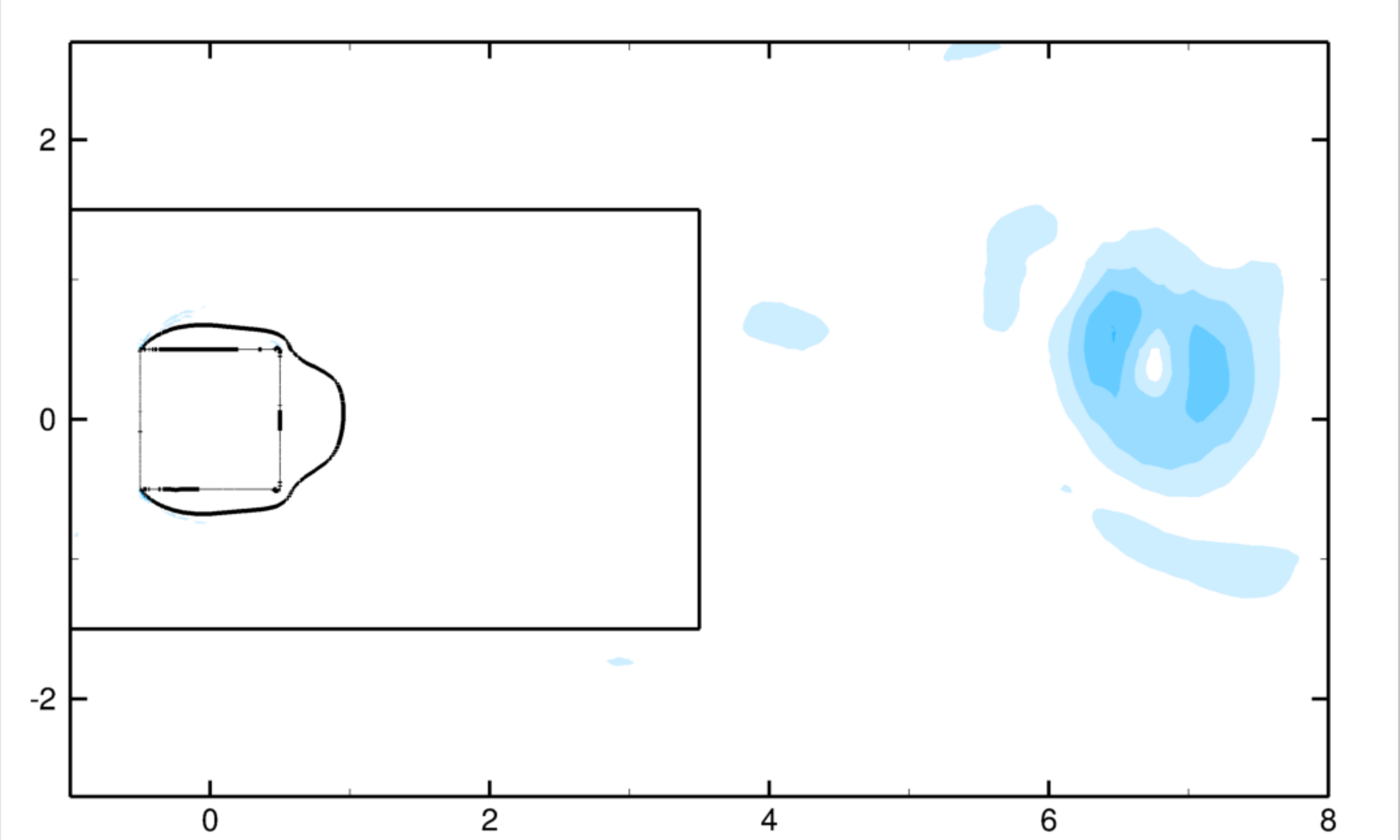} & \includegraphics[width=0.32\columnwidth]{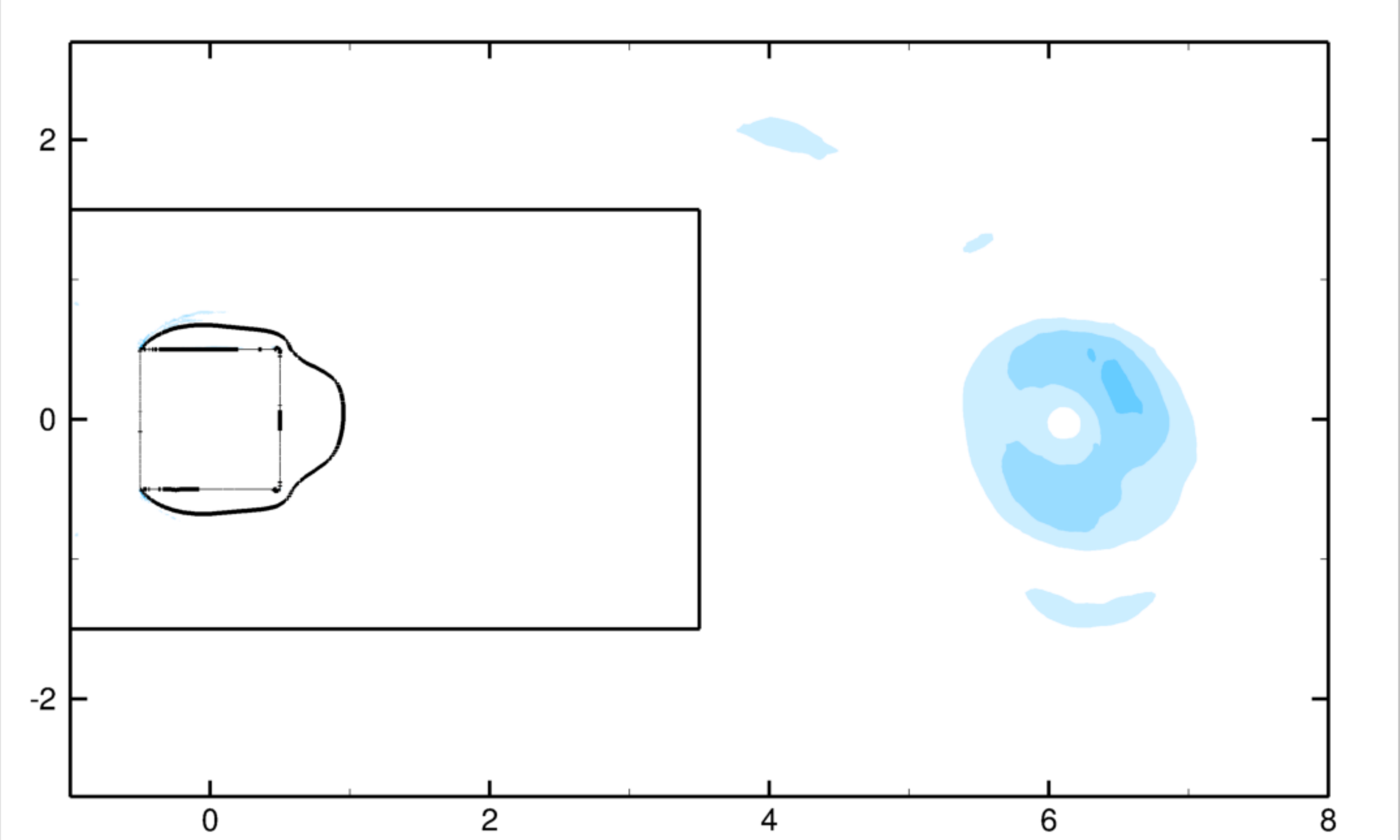} 
\end{tabular}
\begin{tabular}[t]{l}
(j) \\
\includegraphics[width=0.95\columnwidth]{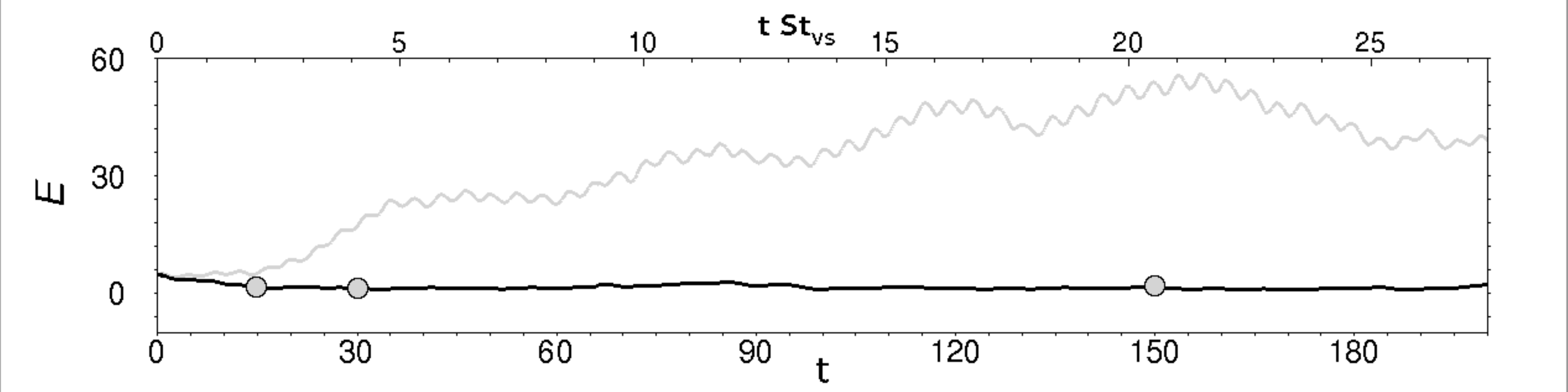}
\end{tabular}
\caption{
  Instantaneous vorticity fields of (a-c) the reference flow $\boldsymbol{\bar{u}}_{r}$ and (d-f) the estimated flow $\boldsymbol{\bar{u}}$ with nudging relying on the data-set VS+KH(a) at times (a,d) $t=15$, (b,e) $t=30$ and (c,f) $t=150$. (g-i) Instantaneous error fields $e(\boldsymbol{x},t)$ between the reference and nudged flows (see eq. \ref{eqn:instantaneous-error}). (i) Temporal evolution of the global error $E(t)$ (see eq. \ref{eqn:total-error}) between the nudged and reference flow (black curve) and between the URANS and reference flows (grey curve). The bottom axis reports the (nondimensional) time while the top axis reports the number of low-frequency cycles determined as $t/\tau_{VS}=t \ St_{VS}$ with $St_{VS} = 0.137$. The grey circles indicates time $t=15$, $30$ and $150$}
\label{fig:TotalErrors}
\end{figure}

This improvement in the prediction of the fundamental frequency $St_{VS}$ is associated with a better synchronisation of the estimated flow with the reference one. This effect, which could have already been identified by comparing figure \ref{fig:standardURANS}(c) with \ref{fig:reference}(a), is clearly visible in figure \ref{fig:TotalErrors} which displays three instantaneous snapshots of the reference and nudged flows in figures \ref{fig:TotalErrors}(a-c) and  \ref{fig:TotalErrors}(d-f), respectively. The large scale vortices are now perfectly synchronised, in contrast with standard URANS and figure \ref{fig:unsynchronisation}). The small-scale structures seem also satisfactorily captured in the nudging region, but not further downstream. This defect, which is also identifiable through the examination of the spatial error distribution in figures \ref{fig:TotalErrors}(g-i), may be due to imperfections of the turbulence model, leading to excessive dissipation of the von-Kármán vortices. Nevertheless, the induced discrepancies remain significantly smaller than the ones for standard URANS in figure \ref{fig:unsynchronisation}(g-i). The large improvement in the estimated flow through the nudging approach is further confirmed by considering the temporal evolution of the global error $E(t)$ (defined in eq. \ref{eqn:total-error}) as illustrated in figure \ref{fig:TotalErrors}(j). This error is strongly reduced for the nudged simulation (black curve) compared with the standard one (grey curve). In the former case, $E(t)$ decreases during an initial transient and eventually maintains a fairly constant low level. This is in contrast with the standard URANS results, for which $E(t)$ significantly increases and then oscillates due to insufficient synchronisation with respect to the reference solution, as discussed in \S\ref{sec:DNSvsURANS}.\\


\begin{figure}
\vspace{0.25cm}
\centering
\begin{tabular}[t]{@{}l@{}r}
(a) & \imagetop{\includegraphics[width=0.8\columnwidth]{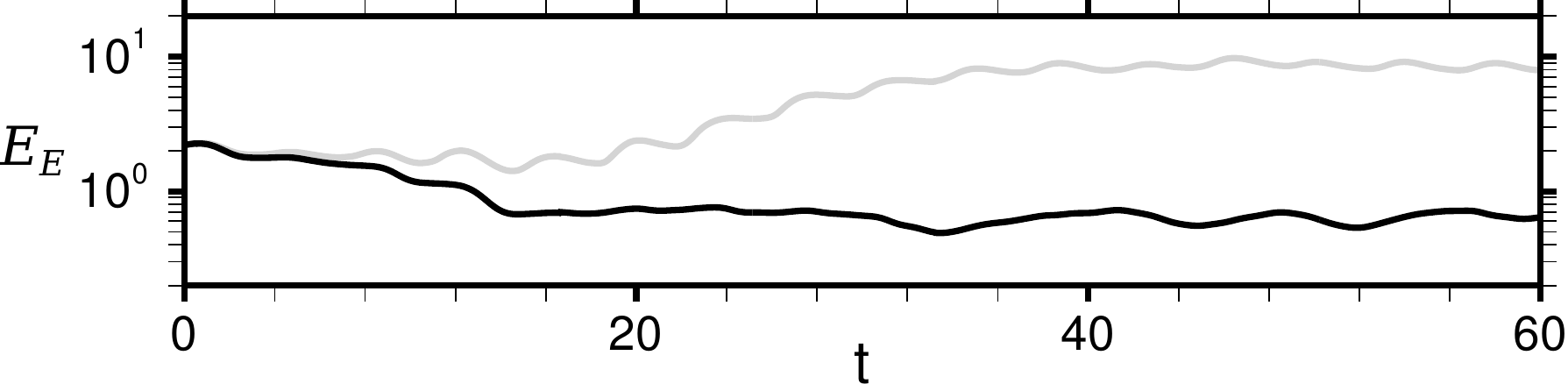}}  \\
(b) & \imagetop{\includegraphics[width=0.8\columnwidth]{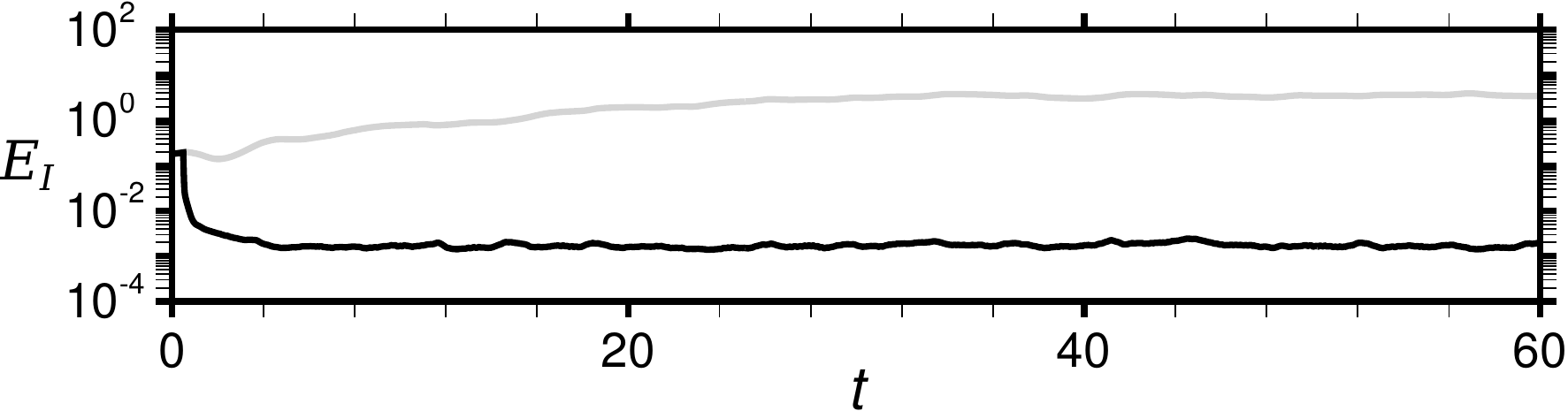}}  
\end{tabular}
  \caption{
  Temporal evolution of (a) external error $E_{E}$ and (b) internal error $E_{I}$ (see (\ref{eqn:internal-external-error})) for standard URANS (grey lines) and nudged URANS with the data-set VS+KH(a) (black lines).}
\label{fig:DecomposedErrors}
\end{figure}

In order to get a better insight in the extrapolation capabilities of nudging, the evaluation of $E(t)$ is decomposed according to
\begin{eqnarray}\label{eqn:internal-external-error}
    E(t)^{2} = E_{I}(t)^{2} + E_{E}(t)^{2} 
    =\int_{\Omega_I} e(\boldsymbol{x},t)^{2} \, d\boldsymbol{x} + \int_{\Omega_E} e(\boldsymbol{x},t)^{2} \, d\boldsymbol{x}.
\end{eqnarray}
This decomposition allows to distinguish between an internal error $E_{I}(t)$ defined over the nudging domain $\Omega_{I}$ (see figure \ref{fig:scketch-nudging}), which is directly influenced by the measurements, and an external error $E_{E}(t)$ defined over the complementary domain $\Omega_{E}=\Omega \setminus \Omega_{I}$. 
Both error components $E_{E}$ and $E_{I}$ are shown as functions of time in figures \ref{fig:DecomposedErrors} (a) and (b), respectively.  

Initially, the internal and external errors in the estimated flow are of the same order of magnitude. The internal error $E_{I}$ for nudged URANS (black curve) achieves a large reduction by three orders of magnitude in about $5$ convective times, which indicates an excellent agreement with the reference flow in $\Omega_{I}$. More time is necessary to decrease the external error $E_{E}$ as the corrected flow in the nudging region has to be convected in the complementary region $\Omega_{E}$. The external error remains however significantly larger than the internal one and thus dominates the total error $E$. The reduction in the external error compared to standard URANS results (grey curve) is still noticeable (by one order of magnitude), which confirms the significant benefits of nudging even outside of the nudging region $\Omega_{I}$.

So far, the potential of nudging the URANS equations for unsteady data assimilation using a measurement data-set with high spatio-temporal resolution has been demonstrated. 
In the following, we will study the performance of the nudging procedure with respect to the spatial and temporal resolutions of the measurement data as characterised by $\Delta s$ and $\Delta t$, respectively. 

\subsection{Nudging the low-frequency large-scale structures (data-set group VS)}\label{sec:results_low_frequency}

We now analyse the performance of the nudging approach to specifically improve the accuracy of low-frequency phenomena ($St<1$) using data-sets of lower resolution. More specifically, we now consider data-set group VS in table \ref{tab:datasetgroup} which is characterized by the temporal sampling $\Delta t = 0.209$, this value being $10$ times larger than for the data-set group VS+KH used in \S \ref{sec:results_concept}. Moreover, for most of the data-sets in group VS, the spatial sampling $\Delta s$ is also decreased compared to previous results. As summarised in table \ref{tab:arrays}, the investigated values of $\Delta s$ amount to $2 \leq N_s \leq 128$, where $N_s=\lambda_x/\Delta s $ is the number of nudging points per characteristic wavelength $\lambda_x \sim 4$ of the large-scale vortices.
\begin{figure}
\vspace{0.25cm}
\centering
\begin{tabular}[t]{lcl}
(a) & & (b)      \\
\imagetop{\includegraphics[width=0.42\columnwidth]{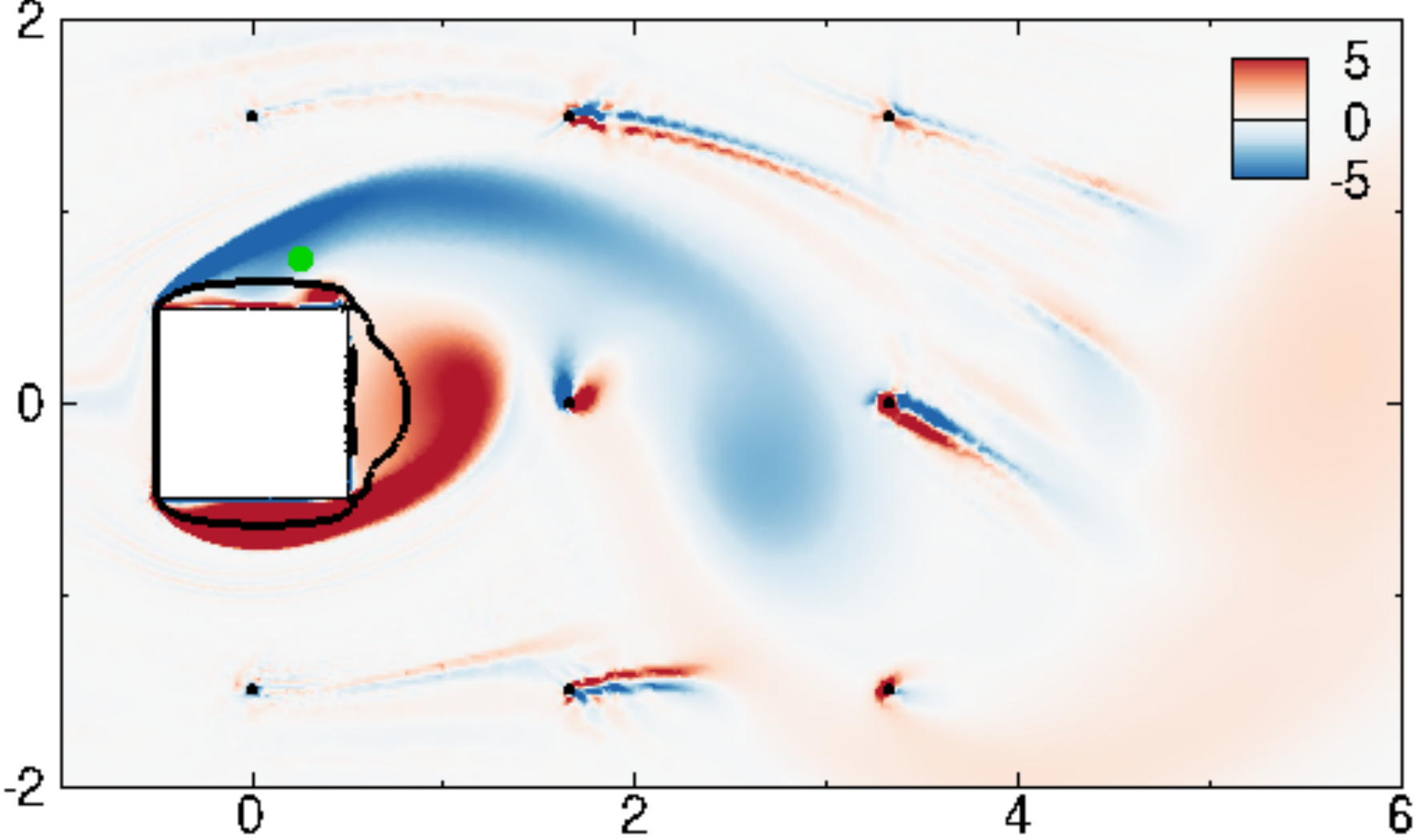}} & & \imagetop{\includegraphics[width=0.3\columnwidth]{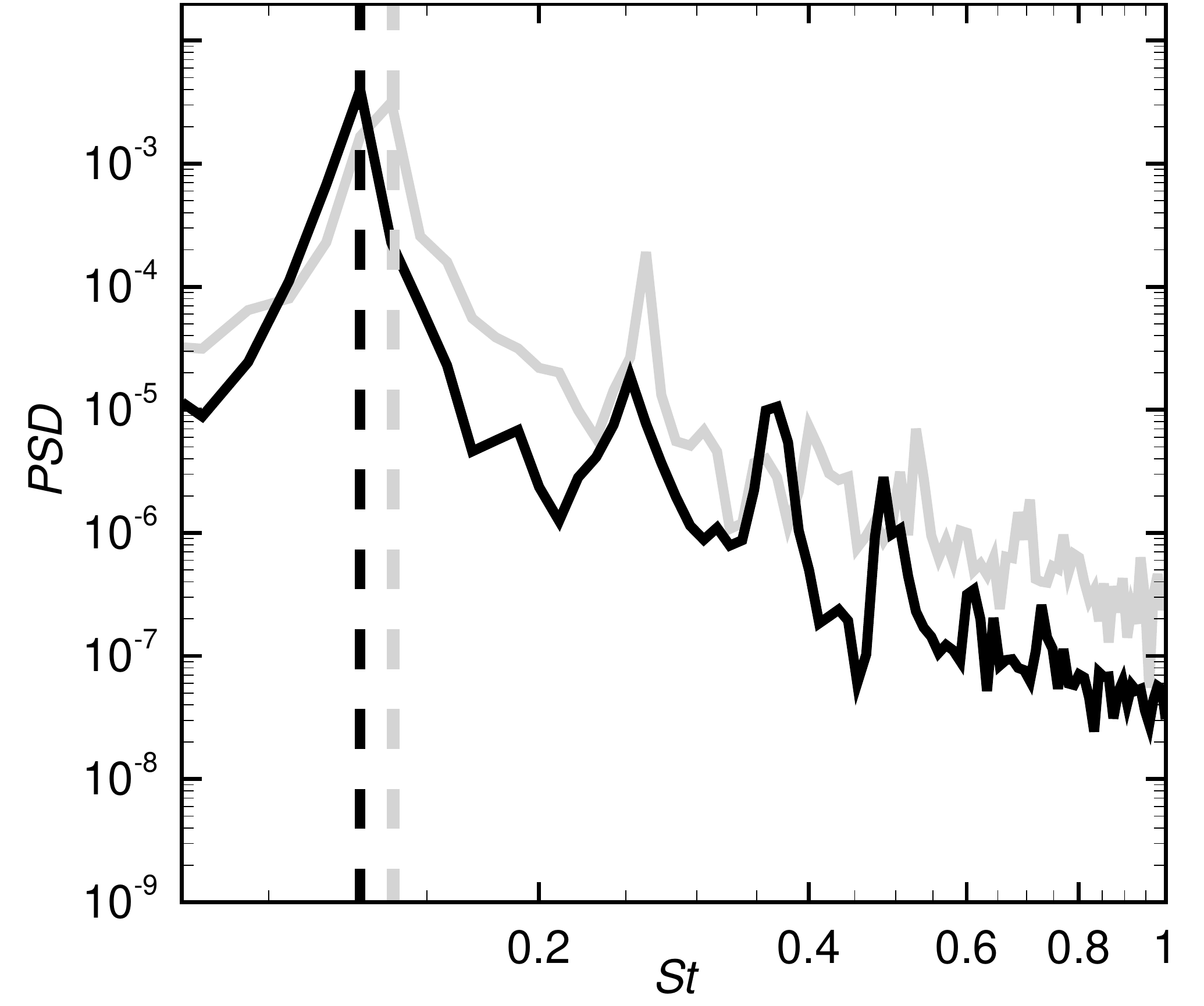}} \\
(c)  & & (d)      \\
\imagetop{\includegraphics[width=0.42\columnwidth]{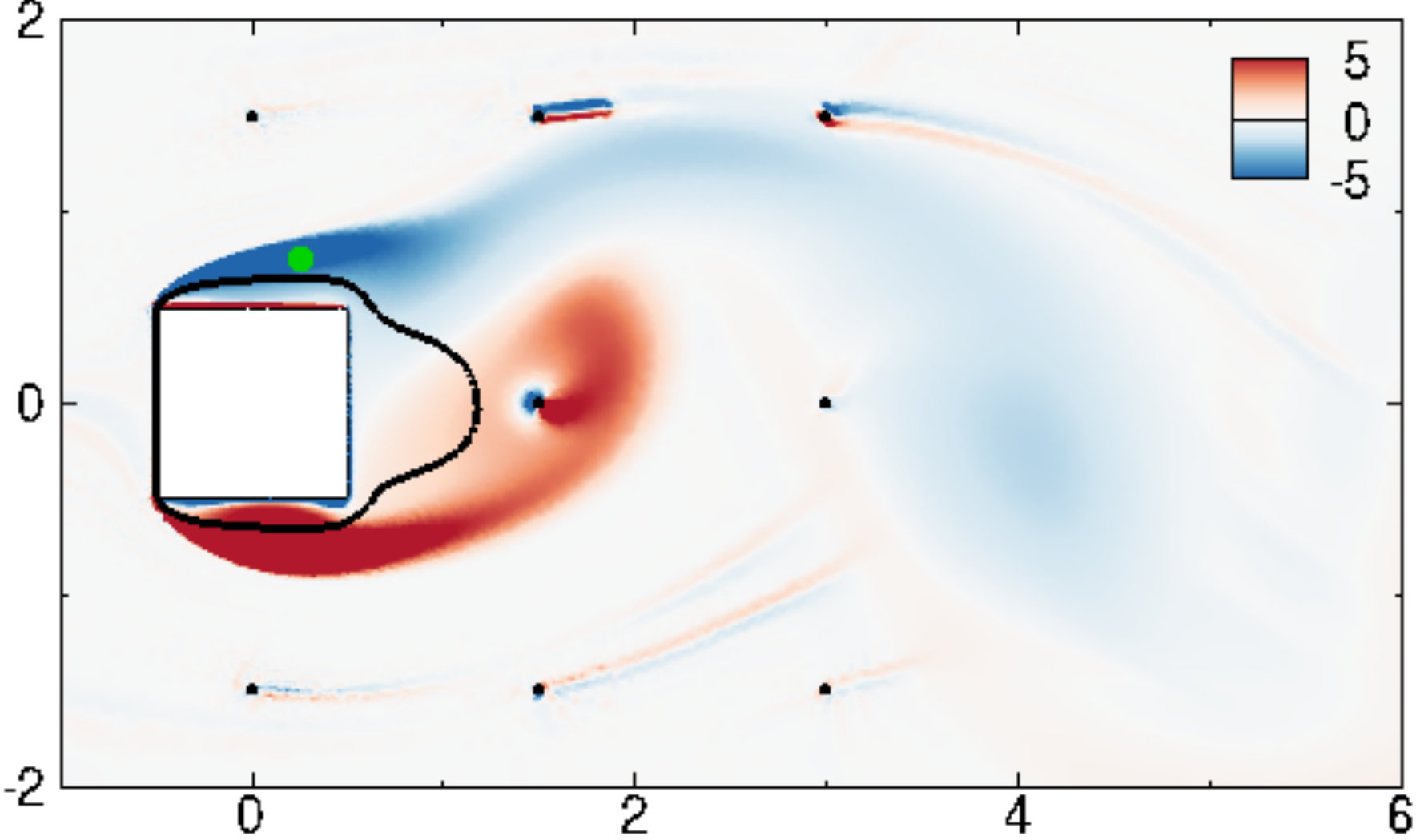}} & & \imagetop{\includegraphics[width=0.3\columnwidth]{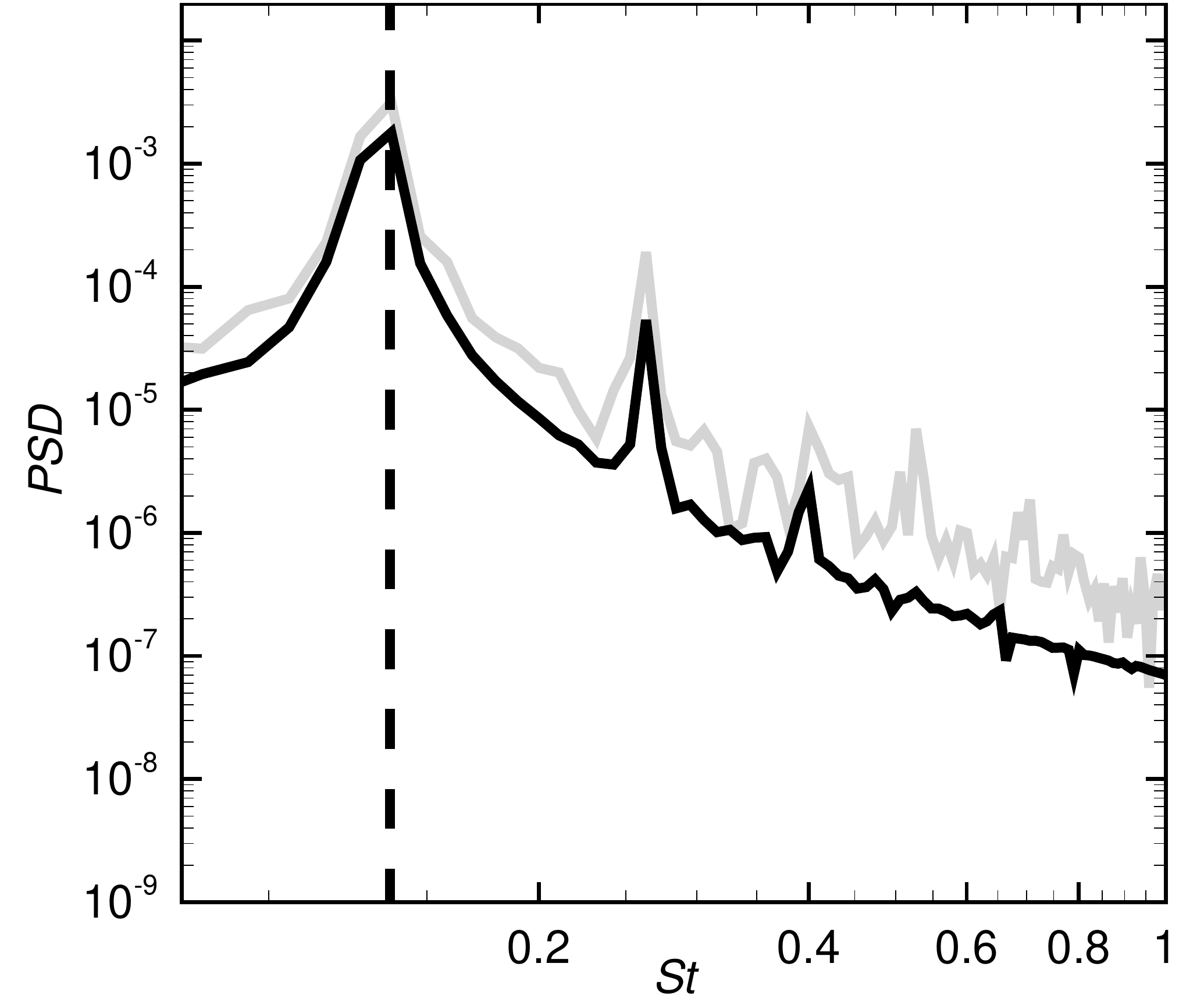}} \\
(e) & & (f)   \\
\imagetop{\includegraphics[width=0.42\columnwidth]{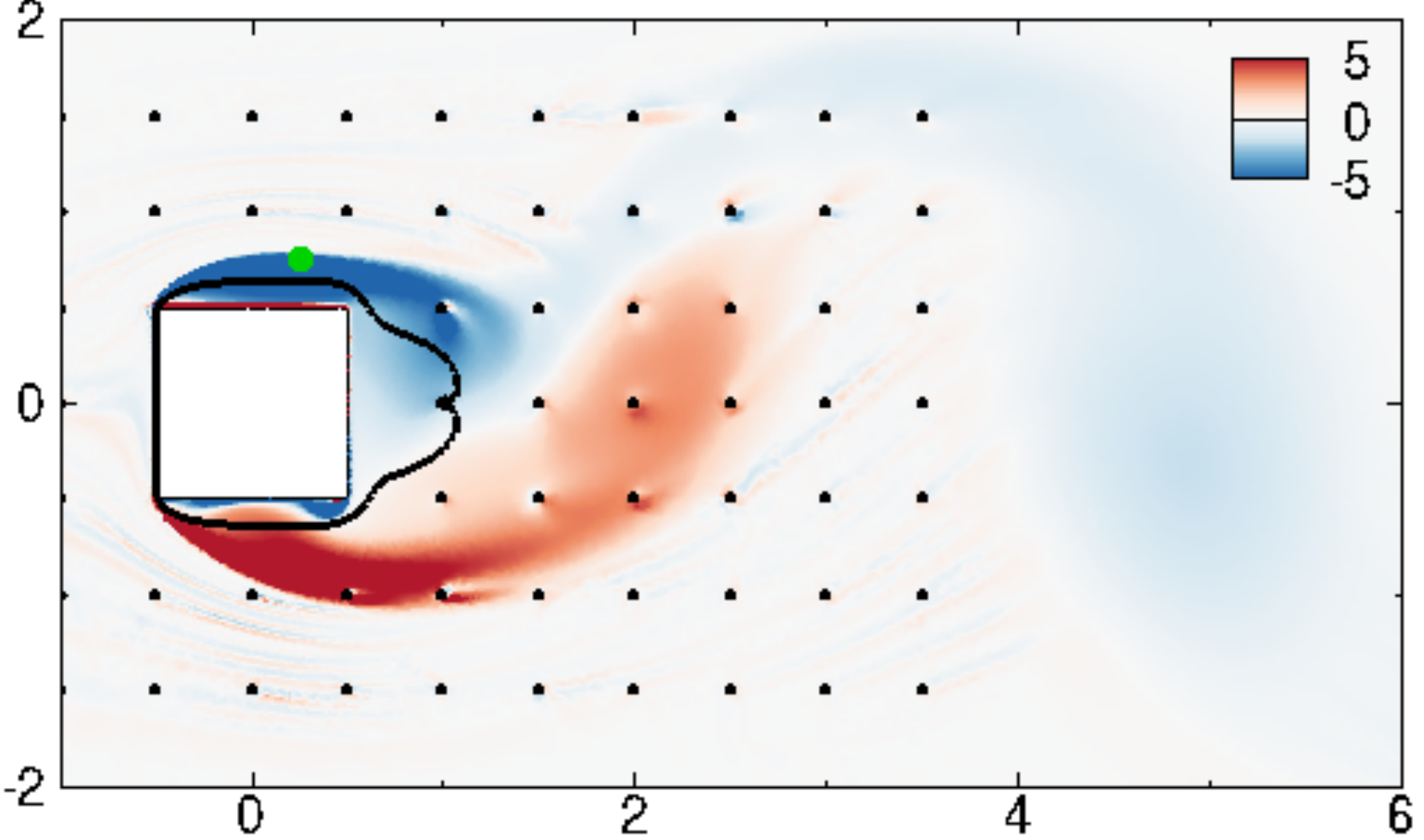}} & & \imagetop{\includegraphics[width=0.3\columnwidth]{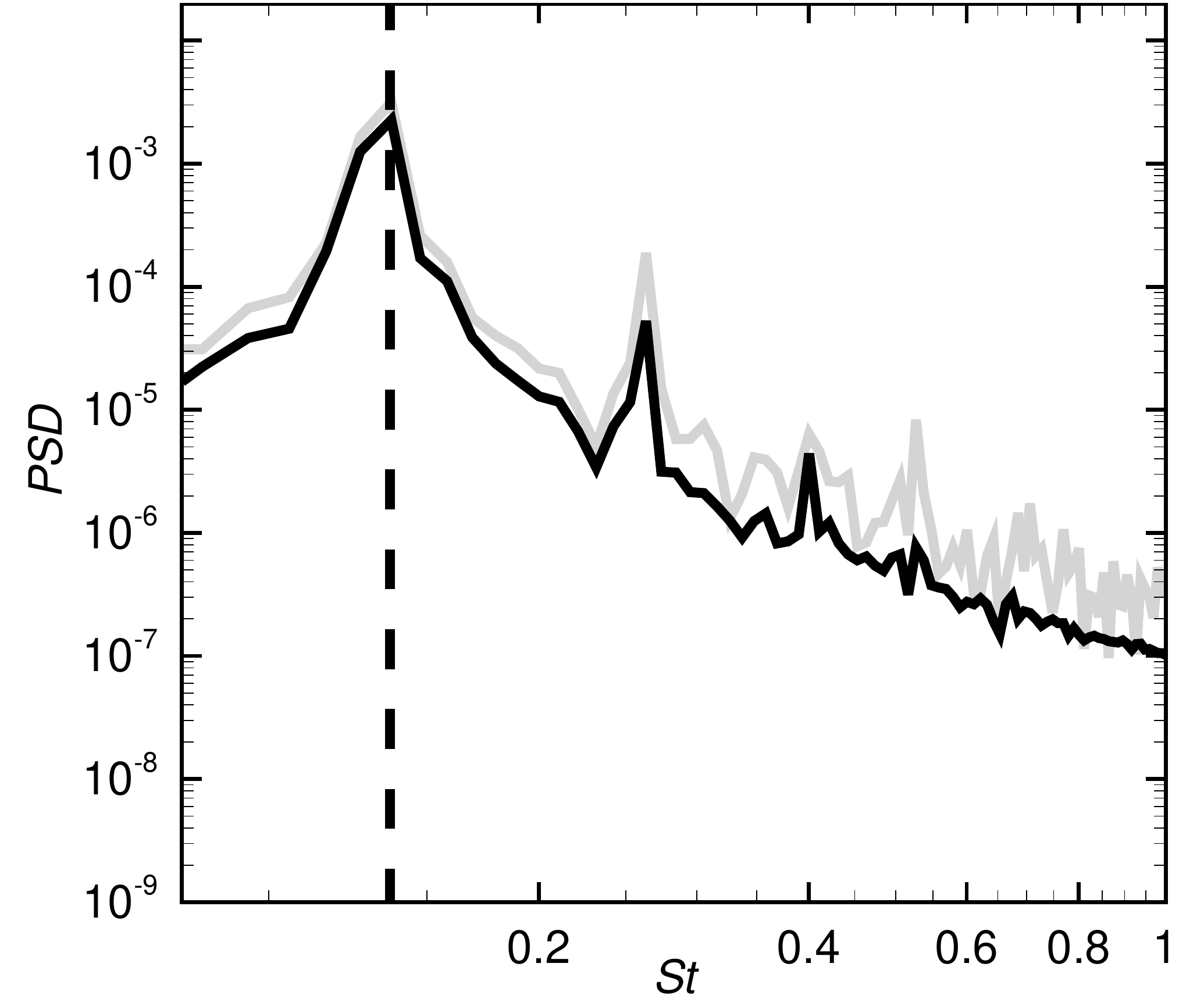}} \\
(g) & & (h)   \\
\imagetop{\includegraphics[width=0.42\columnwidth]{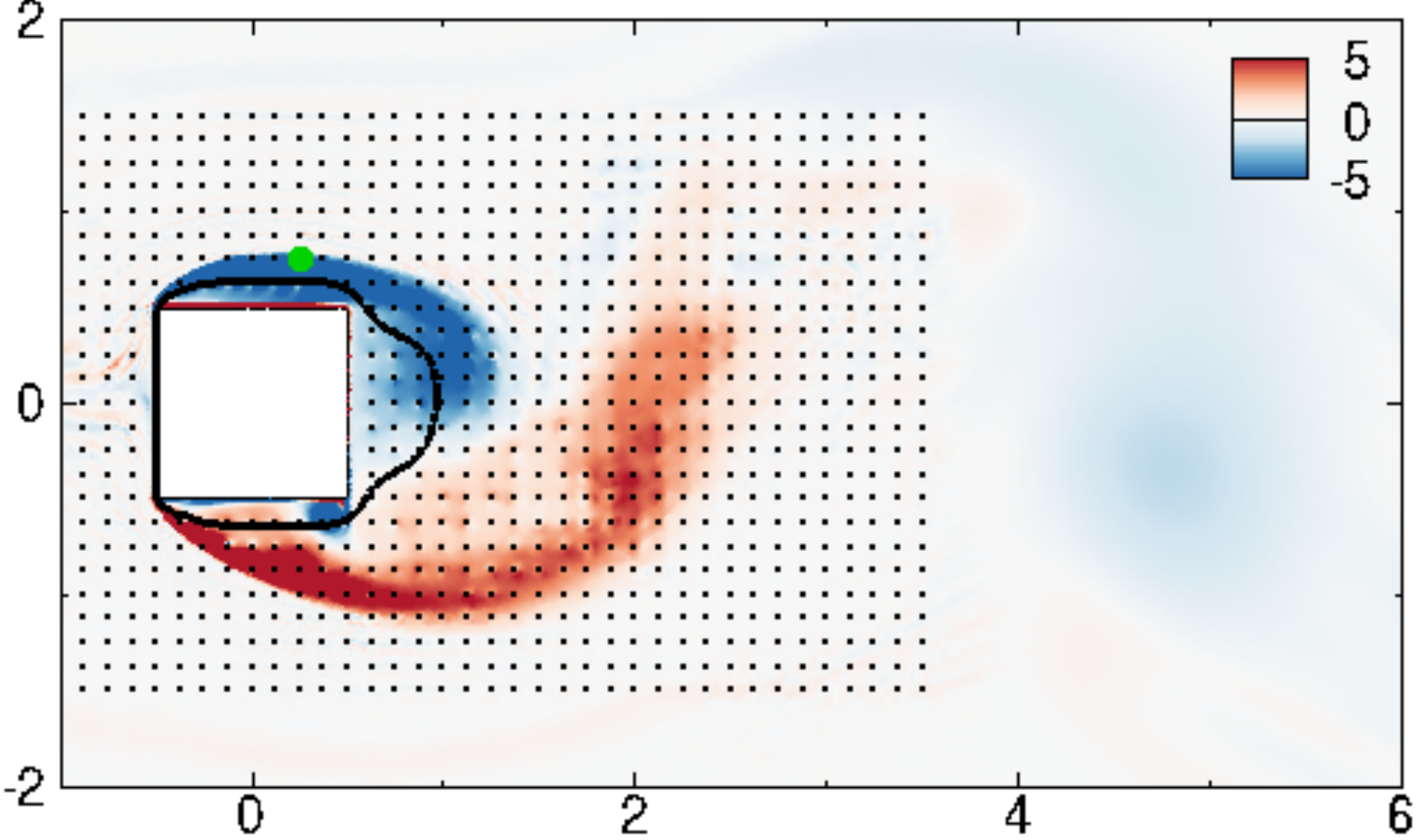}} & & \imagetop{\includegraphics[width=0.3\columnwidth]{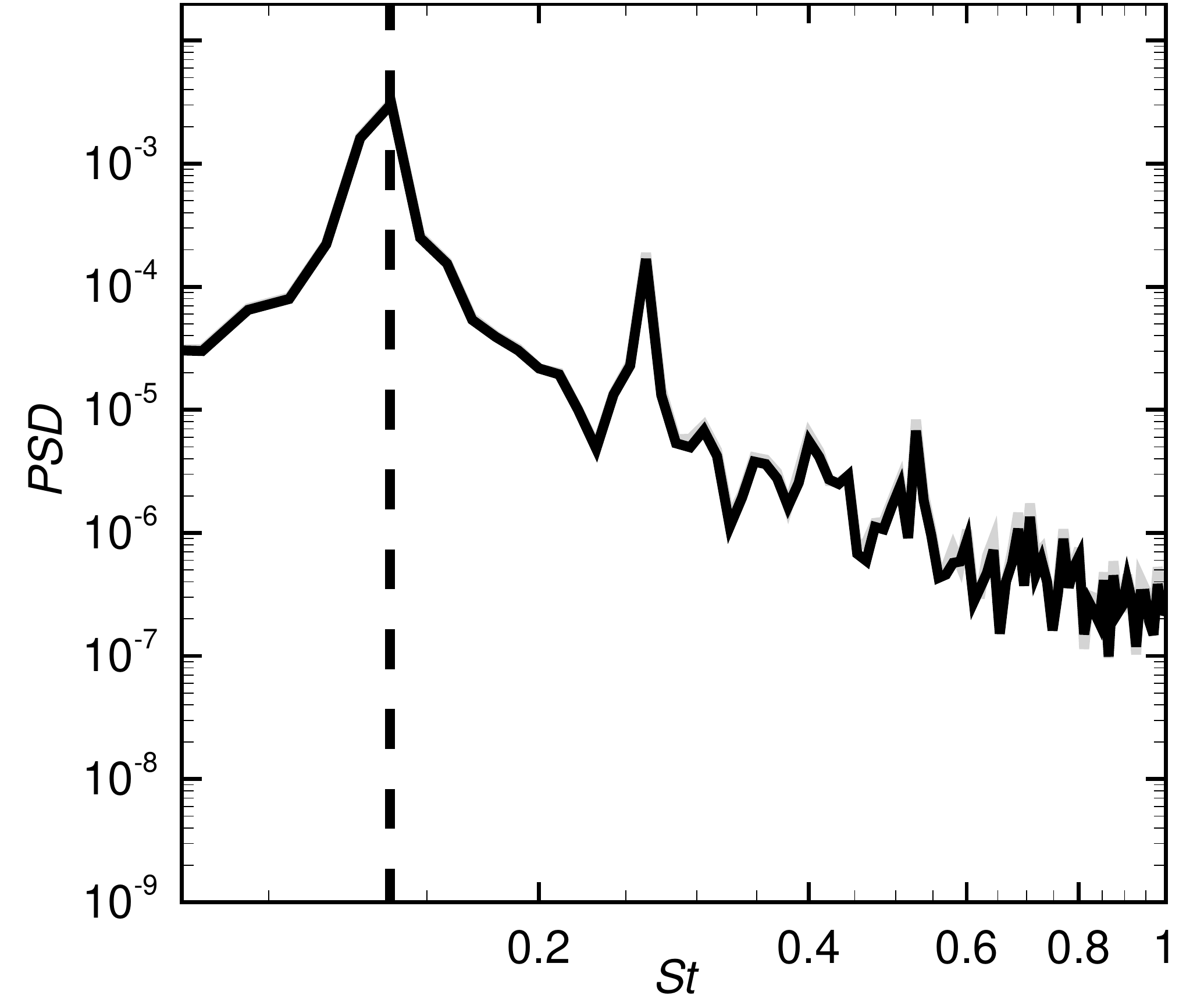}} \\
\end{tabular}
  \caption{
Results for nudged URANS simulations in data-set group VS with spatial spacings (a-b) $\Delta s = 1.66$, (c-d) $\Delta s = 1.5$, (e-f) $\Delta s = 0.5$, and (g-h) $\Delta s = 0.125$. (a,c,e,g): Instantaneous spanwise vorticity field $\bar{\omega}_{z}$ at $t=50$, where black iso-curves denote $\left< \overline{u} \right> =0$, while dots refer to nudging points. (b,d,f,h): Fourier  spectrum  of  the  streamwise  velocity  at the  green  monitor  point  in  (a,c,e,g)  (full black  line), the reference spectrum to recover is also reported (full grey curve). Vertical dashed lines indicate the low-frequency peak associated with large-scale vortex-shedding.
}
\label{fig:example3c}
\end{figure}

Figure \ref{fig:example3c} shows the vorticity fields at $t=50$ and the Fourier spectra of the streamwise velocity at point $(x,y)=(0.25,0.75)$, similarly as in figures \ref{fig:reference} and \ref{fig:standardURANS}, which are obtained for nudged URANS simulations with decreasing values of the spatial sampling $\Delta s$. In the Fourier spectra, results corresponding to nudged URANS are displayed with black curves, while the grey curves correspond to results of the reference flow when sampled at $\Delta t = 0.209$, i.e. the temporal sampling of the present data-set. This spectrum is therefore different from the reference spectrum in figures \ref{fig:reference} and \ref{fig:standardURANS} that corresponds to a temporal sampling ten times smaller, ($\Delta t = 0.021$). Their comparison clearly shows the filtering effect when down-sampling the reference flow. 

Let us first examine results obtained with the largest spatial samplings and displayed in figure \ref{fig:example3c}(a-b) for $\Delta s = 1.66$ and in figure \ref{fig:example3c}(c-d) for $\Delta s = 1.5$. For $\Delta s = 1.66$, the nudged simulation is not synchronized with the measurement data. The largest amplitude peak is obtained for a frequency that is slightly smaller than in the reference simulation (see figure \ref{fig:example3c}(b)) and is the same as in the standard URANS results ($St_{VS}=0.126$ versus $St_{VS}=0.137$ in the reference simulation). Note that, in the displayed vorticity field, we clearly observe undesired small-scale wake-like structures around some nudging points. By decreasing the spatial sampling to $\Delta s = 1.5$, the vortex-shedding frequency of the nudged simulation already locks to the reference frequency (figure \ref{fig:example3c}(d)). Furthermore, this appears to be true not only for the fundamental frequency $St_{VS}$, but also for its second and third harmonics. For the present set-up, the smallest number of measurements points that is necessary to lock the low-frequencies that are associated to large-scale vortex shedding is thus around only $N_s \sim 2.5$ points per characteristic wavelength. However, despite this frequency locking, it may be noticed that the estimated flow is still not perfectly in phase with the reference one for $\Delta s=1.5$, as further discussed in the following. Moreover, the largest amplitude peak is under-estimated, as well as the power spectral densities at higher frequencies. A better flow estimation is obtained by further decreasing the spatial sampling to $\Delta s = 0.5$ (figure \ref{fig:example3c}(e-f)). The large-scale vortices are now well synchronized with the reference flow (figure \ref{fig:reference}(a)) and the dominant peak in the spectrum is almost perfectly recovered. Moreover, the spurious small-scale structures around nudging points that were observed for larger spatial samplings are not  visible anymore. Decreasing the spacing between nudging points to $\Delta s = 0.125$ adds more small-scale features in the nudged URANS simulation, as illustrated by figure \ref{fig:example3c}(g), which gets even closer to the reference snapshot of figure \ref{fig:reference}(a). In addition, the corresponding spectrum in figure \ref{fig:example3c}(h) agrees perfectly well with the down-sampled reference one. 
\begin{figure}
\vspace{0.25cm}
\centering
\begin{tabular}[t]{l}
(a) \\
\imagetop{\includegraphics[width=0.8\columnwidth]{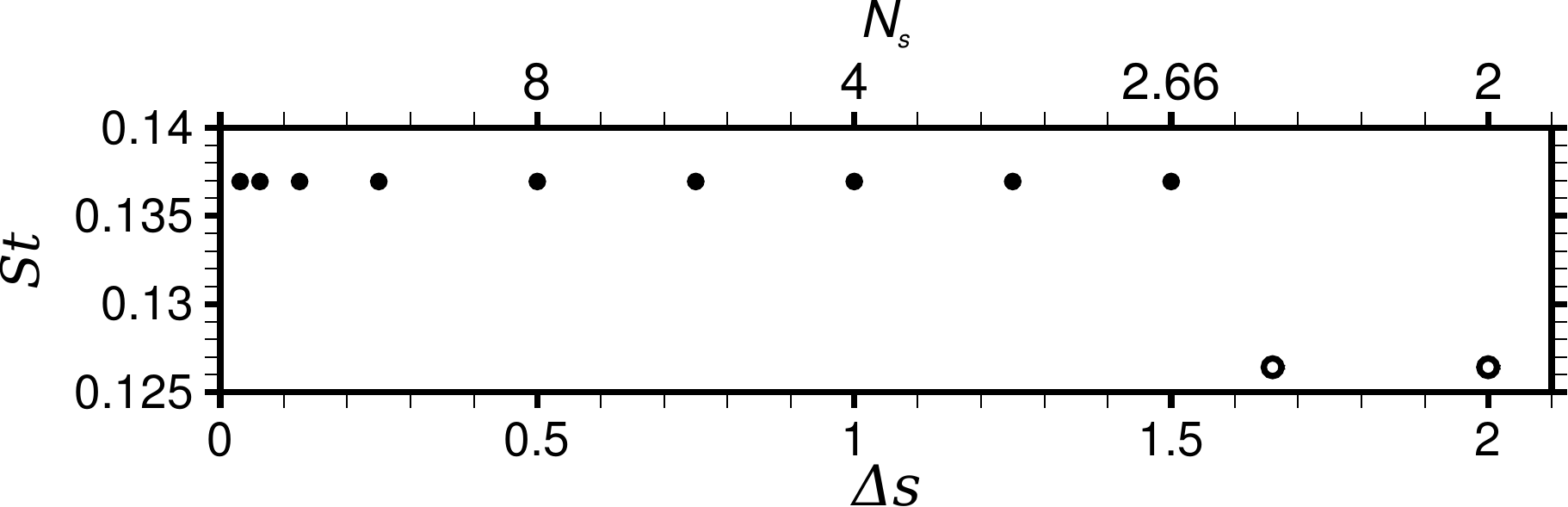}} \\
(b) \\
\imagetop{\includegraphics[width=0.8\columnwidth]{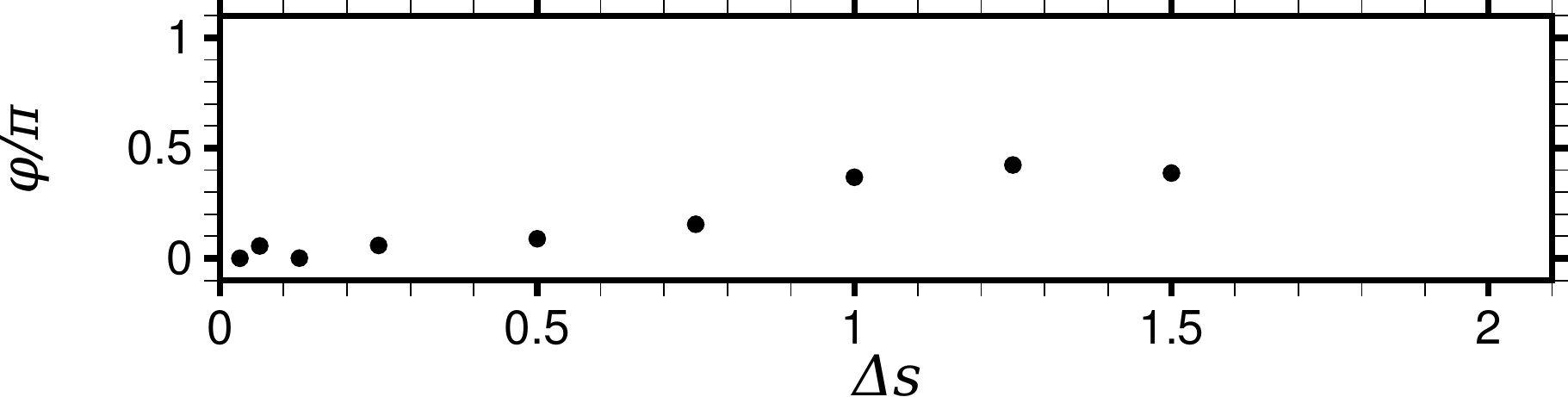}} \\
\end{tabular}
  \caption{
  (a) Dominant Strouhal number and (b) phase angle with respect to reference results as a function of the spacing between nudging points $\Delta s$, or equivalently the number $N_{s}$ of nudging points per characteristic wavelength, for nudged URANS simulations of data-set group VS. Results are obtained from the Fourier analysis of the signal at the same monitor point as in figure \ref{fig:example3c}. Filled symbols correspond to the case where the estimated shedding frequency is $St_{VS}=0.137$, as in DNS results, while open symbols correspond to $St_{VS}=0.126$, as in standard URANS.
  }
\label{fig:FreqPhase}
\end{figure}

To better assess the frequency locking and phasing effects when varying the number of nudging points, we display in figures \ref{fig:FreqPhase}(a) and (b) the frequency of the largest amplitude peak $St_{VS}$ and the phase shift $\phi$, respectively, as a function of $\Delta s$ (and $N_s$). As noticed above, the fundamental frequency of the nudged simulations is equal to that in standard URANS results ($St_{VS}=0.126$) for low spatial resolution ($N_s<2.5$) and locks to the frequency of the reference flow ($St_{VS}=0.137$) for higher spatial spatial resolution ($N_s>2.5$). This frequency locking does not guarantee a perfect flow synchronization. Indeed, for $2.5 \le Ns \le 5.75$, we observe a significant phase shift between the nudged and reference temporal evolutions of the streamwise velocity at the monitor point in figure \ref{fig:example3c}, which vanishes only for higher spatial resolution ($N_s \ge 8$).

\begin{figure}
\vspace{0.25cm}
\centering
\begin{tabular}[t]{l}
(a) \\
\imagetop{\includegraphics[width=0.8\columnwidth]{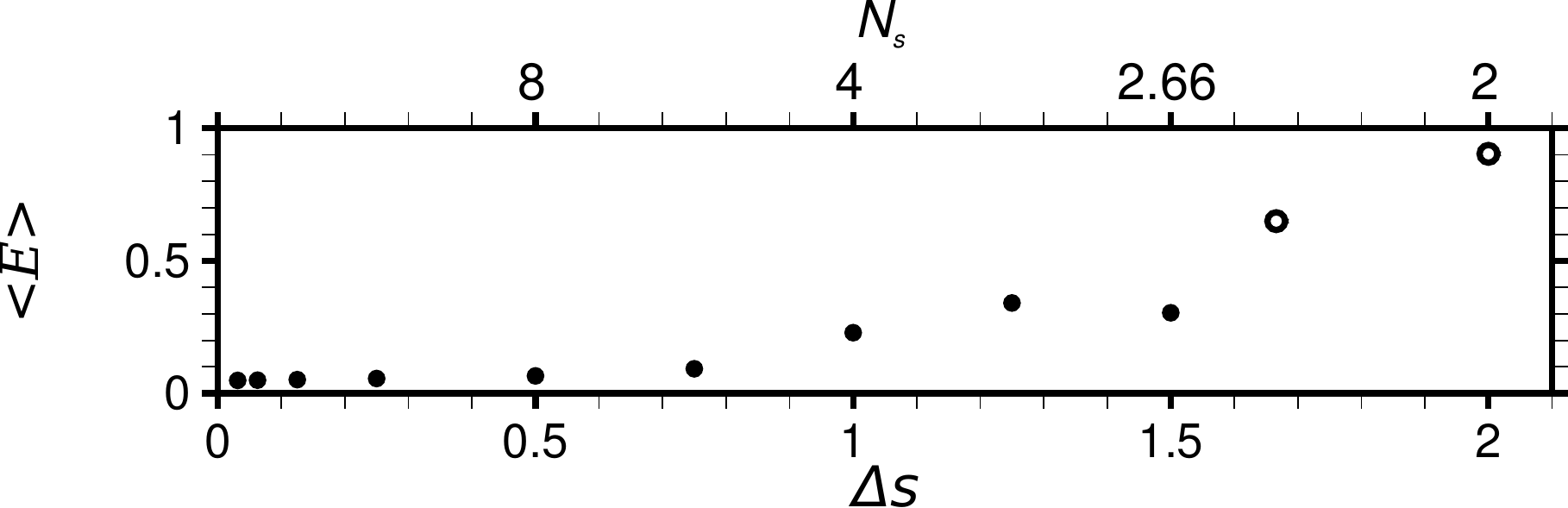}} \\
(b) \\
\imagetop{\includegraphics[width=0.8\columnwidth]{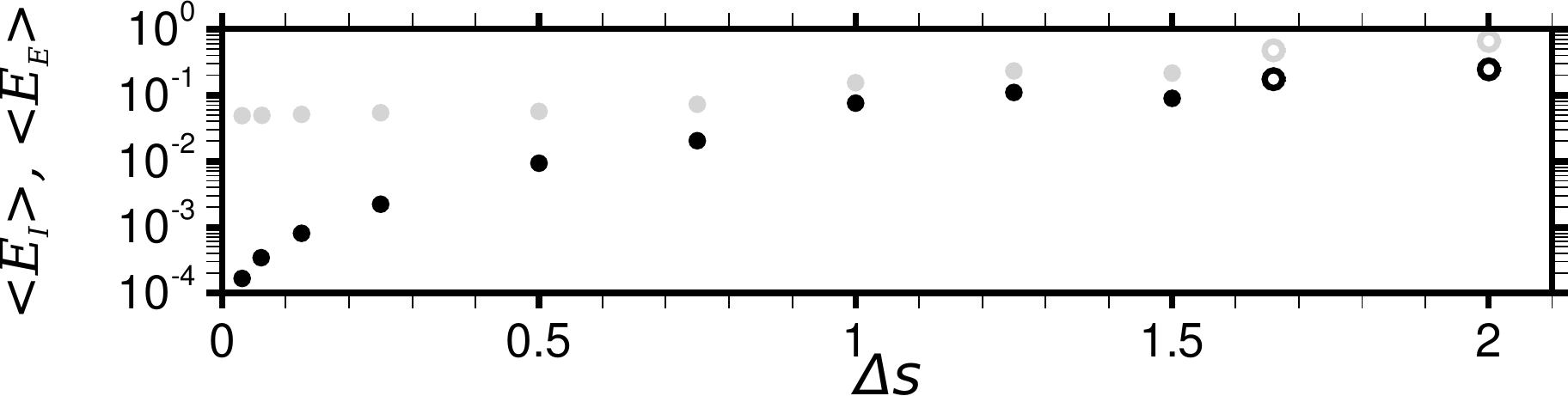}} \\
\end{tabular}
  \caption{Time-averaged errors as a function of the spacing between nudging points $\Delta s$, or equivalently the number $N_{s}$ of nudging points per characteristic wavelength, for nudged URANS simulations of data-set group VS. (a) Total error $\langle E \rangle$ in (\ref{eqn:total-error}), which may be decomposed as the sum of (b) internal $\langle {E_I} \rangle$ (black symbols) and external errors $\langle {E_E} \rangle$ (grey symbols). The reported values are normalised by the total time-averaged error for the standard URANS solution. Filled symbols correspond to the case where the estimated shedding frequency is $St_{VS}=0.137$, as in DNS results, while open symbols correspond to $St_{VS}=0.126$, as in standard URANS.
  }
\label{fig:error_deltas}
\end{figure}

The effect of the spatial resolution of measurement points on the time-averaged global error $\langle {E} \rangle$ in (\ref{eqn:total-meanerror}) is now examined through figure \ref{fig:error_deltas}(a). In this figure, $\langle {E} \rangle$ is normalised by its value for standard URANS results. For the smallest number of nudging points $N_s=2 \, (\Delta s=2)$, the normalized error is slightly smaller than one, indicating a very weak improvement compared to the standard simulation. When increasing to $N_s=2.66$, i.e the smallest number of nudging points leading to a frequency locking, the normalized error decreases to $0.3$, which corresponds to an improvement of $70\%$ compared to the standard URANS results. Interestingly, for $2.66 \le N_s \le 5.75$, the evolution of the error follows that of the phase shift shown in figure \ref{fig:FreqPhase}(b). In this range of spatial resolution, the decrease of the error is thus strongly related to the decrease of the phase shift noticed before. Finally, for $N_s \ge 8$, the time-averaged error reaches a constant value that is close to $0.05$ times the standard URANS error. 

To assess the extrapolation capability of the nudged simulations, $\langle {E} \rangle$ is decomposed into internal and external components, $\langle {E}_I \rangle$ and $\langle {E}_E \rangle$, corresponding to the error inside and outside the nudging region, respectively, similarly as in \S\ref{sec:results_concept}. The time-averaged internal and external errors, normalised by the total error of the standard simulation, are displayed as a function of $N_s$ in figure \ref{fig:error_deltas}(b) using black and grey dots, respectively. We clearly note that the external error is the dominant component, whatever the number of nudging points, as for the data-set VS+KH(a) discussed in \S\ref{sec:results_concept}. For a small number of nudging points ($N_s < 5.75$), these two errors are of similar order of magnitude, and they decrease in a similar way when increasing the number of nudging points. This is related to the improved flow synchronisation (frequency locking and phase-shift decrease) which has been previously noticed for that range of spatial resolution and occurs everywhere in the flow. Once the number of nudging points is large enough to perfectly synchronise the flow  ($N_s > 5.75$), the internal error strongly decreases unlike the external error that reaches a constant value around $5\%$ of the total error obtained by standard URANS simulation. This constant value is at the origin of the plateau reached by the total error in figure \ref{fig:error_deltas}(a). This irreducible error may be attributed to (turbulence) model errors which limit the extrapolation capability of the nudging approach. Its interpolation capability, i.e. the ability to estimate the flow in between nudging points, is nevertheless very good as the internal error decreases to $10^{-4}$. \\

\begin{figure}
\vspace{0.25cm}
\centering
\begin{tabular}[t]{l}
(a) \\
\imagetop{\includegraphics[width=0.9\columnwidth]{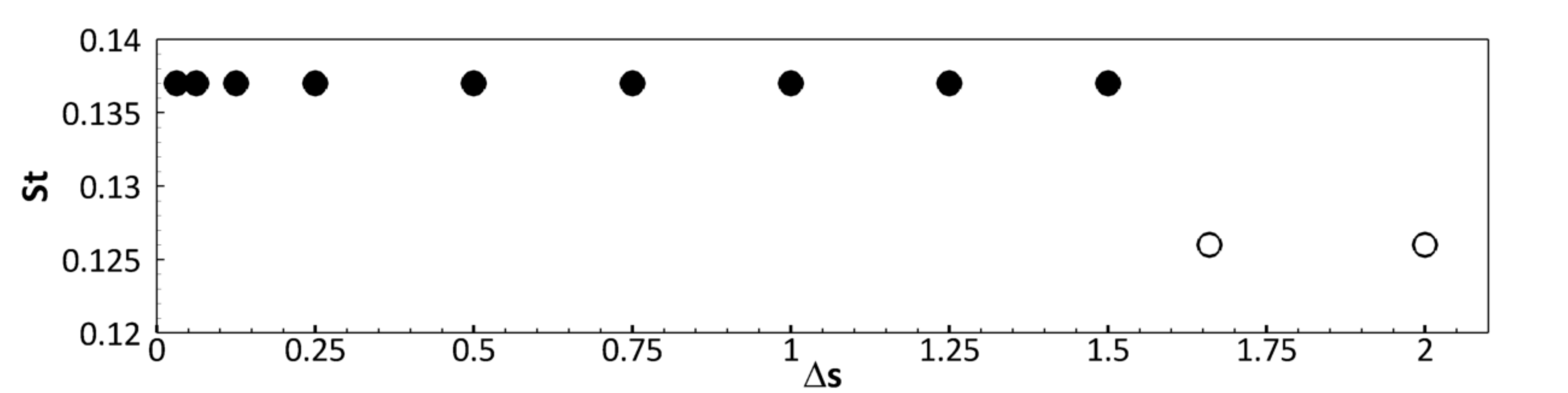}} \\
(b) \\
\imagetop{\includegraphics[width=0.9\columnwidth]{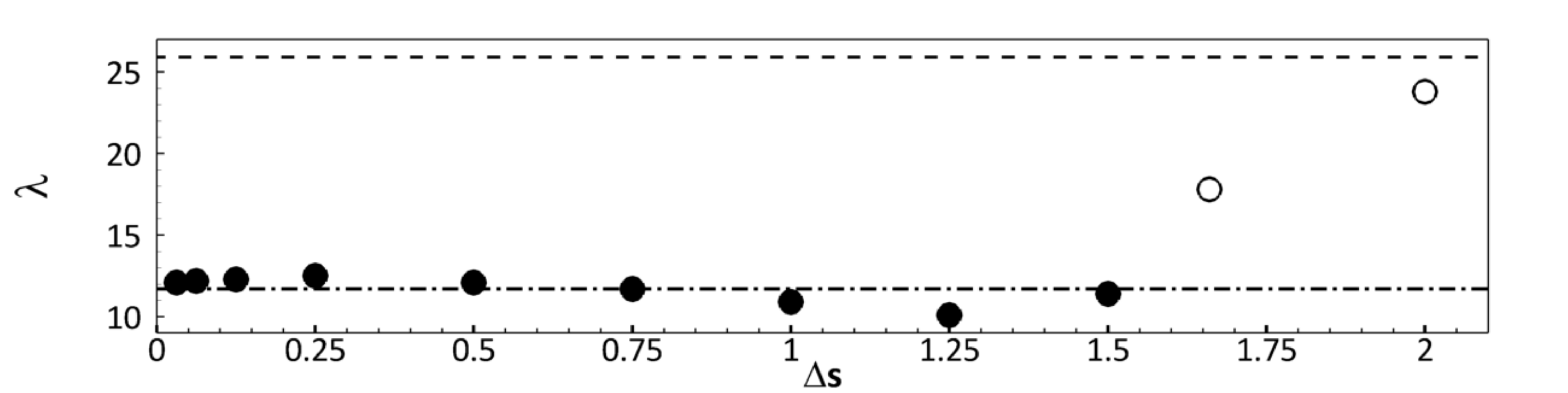}} \\
\end{tabular}
  \caption{(a) Frequency and (b) kinetic energy of the dominant SPOD modes corresponding to vortex-shedding as a function of the spatial sampling $\Delta s$ for nudged URANS simulations of data-set group VS. The dashed and dash-dotted horizontal lines in (b) correspond to energy of the SPOD mode for the URANS and reference data set, respectively. The spatial structure of few modes are displayed in figure \ref{fig:SPOD_modes_LowFreq}. Filled symbols correspond to the case where the estimated shedding frequency is $St_{VS}=0.137$, as in DNS results, while open symbols correspond to $St_{VS}=0.126$, as in standard URANS.}
\label{fig:SPOD_deltas}
\end{figure}

\begin{figure}
\vspace{0.25cm}
\begin{tabular}[t]{ll}
(a) & (b) \\
\imagetop{\includegraphics[width=0.48\columnwidth]{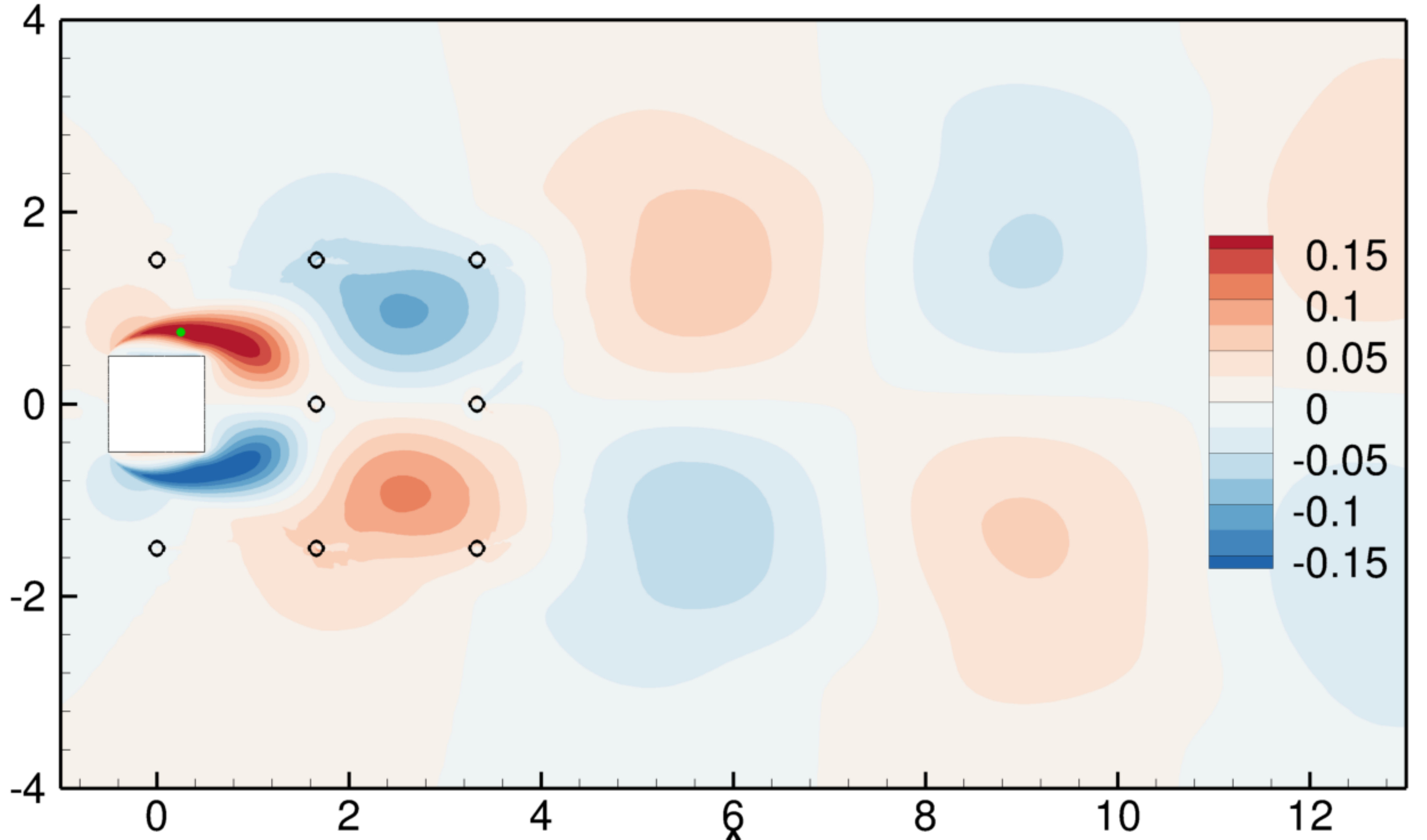}} & 
\imagetop{\includegraphics[width=0.48\columnwidth]{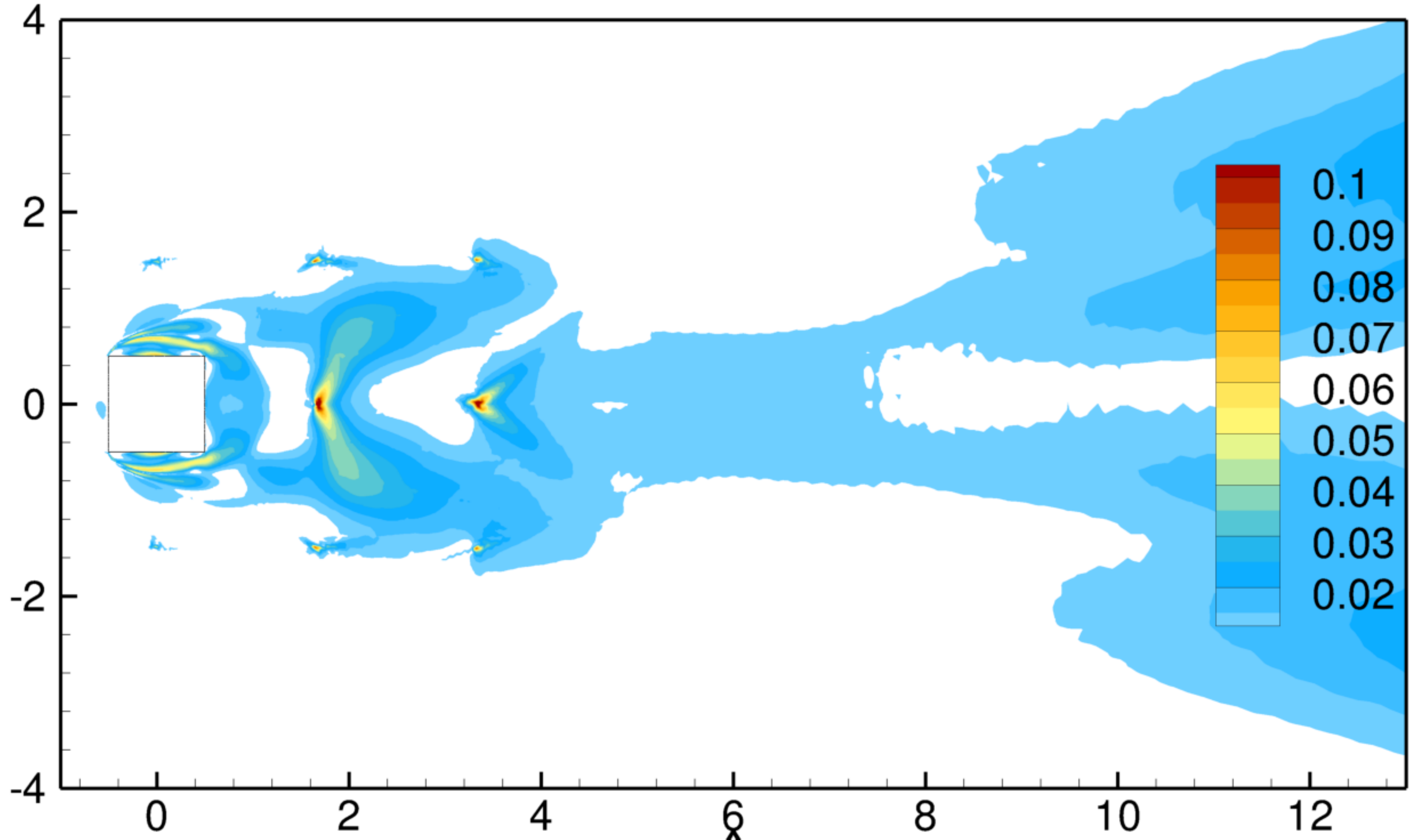}} \\
(c)  & (d)  \\
\imagetop{\includegraphics[width=0.48\columnwidth]{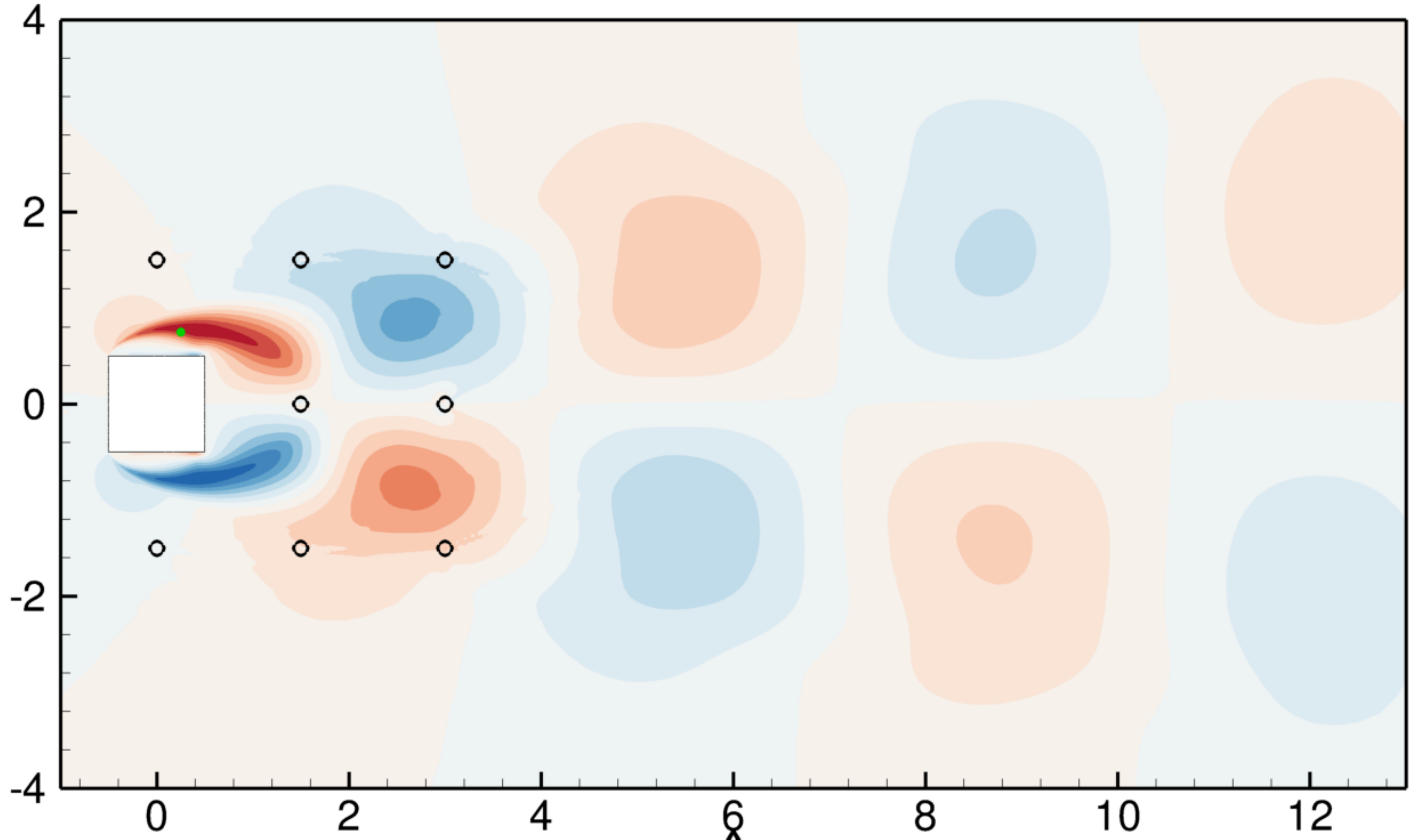}} & 
\imagetop{\includegraphics[width=0.48\columnwidth]{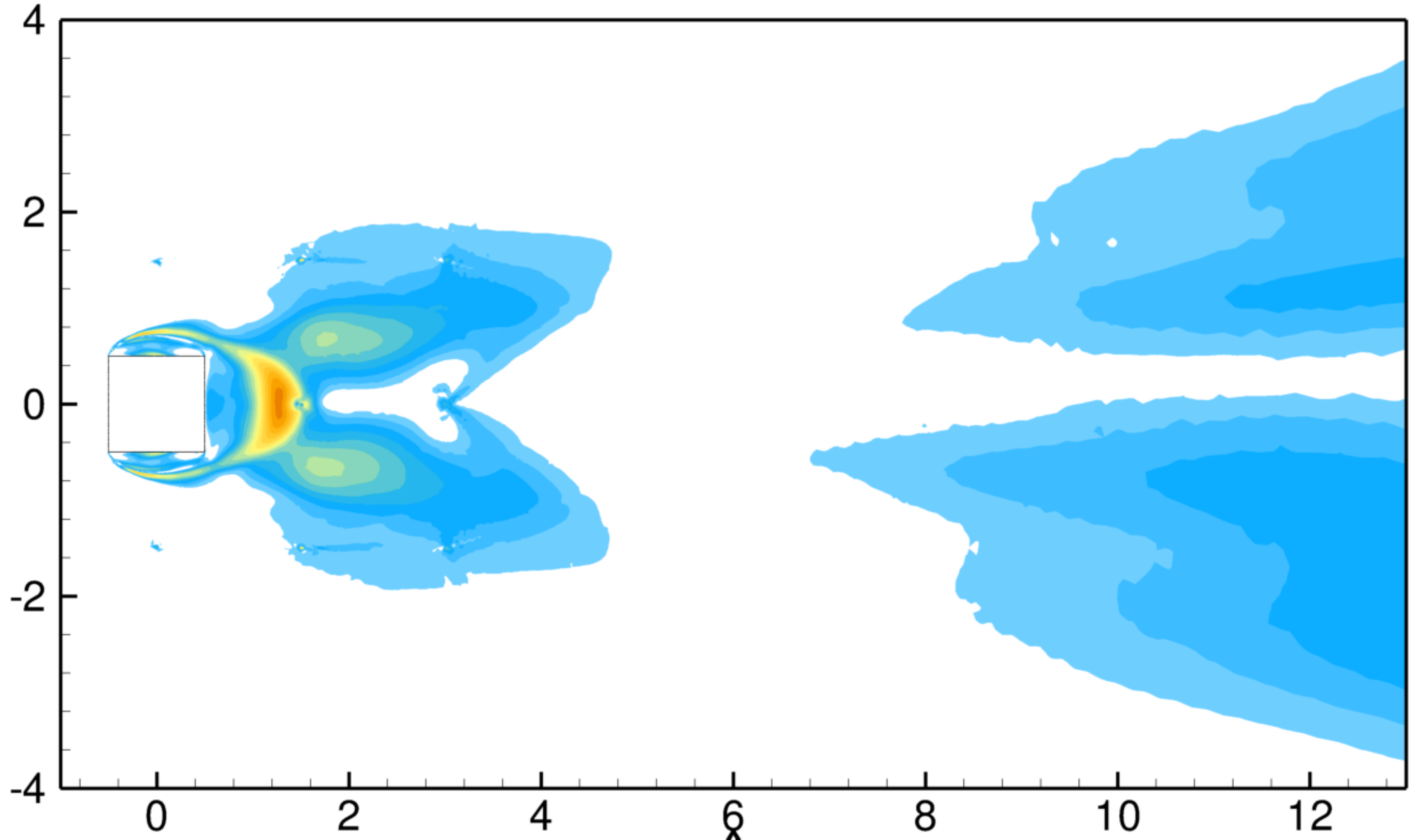}} \\
(e)  & (f)\\
\imagetop{\includegraphics[width=0.48\columnwidth]{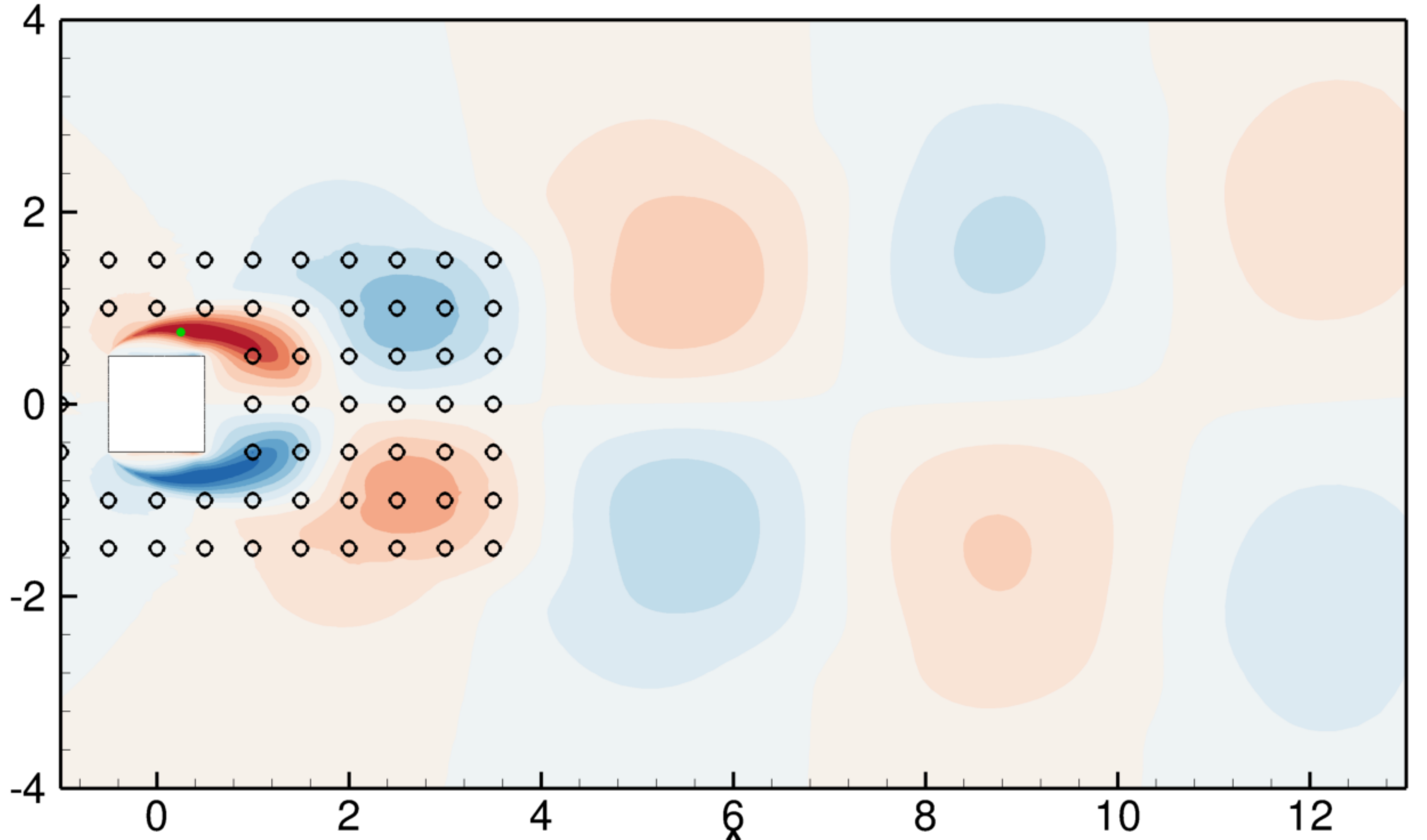}} & 
\imagetop{\includegraphics[width=0.48\columnwidth]{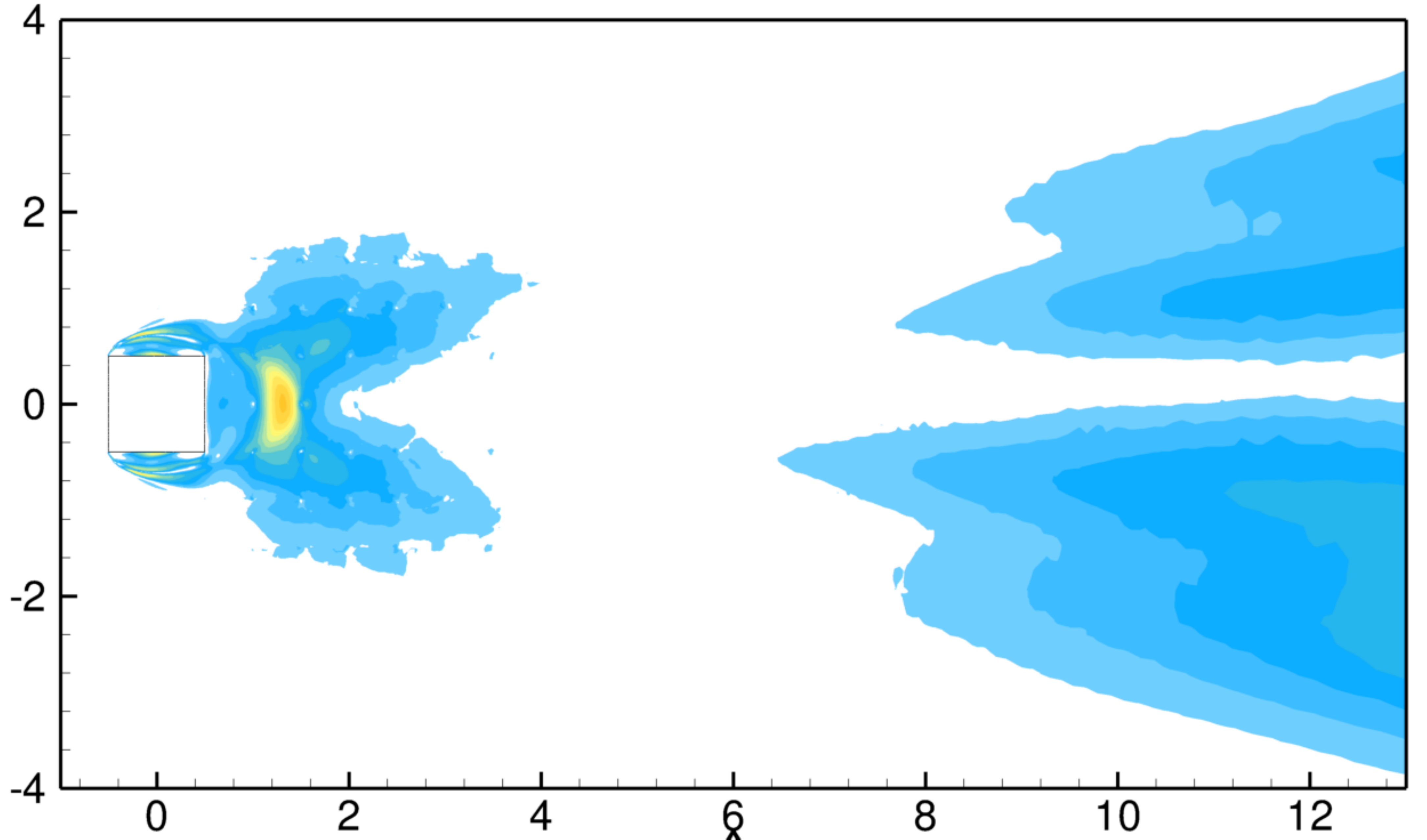}} \\
(g)  & (h)\\
\imagetop{\includegraphics[width=0.48\columnwidth]{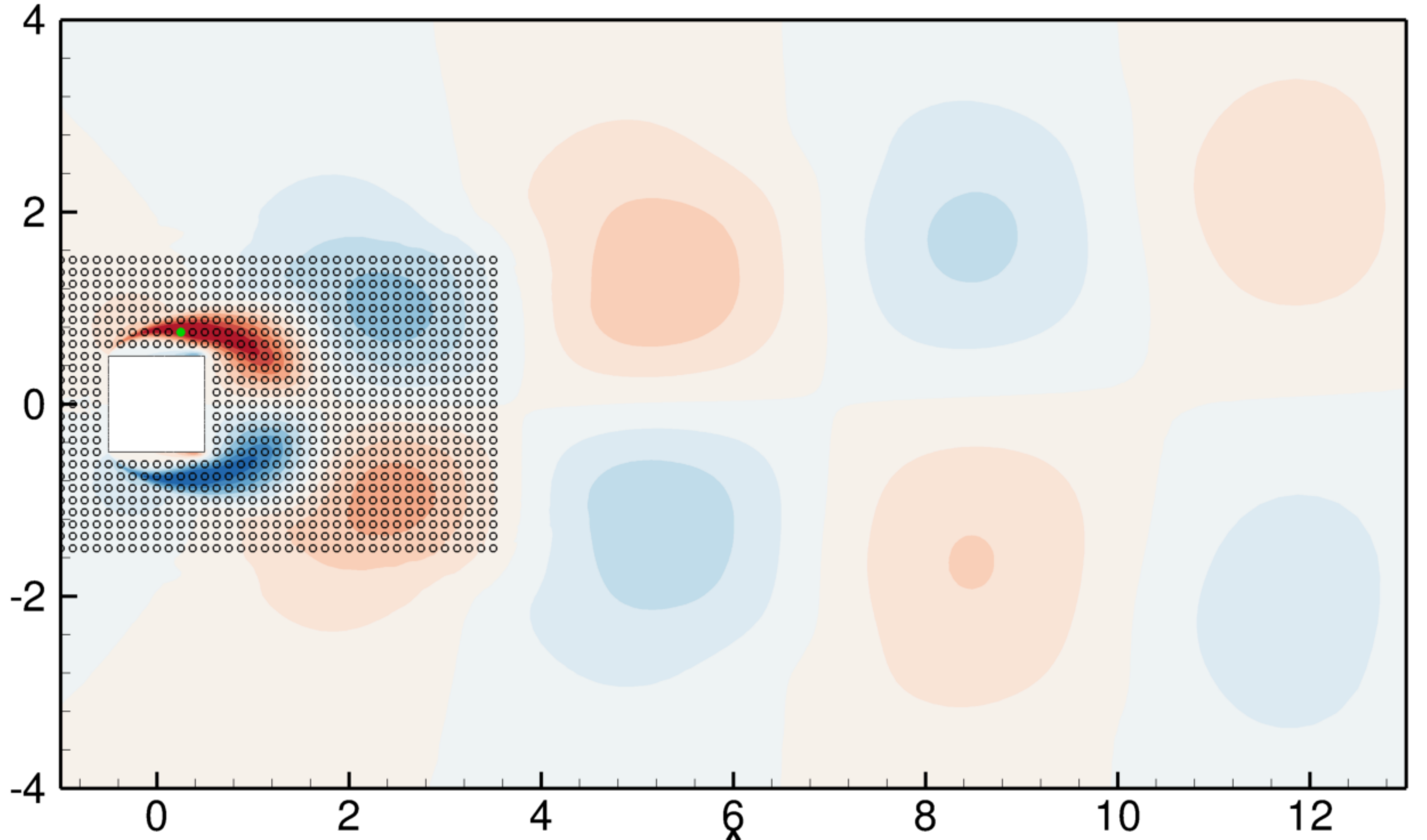}} & 
\imagetop{\includegraphics[width=0.48\columnwidth]{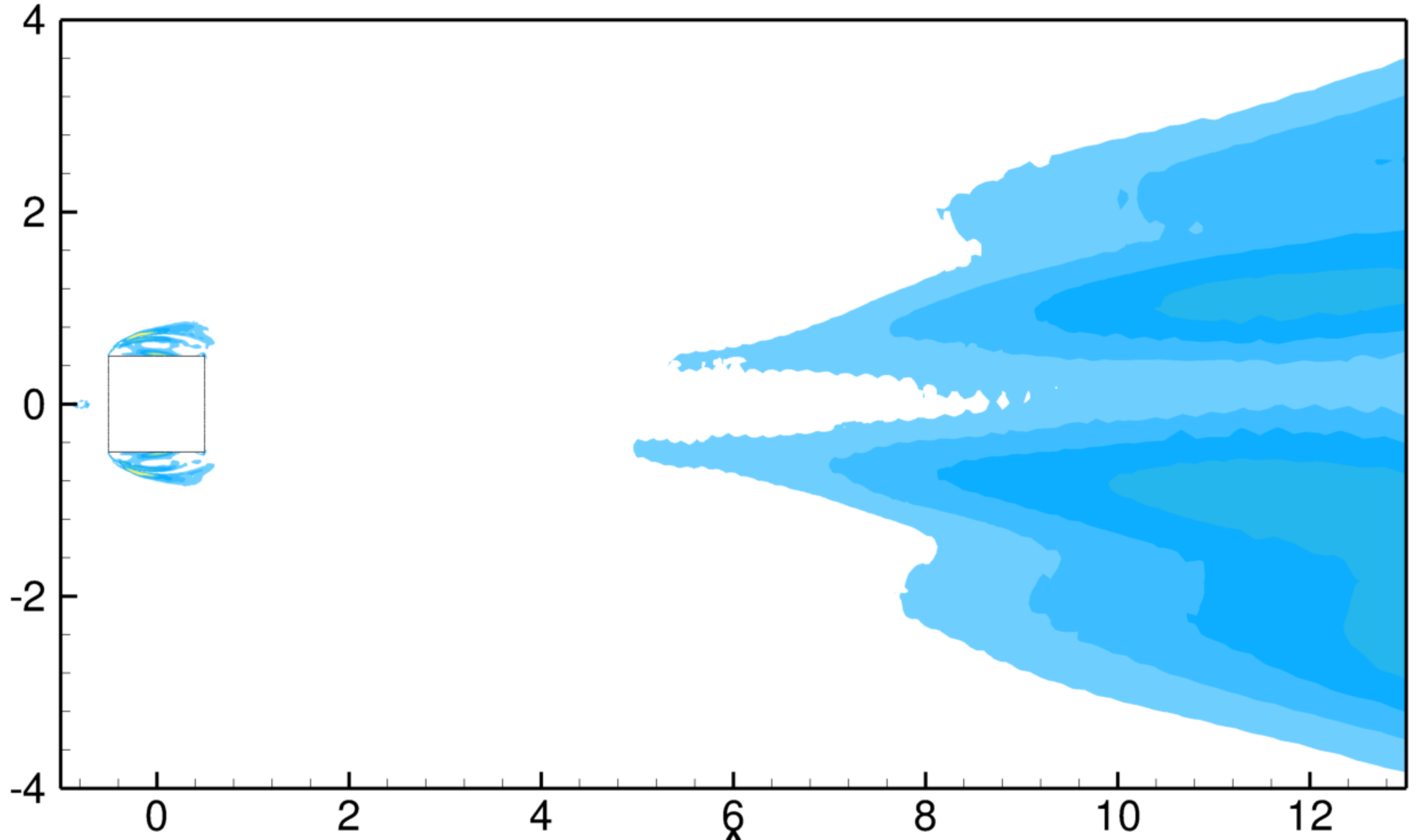}} \\
\end{tabular}
  \caption{
  SPOD analysis of the vortex-shedding phenomenon for nudged URANS simulations in data-set group VS with spatial spacings (a-b) $\Delta s = 1.66$, (c-d) $\Delta s = 1.5$, (e-f) $\Delta s = 0.5$, and (g-h) $\Delta s = 0.125$. (a,c,e,g): Streamwise velocity contours of the imaginary part of the dominant SPOD mode for the fundamental frequency $St_{VS}$. While the frequency of the mode in (a) is $St_{VS}=0.126$, remaining modes oscillate at $St=0.137$. Modes are phased so that each of them reaches its maximum value at the green monitor point. (b,d,f,h): Corresponding modal discrepancy field $\Phi_e$ in (\ref{eqn:modal-error}).
   }
\label{fig:SPOD_modes_LowFreq}
\end{figure}

Having carefully characterised in time the flow estimations for the data-set group VS, we now analyse these results in the frequency space through SPOD, as performed in previous sections. For the present data-set, SPOD is computed based on overlapping bins that cover about $63$ time units corresponding to approximately $9$ vortex-shedding cycles. The frequency $St_{VS}$ and eigenvalue $\lambda_{VS}$ (kinetic energy) corresponding to the largest amplitude peak (see figure \ref{fig:SPOD_Spec_DNS_VS}(b)) are shown in figure \ref{fig:SPOD_deltas}(a) and (b) respectively, as a function of the spacing between measurements $\Delta s$. For $\Delta s > 1.5$, the frequency of the nudged mode is equal to that of the URANS mode ($St_{VS}=0.126$), while for $\Delta s \le 1.5$, it locks to the frequency of the reference mode ($St_{VS}=0.137$), in agreement with previous observations deduced from a pointwise temporal signal (figure \ref{fig:FreqPhase}). Figure \ref{fig:SPOD_deltas}(a) thus confirms the global aspect of frequency locking or $\Delta s \le 1.5$.
Examining now the estimated kinetic energy $\lambda_{VS}$ of dominant SPOD mode at $St_{VS}$ in figure \ref{fig:SPOD_deltas}(b), it is almost equal to that of the reference mode (dash-dotted line) once the frequency is locked to the reference frequency. Using $N_s=2$ points per wavelength ($\Delta s=2$ in the figure), the kinetic energy is still close to that of the URANS mode (dashed line). But a slight increase to $N_s=2.66$ points ($\Delta s=1.5$) yields an almost perfect agreement with the reference value. Further increasing the number of nudging point (decreasing their spacing $\Delta s$) does not significantly improve the estimation of $\lambda_{VS}$. The spatial structure of the corresponding mode is shown in figures \ref{fig:SPOD_modes_LowFreq}(a,c,e,g) for several spatial spacings $\Delta s$. 
In addition, the corresponding spectral error fields $\Phi_e$ in (\ref{eqn:modal-error}) are reported in figures \ref{fig:SPOD_modes_LowFreq}(b,d,f,h). For the SPOD mode associated to the unlocked frequency (figure \ref{fig:SPOD_modes_LowFreq}(a-b)), the largest errors are obtained close to some of the nudging points downstream of the cylinder. They are associated to errors in the cross-stream velocity (not displayed here) since the streamwise velocity is zero on the symmetry axis $y=0$. These specifics defects vanish once the frequency is locked to the reference frequency (figures  \ref{fig:SPOD_modes_LowFreq}(c-d)). Largest errors are now obtained in between nudging points, in a (red) pocket immediately downstream of the cylinder ($x<1.5$) and in two symmetric (yellow) pockets around $x=2$. Both the shape and the intensity of the errors are actually similar to those for standard URANS (figure \ref{fig:SPOD_Spec_DNS_VS}(d)). Increasing the number of nudging points to $\Delta s=0.5$ (figures \ref{fig:SPOD_modes_LowFreq}(e-f)) allows for decreasing these errors in the near wake. One may directly identify for the mode in figure \ref{fig:SPOD_modes_LowFreq}(e) a change in the structures around $2 \leq x \leq 4$ compared to previous cases, these structures being now closer to the ones in the reference mode (figure \ref{fig:SPOD_Spec_DNS_VS}(a)). However, modal errors are significantly reduced only for spacings $\Delta s \leq 0.25$, as for the case $\Delta s = 0.125$, which is illustrated in figures \ref{fig:SPOD_modes_LowFreq}(g-h). The residual errors in the far wake ($x>8$) are barely modified when further increasing the number of nudging points. Accurately recovering the detailed shape of the dominant SPOD mode associated to vortex shedding thus seems to require more spatially dense measurements ($N_s\geq 16$ nudging points per characteristic wavelength) compared to the estimation of its frequency and kinetic energy/magnitude ($N_s \sim 2.5$). However, it may be emphasised that the estimation of the shape of such a mode, which is already acceptable with standard URANS (see figure \ref{fig:SPOD_Spec_DNS_VS}), may be considered as only secondary to the correct prediction of associated frequency and kinetic energy. This is supported by the low values for the time-averaged global errors $\left \langle E \right \rangle$ that are already reached for $N_s \geq 2.66$, i.e. after frequency locking and phasing are achieved, and also with a correct estimation of $\lambda_{VS}$, as discussed previously.

\begin{figure}
\vspace{0.25cm}
\centering
\begin{tabular}[t]{ll}
(a) & (b) \\
\imagetop{\includegraphics[width=0.4\columnwidth]{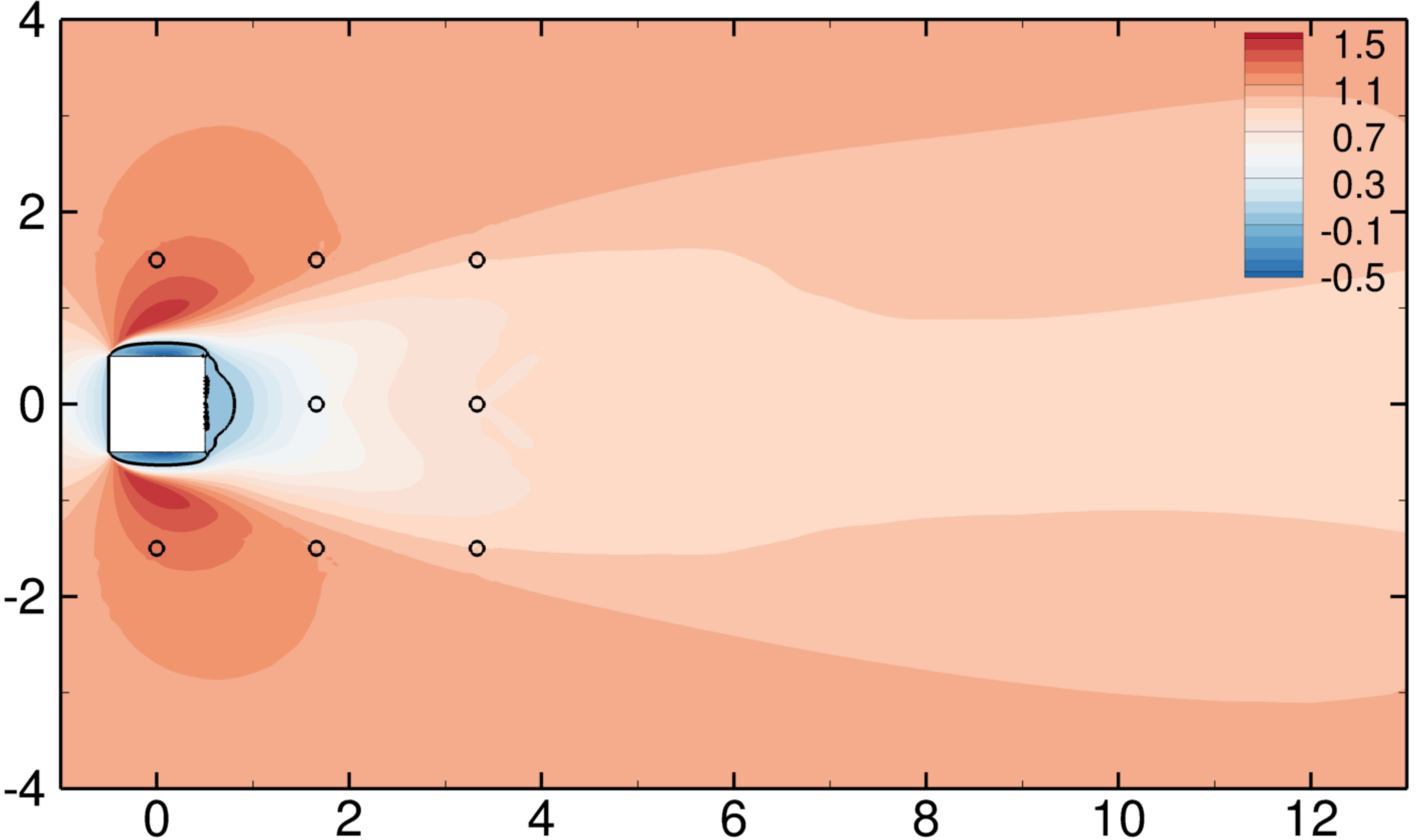}} & 
\imagetop{\includegraphics[width=0.4\columnwidth]{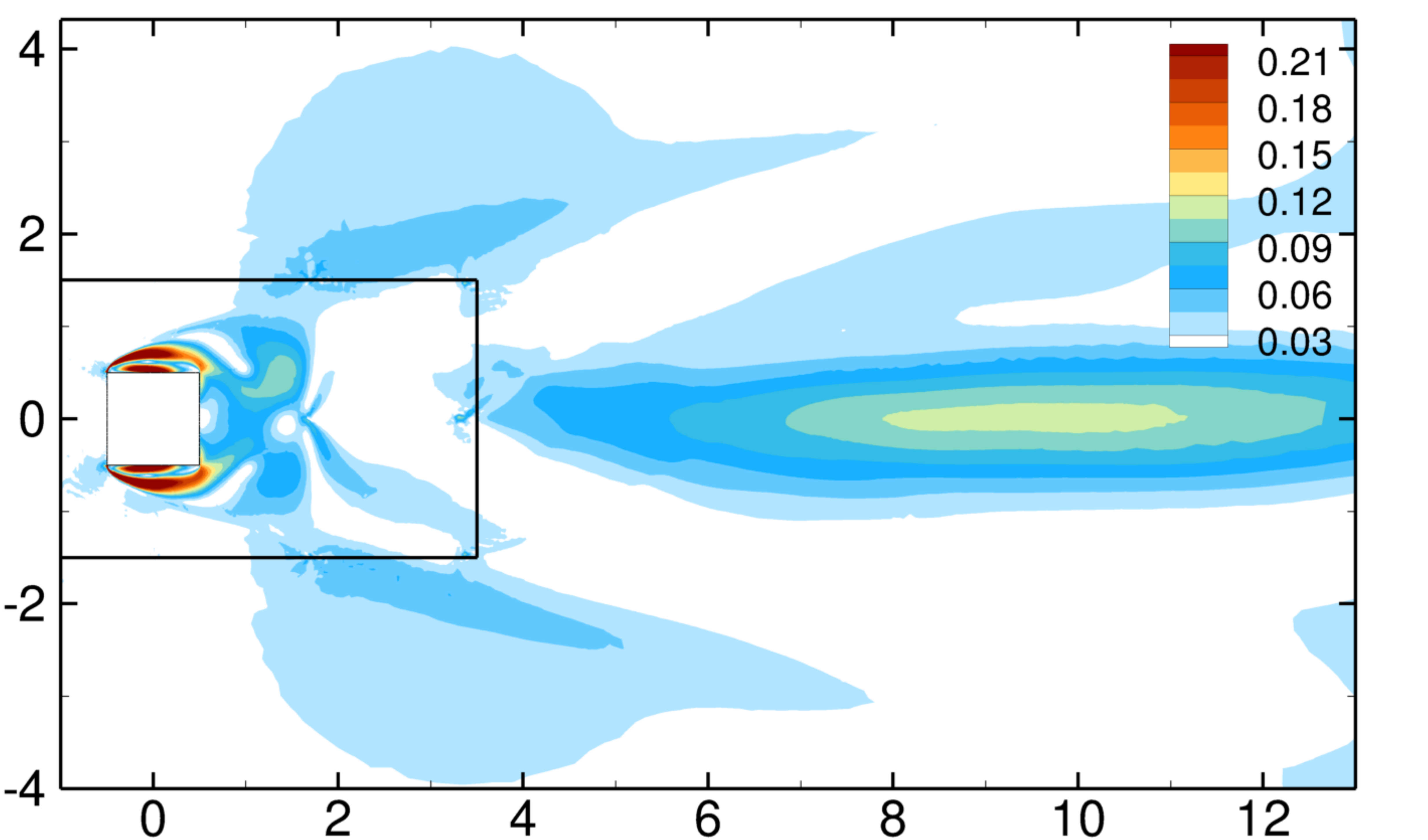}} \\
(c) & (d) \\
\imagetop{\includegraphics[width=0.4\columnwidth]{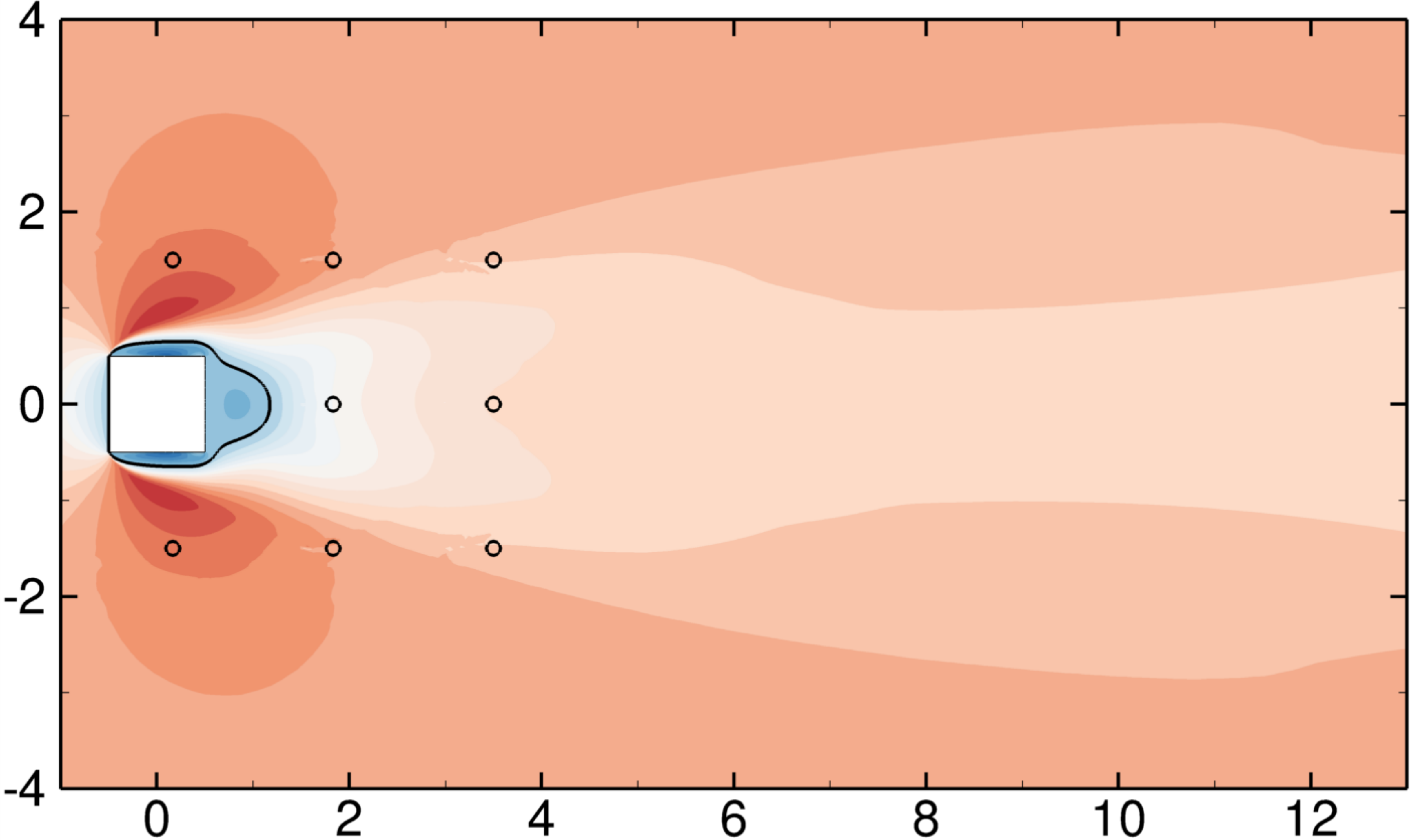}} & 
\imagetop{\includegraphics[width=0.4\columnwidth]{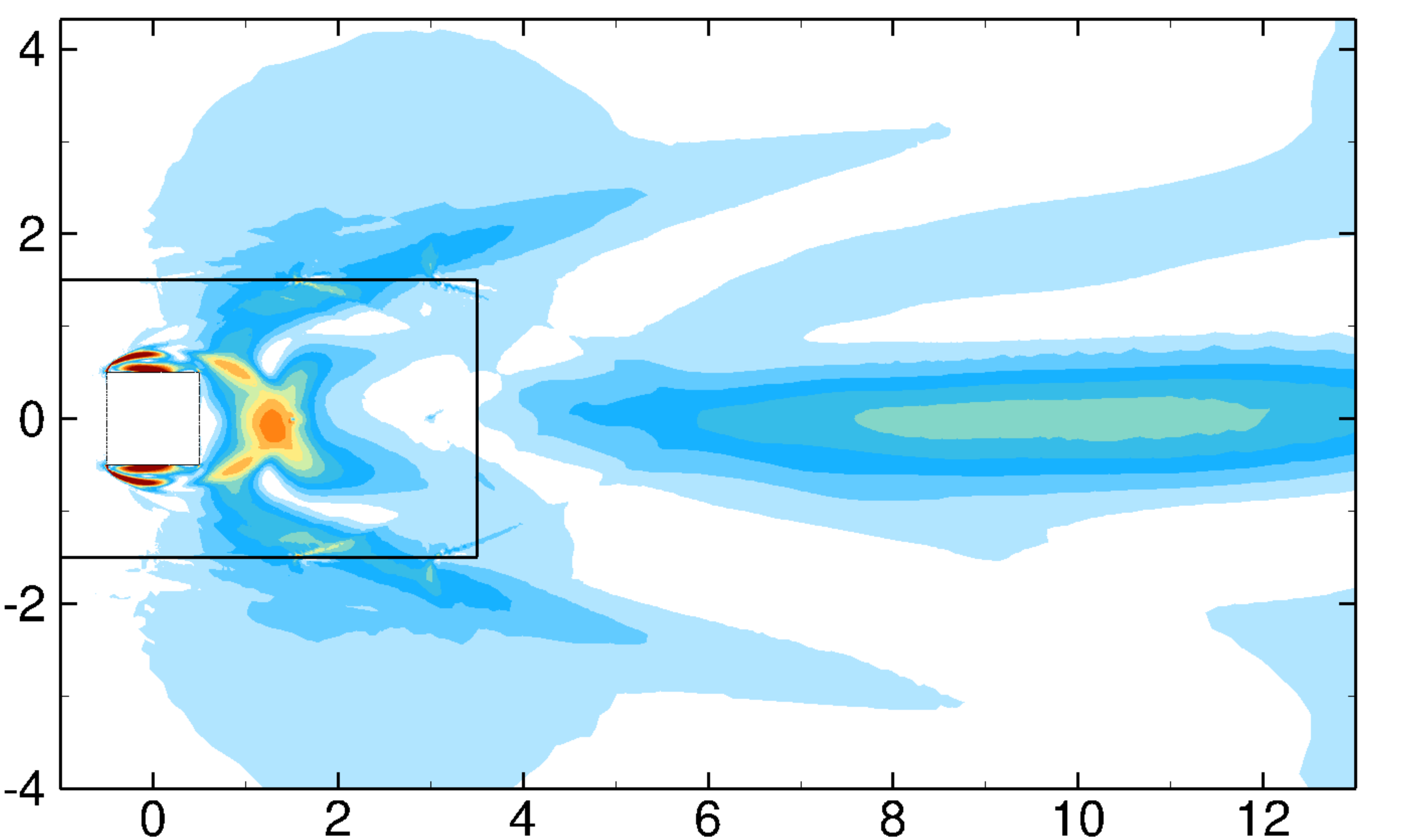}} \\ 
(e)  & (f)  \\
\imagetop{\includegraphics[width=0.4\columnwidth]{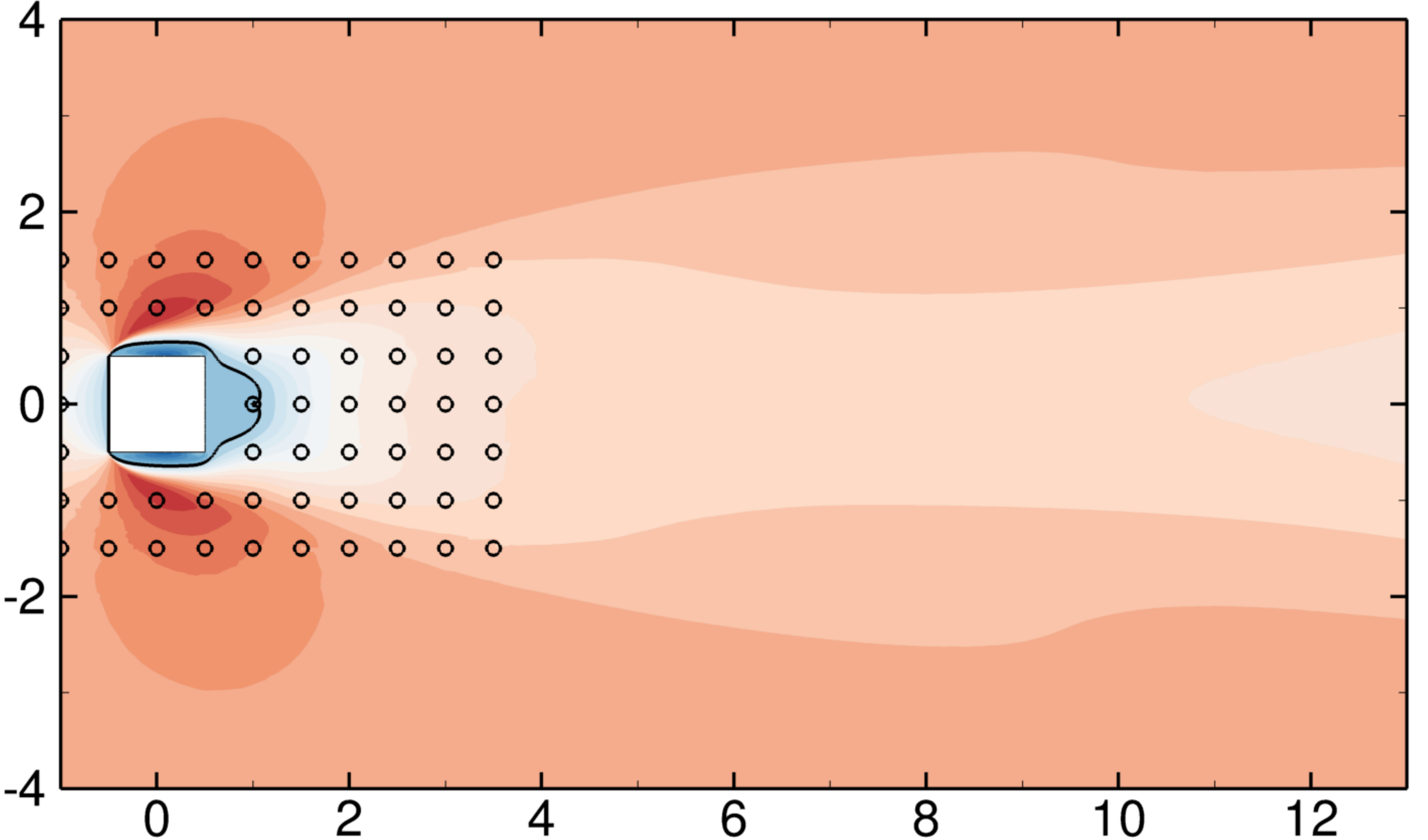}} & 
\imagetop{\includegraphics[width=0.4\columnwidth]{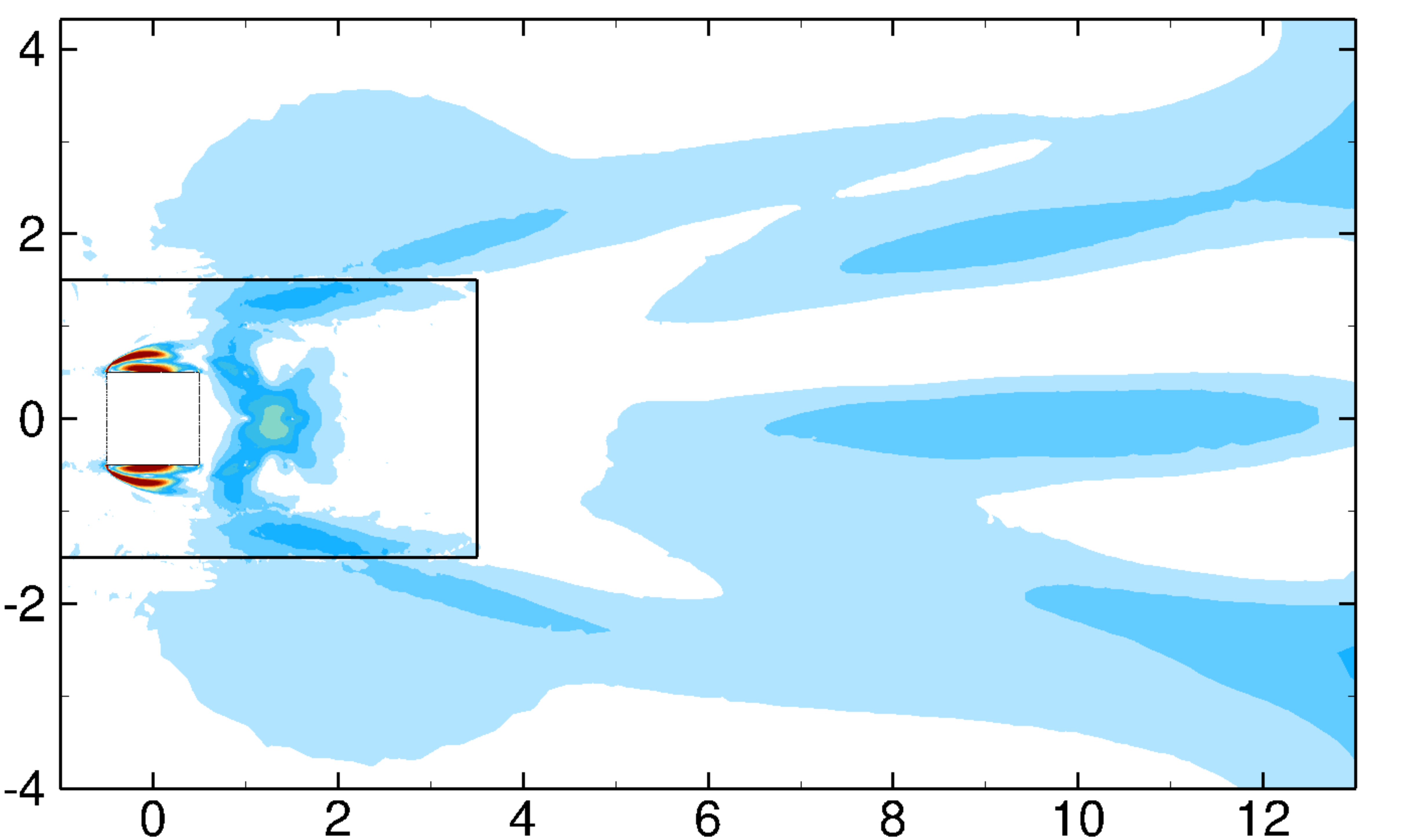}} \\
(g)  & (h)\\
\imagetop{\includegraphics[width=0.4\columnwidth]{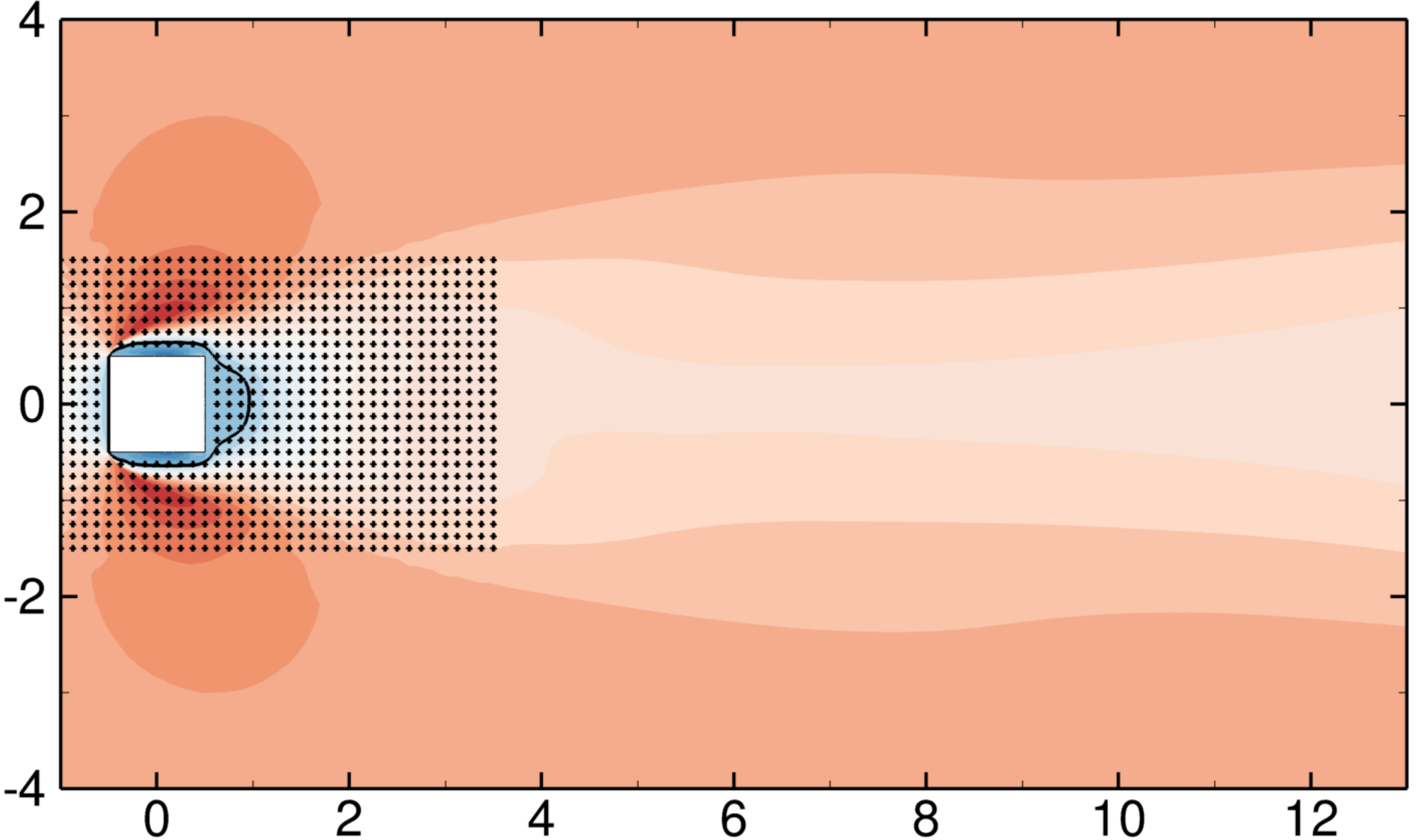}} & 
\imagetop{\includegraphics[width=0.4\columnwidth]{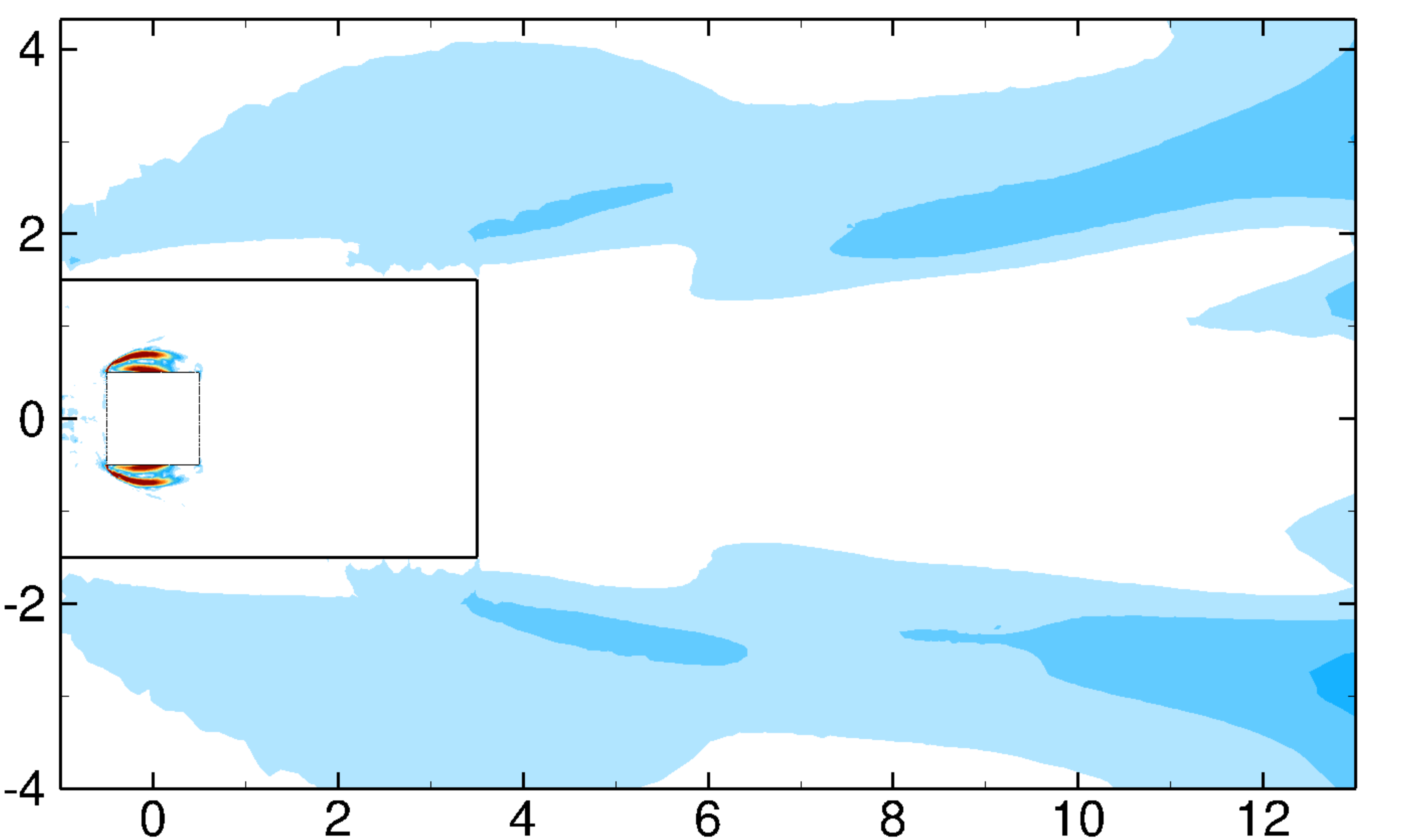}} 
\end{tabular}
  \caption{
    Mean-flow results for nudged URANS simulations in data-set group VS with spatial spacing (a,b) $\Delta s = 1.66$, (c,d) $\Delta s = 1.5$, (e,f) $\Delta s = 0.5$ and (g,h) $\Delta s = 0.125$. (a,c,e,g): Streamwise component of the mean flow. Black curves correspond to the isocontours of zero streamwise velocity. (b,d,f,h): Mean-flow error field $e_{\left\langle \bar{\boldsymbol{u}}\right\rangle}$ in (\ref{eqn:total-meanerror}). The extent of the nudging region is delineated by black lines.}
\label{fig:MeanErrorFieldUnsteady}
\end{figure}

Finally in this section, the quality of the mean flow as estimated by the nudged URANS simulations for data-set group VS is discussed. The streamwise mean velocity field is depicted in figures \ref{fig:MeanErrorFieldUnsteady}(a,c,e,g) for several values of $\Delta s$ and can be compared to the reference and URANS mean flows displayed in figures \ref{fig:MeanField}(a) and (b), respectively. The corresponding error fields $e_{\left\langle \bar{\boldsymbol{u}}\right\rangle}$ in (\ref{eqn:total-meanerror}) are reported in figures \ref{fig:MeanErrorFieldUnsteady}(b,d,f,h). For the lowest spatial resolution ($\Delta s=1.66$), significant improvement is already observed in the far wake outside of the nudging region, which is remarkable as the fundamental frequency $St_{VS}$ is not locked to the reference one in this case. Once it is locked for $\Delta s=1.5$ (figures  \ref{fig:MeanErrorFieldUnsteady}(c-d)), the error increases in the near wake, similarly as the spectral error of the SPOD mode in figure \ref{fig:SPOD_modes_LowFreq}(d) for the same spacing. This mean-flow error can be decreased by increasing the number of nudging points. Significant improvements may already be identified for $\Delta s=0.5$ (figures \ref{fig:MeanErrorFieldUnsteady}(e-f)). For the largest spatial resolution ($\Delta s=0.125$) displayed in figure \ref{fig:MeanErrorFieldUnsteady}(g-h), the mean-flow error drops to below $3\%$ in the near and far wake. Only the error in the shear layers cannot be further decreased when increasing the number of nudging points. 

\begin{figure}
\vspace{0.25cm}
\centering
\begin{tabular}[t]{l}
(a) \\
\imagetop{\includegraphics[width=0.9\columnwidth]{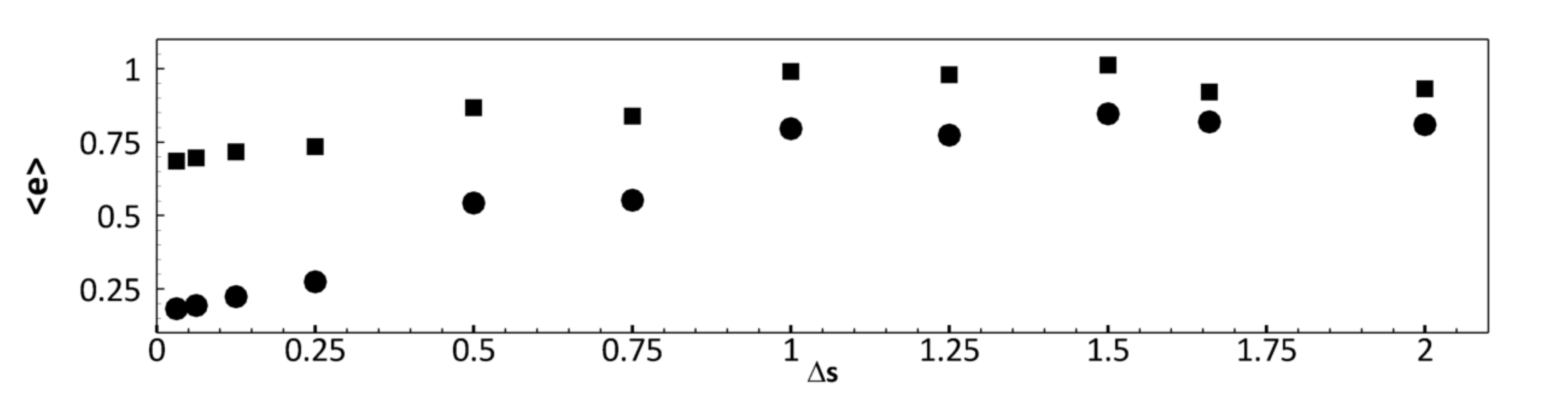}} \\
(b) \\
\imagetop{\includegraphics[width=0.9\columnwidth]{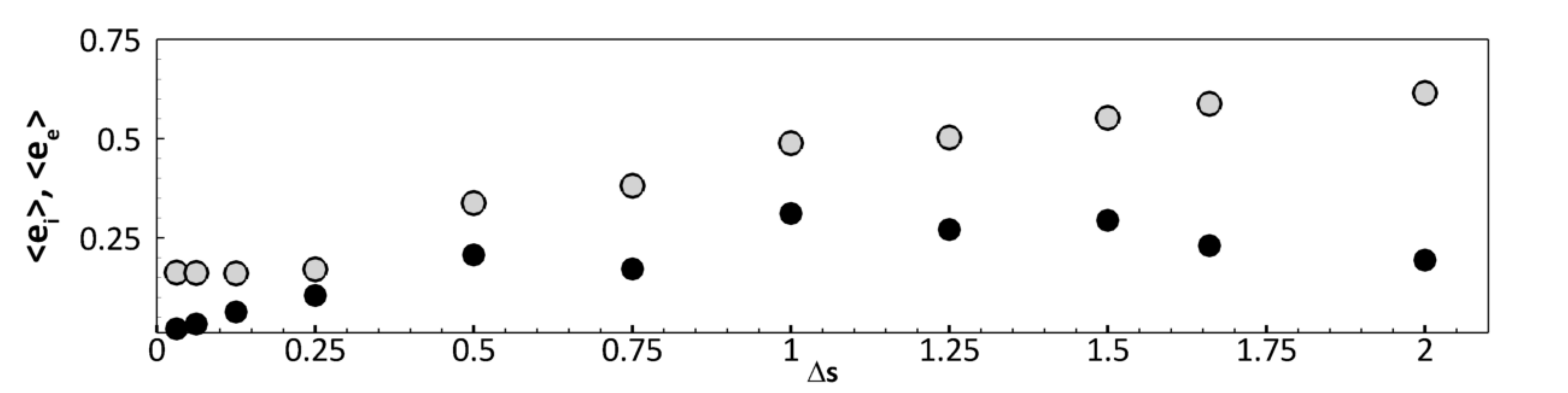}} \\
\end{tabular}
  \caption{
  (a) Spatially-integrated mean-flow error $\int_{\Omega} e_{\langle \bar{\boldsymbol{u}} \rangle}  (\boldsymbol{x}) dS$ of the nudged URANS simulations of data-set group VS as a function of the spatial sampling $\Delta s$. This integrated error is evaluated over a spatial domain $\Omega$ that extends in the streamwise direction between $-1.5 \le x \le 12.5$ and is restricted in the cross-stream direction to (squares) $-4 \le y \le 4$ or (circles) $-1.5 \le y \le 1.5$. The reported values are normalised by the corresponding error of the standard URANS solution. (b) The integrated error in the restricted domain $-1.5 \le y \le 1.5$ is decomposed as the sum of internal (black) and external (grey) errors to the nudging region. }
\label{fig:meanerror_deltas}
\end{figure}

A more quantitative description of the mean-flow error is provided in figure \ref{fig:meanerror_deltas} which displays the spatially-integrated mean-flow error $\int_{\Omega} e_{\langle \bar{\boldsymbol{u}} \rangle}  (\boldsymbol{x}) dS$, where $\Omega$ is the domain of integration. Figure \ref{fig:meanerror_deltas}(a) illustrated this global error, divided its the value for standard URANS results, as a function of $\Delta s$ for two domains having the same streamwise extent ($-1.5 \le x \le 12$) but different cross-stream ones. Squares correspond to the mean-flow error when integrated in a domain of size $-4 \le y \le 4$, which is identical to the $y$-extension used for the figures. In this case, the global mean-flow error seems only slightly decreased $(by \sim 30\%)$ even for the smallest value of $\Delta s$. However, when restricting the domain $\Omega$ to the extension of the nudging region in the cross-stream direction ($-1.5 \le y \le 1.5$), the global mean-flow error (circles) now exhibits a significant decrease for $\Delta s \le 1$, while it remains fairly constant for $\Delta s > 1$. This evolution is better understood by examining the internal (black circles) and external (grey circles) components of this error, which are displayed in figure \ref{fig:meanerror_deltas}(b). For $\Delta  s >1$, the decrease of the external error is actually balanced by the increase of the internal error. These two opposite effects were already visible in figure \ref{fig:MeanErrorFieldUnsteady}(a-d). For $\Delta  s \le 1$, both internal and external mean-flow errors decrease almost monotonically. The external contribution reaches a constant value for $\Delta s \le 0.25$ while the internal one still decreases, which is reminiscent of the evolution of the time-averaged spatially-integrated errors in figure \ref{fig:error_deltas}. To conclude, the nudging approach using the low-frequency data-set also allows for an improvement in the mean-flow prediction, not only inside but also outside of the nudging region. This extrapolation capability is mainly significant for a domain of limited extension in the cross-stream direction, which seems related to the extension of the nudging region and appears to be directly downstream of the latter, according to the  present results.

\subsection{Nudging the high-frequency small-scale structures (data-set group KH)} \label{sec:results_high_frequency} 

As detailed in \S \ref{sec:DNSvsURANS}, DNS results capture the shedding of Kelvin-Helmholtz (KH) vortices in the shear layers of the cylinder. As illustrated by figure \ref{fig:SPOD_Spec_DNS_KH}, a broadband peak is observed around frequency $St=4.384$ in the spectrum obtained by the SPOD analysis, and the dominant SPOD mode is characterized by spatially oscillating structures of wavelength $\lambda_{x}\sim 0.25$ located in the upper shear layer. This high-frequency small-scale structures are in general not captured by standard URANS simulations \citep{Iaccarino2003_ijhff,Palkin2016_ftc} and could also not be obtained in the present study, whatever the mesh and temporal resolutions for the URANS calculations. Therefore, we now investigate the potential of the nudging approach to recover this unsteady phenomenon by using the data-sets of group KH (see tables \ref{tab:datasetgroup} and \ref{tab:arrays}). While data-sets in this group generally correspond to higher temporal and spatial samplings compared to previously discussed data-sets (group VS in \S\ref{sec:results_low_frequency}), it may be noticed that these resolutions are actually moderate compared to the time and spatial scales that are associated to the KH vortices. The time sampling $\Delta t$ indeed corresponds to around $12$ snapshots per KH cycle, while the investigated values of $\Delta s$ amount to a number of nudging points per wavelength of the KH vortices verifying $1 \le  N_s \le 8$. Results are here directly analysed based on a SPOD analysis that is adapted to KH phenomena as performed in \S\ref{sec:DNSvsURANS}.

\begin{figure}
\vspace{0.25cm}
\begin{tabular}[t]{lll}
& (a)   & (b) \\
& \imagetop{\includegraphics[width=0.45\columnwidth]{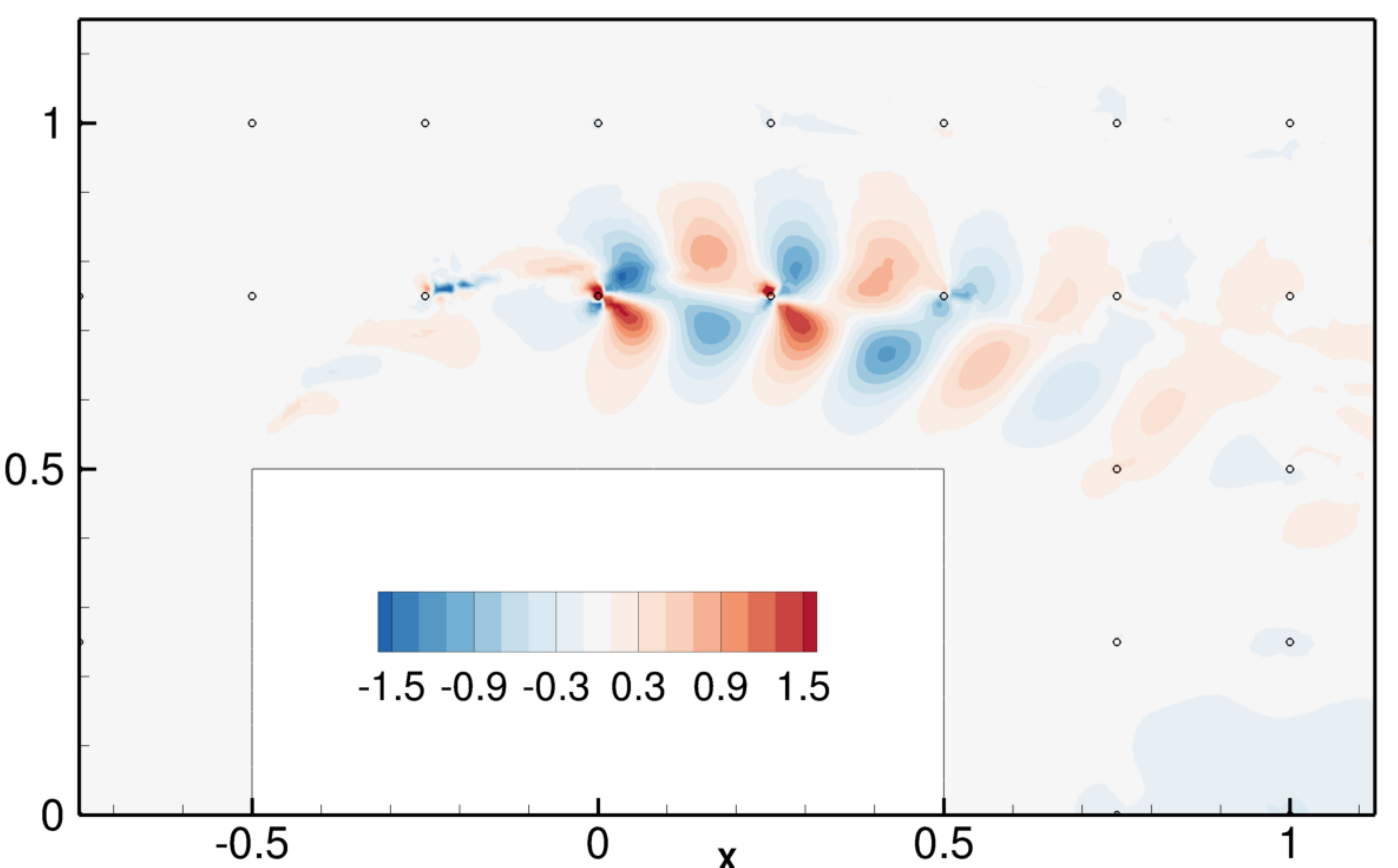}} & \imagetop{\includegraphics[width=0.49\columnwidth]{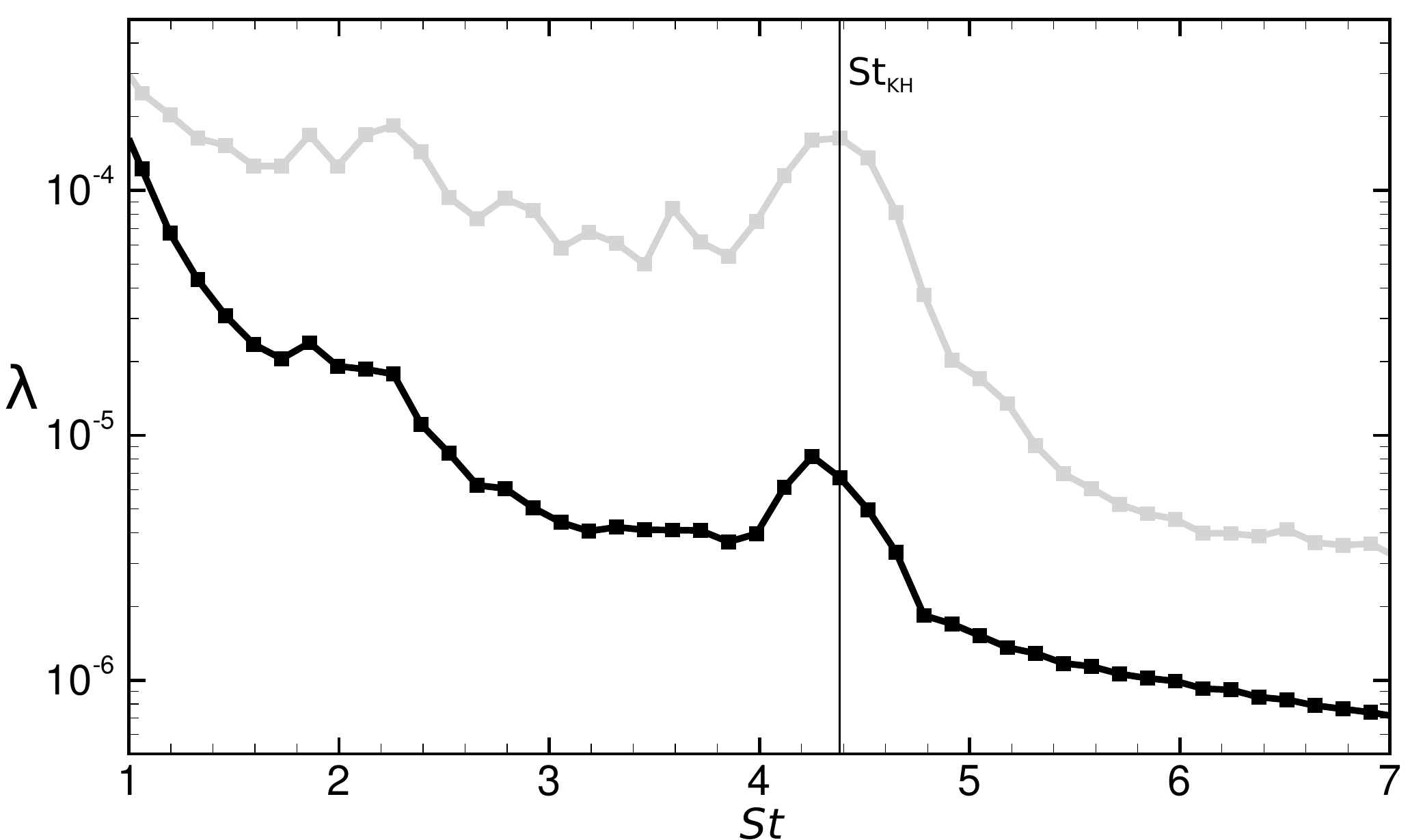}} \\
& (c)  & (d) \\
& \imagetop{\includegraphics[width=0.45\columnwidth]{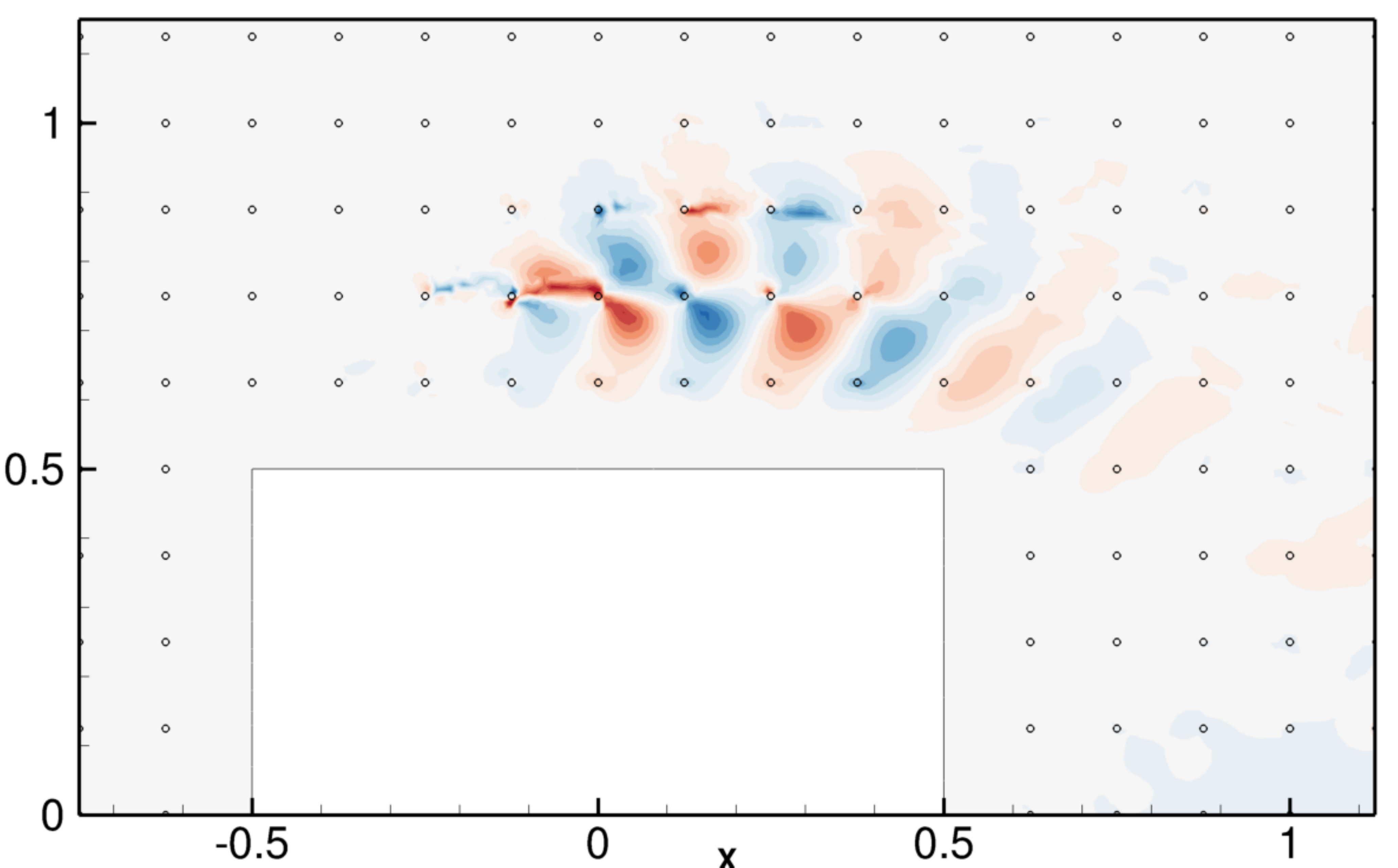}} & \imagetop{\includegraphics[width=0.49\columnwidth]{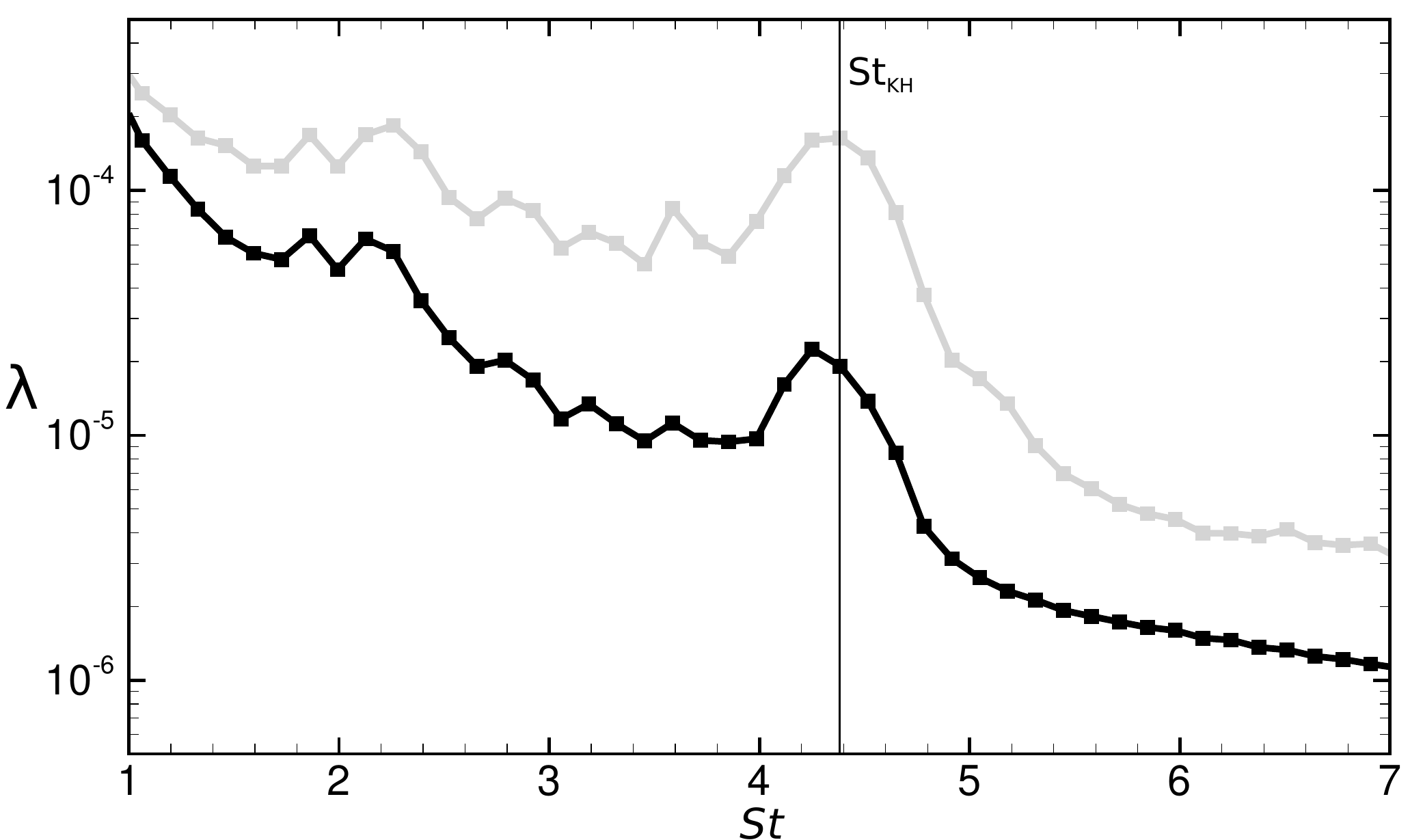}} \\
& (e)  & (f) \\
& \imagetop{\includegraphics[width=0.45\columnwidth]{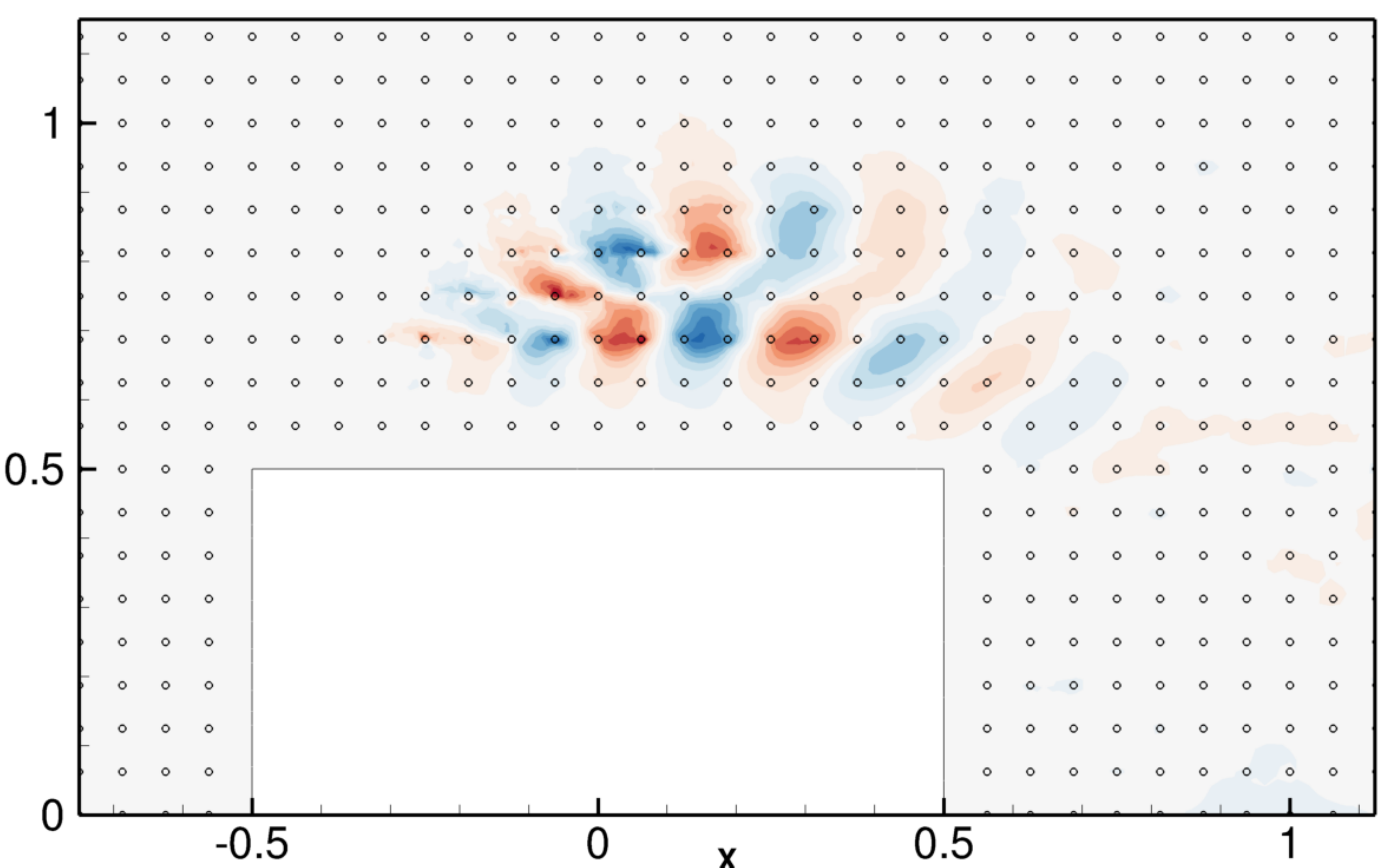}} & \imagetop{\includegraphics[width=0.49\columnwidth]{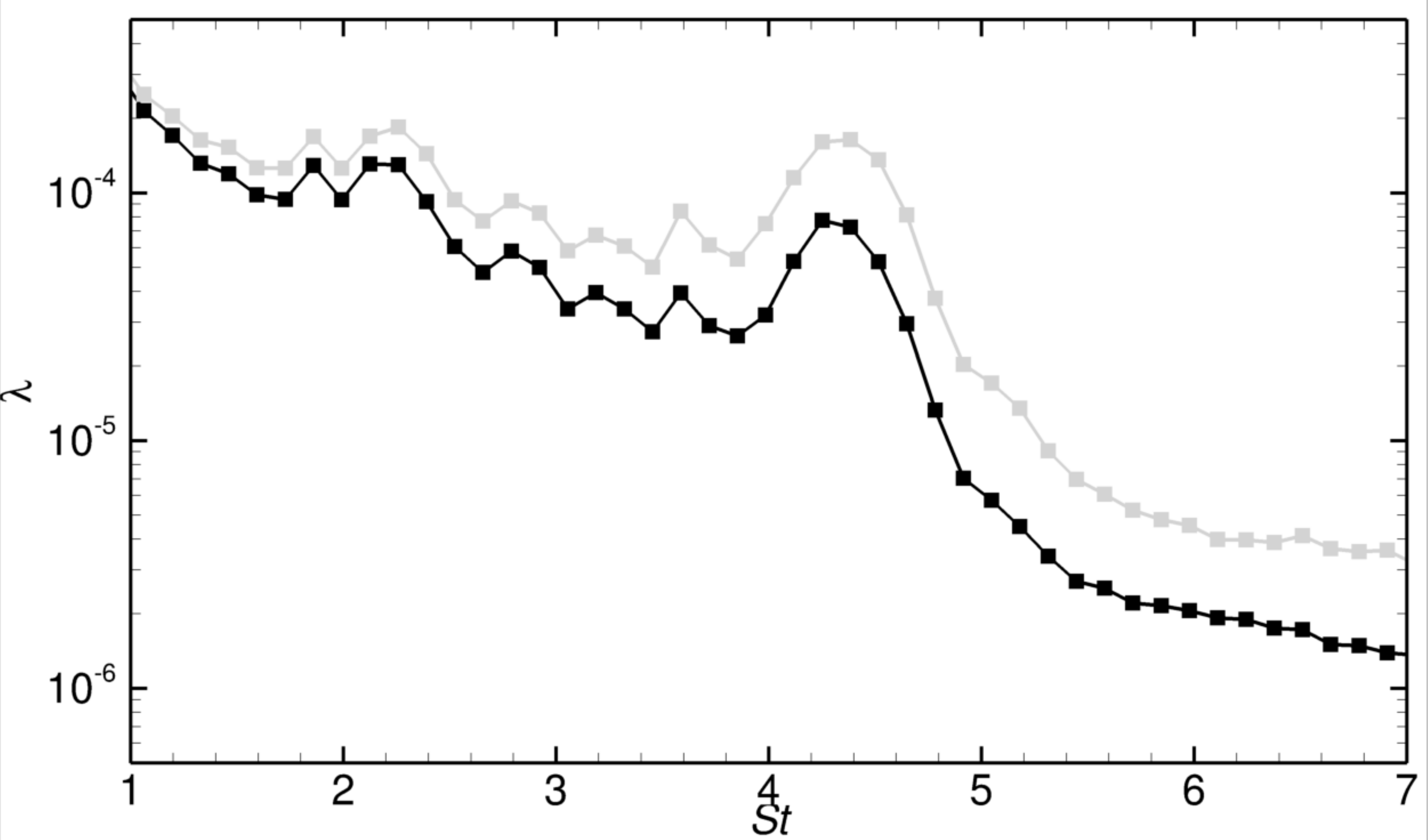}} \\ 
& (g)  & (h) \\
& \imagetop{\includegraphics[width=0.45\columnwidth]{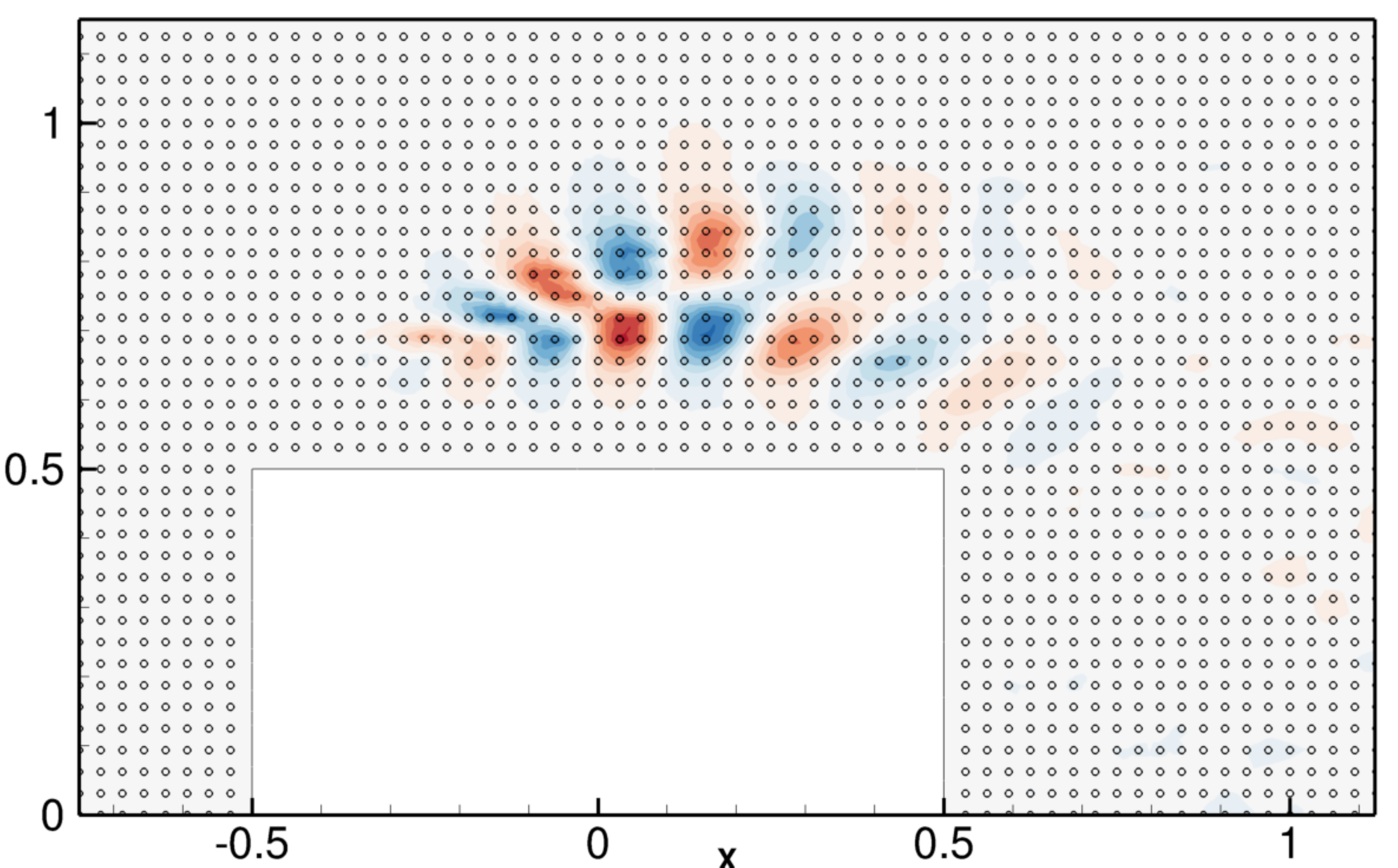}} & \imagetop{\includegraphics[width=0.49\columnwidth]{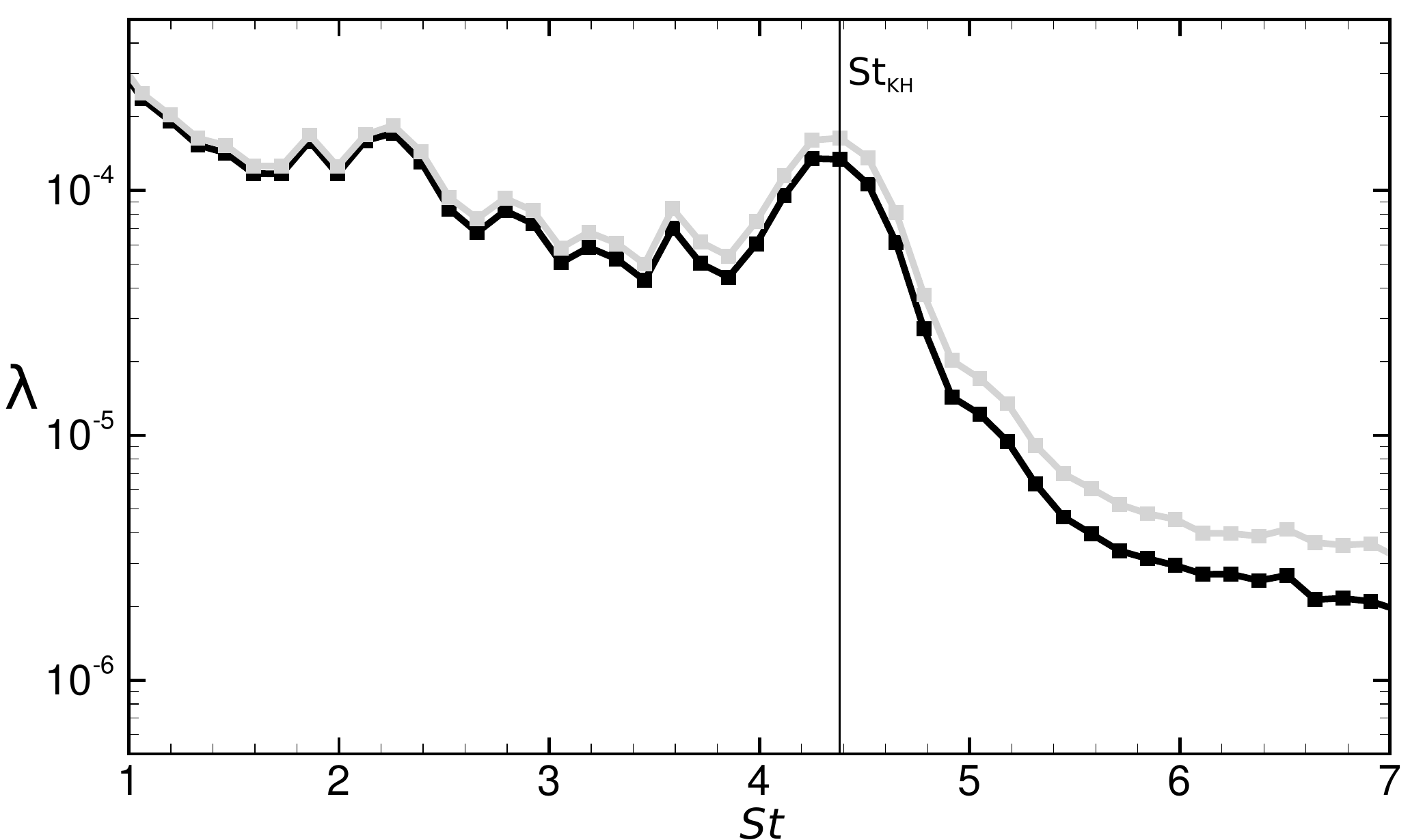}} 
\end{tabular}
  \caption{
   SPOD analysis of Kelvin-Helmholtz phenomena as estimated by nudged URANS simulations for the data-set group KH with (a,b) $N_s=1$ ($\Delta s=0.25$), (c,d) $N_s=2$ ($\Delta s=0.125$), (e,f) $N_s=4$ ($\Delta s=0.0625$) and (g,h) $N_s=8$ ($\Delta s=0.03125$) nudging points per wavelength of the KH instabilities. (a,c,e,g) Real part of the streamwise component of the SPOD mode for $St_{KH}=4.384$ and (b,d,f,h) SPOD spectrum for nudged URANS (black), the DNS results (grey) are also reported for the sake of comparison.}
\label{fig:SPOD_ErrorModesc_HiFreq}
\end{figure}

The real part of of the streamwise component of the dominant SPOD mode at frequency $St_{KH}=4.384$ obtained with nudging is shown in the left-hand-side plots of 
figure \ref{fig:SPOD_ErrorModesc_HiFreq} for the different values of the spatial sampling $\Delta s$ in data-set group KH. SPOD spectra are also displayed in the right-hand-side plots.

The SPOD mode in figure \ref{fig:SPOD_ErrorModesc_HiFreq}(a), already captures fluctuations that are reminiscent of Kelvin-Helmholtz structures visible in the reference SPOD mode in figure \ref{fig:SPOD_Spec_DNS_KH}(a). This is remarkable, as the associated spatial sampling $\Delta s = 0.25$ corresponds to $N_s=1$, i.e. one nudging point per wavelength of coherent vortex structures. However, the spectrum in figure \ref{fig:SPOD_ErrorModesc_HiFreq}(b) indicates significantly under-estimated amplitudes of all dominant modes around $St_{KH}=4.384$. In particular, the amplitude $\lambda_{KH}$ of the mode in figure \ref{fig:SPOD_ErrorModesc_HiFreq}(a) corresponds to only $\sim 4\%$ of the reference value. The nudging approach seems to excite modes that are associated with KH instabilities, but their energy amplification is not large enough to reach sufficient amplitude, probably because of the eddy-viscosity (obtained here with the Spalart-Allmaras model) that dissipates too quickly the energy at that frequency. Increasing the spatial resolution of nudging points leads to increasingly pronounced spectral peaks at $St_{KH}=4.384$, as more energy is introduced to promote KH structures. This can be confirmed in figures \ref{fig:SPOD_ErrorModesc_HiFreq}(c,d) for $\Delta s = 0.125 \, (N_s=2)$. Eventually, the spatial pattern obtained for smaller spatial samplings (see figures \ref{fig:SPOD_ErrorModesc_HiFreq}(e) and (g)) is very similar to the reference SPOD mode (figure \ref{fig:SPOD_ErrorModesc_HiFreq}(a)). The good performance of nudging is also evident in the SPOD spectra in figures \ref{fig:SPOD_ErrorModesc_HiFreq}(f) and (h). $N_s=8$ nudging points per KH wavelength (figures \ref{fig:SPOD_ErrorModesc_HiFreq}(g-h)) thus appear sufficient to satisfactorily predict the kinetic energy of the modes around $St_{KH}=4.384$. In particular, the estimated amplitude $\lambda_{KH}$ corresponds to $\sim 80\%$ of the reference value.


We have shown that applying nudging with sufficiently well-resolved measurement data allows to reproduce KH vortices, which are no captured by the standard URANS equations. It may be emphasised that nudging improves URANS simulations in a 'global' sense and is not simply 'over-ruling' local values in shear layers. This is particularly evident for the case with large spacing between nudging points (see figure \ref{fig:SPOD_ErrorModesc_HiFreq}(a)), where local maxima and minima in SPOD modes associated with KH instabilities do not coincide with nudging positions.

\subsection{Modelling and experimental considerations} \label{sec:criterion}


As discussed in \S\ref{sec:results_low_frequency} and \S\ref{sec:results_high_frequency}, nudging here appears to be able to significantly improve URANS results even when considering moderately dense data with respect to the phenomenon of interest. For both large-scale vortex shedding and Kelvin-Helmholtz instabilities, with a time interval between measurements that is below one tenth of the considered characteristic period, less than $N_s=3$ nudging points per characteristic wavelength is already enough to correct the frequency content of URANS. The kinetic energy that is associated to the (SPOD) modes that oscillate at the frequencies of interest is also satisfactorily recovered with $N_s=3$ and $N_s=8$ for large-scale vortex shedding and Kelvin-Helmholtz phenomena, respectively. The detailed shape of these modes may be further adjusted when increasing the measurement spatial density. If one is only interested in mean quantities, a spatial sampling of $\Delta s=0.25$, or equivalently $N_s=4$ if the cylinder size is here considered as the characteristic length, is sufficient to significantly improve the estimation of the mean flow compared to standard URANS for the present flow configuration.

It may be worth emphasising that the present results have been obtained in conjunction with a specific RANS turbulence model, namely the Spalart-Allmaras one. Through the consideration of other models, one could possibly further decrease the requirements in terms of temporal and spatial samplings of the measurements, in particular concerning the estimation of the Kelvin-Helmholtz instabilities. Indeed, the study of \cite{Palkin2016_ftc} confirmed for a similar flow configuration as the present one, namely the flow past a circular cylinder at $Re=1.4 \times 10^5$, that URANS calculations based on Reynolds stress models may include such high-frequency phenomena, contrary to eddy-viscosity models, thanks to a lower effective viscosity, among others. It could thus be imagined in this case that less nudging points could be required in order to amplify and satisfactorily sustain Kelvin-Helmholtz phenomena. In other words, it could be expected that the correct estimation of the kinetic energy that is associated to these fluctuations requires less than eight points per characteristic wavelength when relying on a Reynolds stress model.

Aside from the choice of the turbulence model in the URANS framework, more generally, it seems worth investigating the use of nudging to enhance the estimation of turbulent flows in conjunction with other modelling approaches such as hybdrid and multiscale techniques \citep{Sagaut2013_icp,Chaouat2017_ftc}. Compared to URANS, methods such as Partially Averaged Navier–Stokes (PANS) \citep{Girimaji2006_jam} or Partially Integrated Transport Modeling (PITM) \citep{Chaouat2005_pof}, among others, may enable to relax the assumption of scale separation between coherent/mean and turbulent phenomena. In addition, such approaches generally provide clearer and explicit definitions of the resolved and modeled scales, and thus may disambiguate which aspects, i.e. scales, in reference/experimental data the estimated flow is supposed to recover.

In the perspective of using the current nudging approach in an actual experimental context, it is interesting to note that the above mentioned temporal and spatial resolutions may be considered as quite realistic in an experimental point of view, i.e. if one were to consider performing a PIV experiment of the present flow. Similar to the current nudging procedure, PIV measurements are also typically characterised by regular spatial sampling of measurements within a rectangular-shaped measurement domain. Current state-of-the-art in high frame rate PIV includes cameras with a typical sensor size of roughly $2000$ pixels. Using standard seeding conditions and current processing algorithms, an interrogation window would consist of 16 pixels with 75\% overlap \citep{Raffel2018_springer}. The final vector spacing would be $\Delta s_{PIV} = 0.03$, which roughly corresponds to the lowest spatial sampling investigated here. Considering a water-flow experiment and a cylinder size of e.g. $20 \ mm$ (leading to a flow velocity of roughly $1.1 \ m/s$), the frequency of shear-layer instabilities would be roughly equal to $230 \ Hz$. The present requirement of acquisition to be performed at a roughly ten times higher frequency (thus $2.3 \ kHz$) is thus also reachable, as a high frame rate PIV system can nowadays operate with full laser energy and camera resolution at frequencies of $2-3 \ kHz$ minimum.

Note that this remains of course a rapid and preliminary verification. Turning to an actual experiment would raise a number of additional challenges (local spatial averaging due to the PIV interrogation windows, measurement noise, etc.), which will be addressed in future steps. However, it is already quite promising that the amount of data necessary for the nudging method to yield excellent results is at first order compatible with current experimental capabilities.

\section{Conclusion}\label{sec:conclusions}

Data assimilation for the prediction of the turbulent flow past a square cylinder has been performed through nudging and in conjunction with URANS modelling. Through a careful examination of mean and dynamical flow features in both physical and spectral spaces, this study has confirmed the potential of nudging in the state estimation of a complex flow that exhibits a wide range of spatio-temporal scales. Despite its conceptual simplicity, nudging was indeed able to significantly enhance the URANS estimation based on sparse velocity observations from DNS. With very few measurements per characteristic wavelength, nudging allowed to improve the characterisation of large-scale vortex shedding. Even more interestingly, nudging was able to overcome the significant modelling limitations of URANS in the prediction of high-frequency Kelvin-Helmholtz instabilities. The latter were successfully recovered through nudging, while they were absent from the present baseline URANS results, based on a still moderate number of measurements per characteristic wavelength. SPOD and mean-flow analyses confirmed that nudging allowed a global improvement in the flow prediction, and in particular outside of the region that is directly driven by the measurements. As nudging virtually induces no supplementary computational cost, and that URANS may be considered as a relatively affordable simulation method, the use of nudging in conjunction with URANS thus appears as a particularly cost-efficient data assimilation approach for the high-fidelity unsteady characterisation of complex turbulent flows.

Future studies could be dedicated to the consideration of other high-Reynolds number turbulent flows. In particular, it could be of interest to investigate configurations which do not necessarily exhibit low-frequency quasi-periodic phenomena in addition to broadband fluctuations at significantly higher frequencies, as in the present case, and which could therefore be more challenging for URANS. This could motivate the consideration of more advanced modelling approaches in conjunction with nudging. Perhaps more importantly, future work could deal with the consideration of actual experimental data. As discussed in this study, this may be encouraged by the fact that the identified requirements of nudging in terms of spatio-temporal resolution of the measurements seem compatible with the possibilities that are offered by optical techniques.

\section*{Acknowledgements}

This work was partly supported by the HOMER project from the European Union's Horizon 2020 research and innovation program under grant agreement No. 769237.

\section*{Declaration of interests}

The authors report no conflict of interests.

\appendix
\section{Influence of nudging parameter $\alpha$}\label{sec:alphastudy}
\begin{table}
    \centering
\begin{tabular}{ccccccccccc}
$\alpha$ & $0$ & $10^{-5}$ &$10^{-4}$ & $0.001$ & $0.01$ & $0.1$ & $1$ &  $1000$  \\ 
\hline
$\left\langle E \right\rangle$ & $30$ & $29.6$ & $26$ & $4.8$ & $1.86$ & $1.58$ & $1.52$ &  $1.52$ \\ 
$\left\langle E_I \right\rangle$ & $7.8$ & $7.8$ & $7.0$ & $1.24$ & $0.090$ & $0.032$ & $0.026$ &  $0.024$ \\ 
$\left\langle E_E \right\rangle$ & $22$ & $21.6$ & $19.0$ & $3.4$ & $1.78$ & $1.54$  & $1.5$ &  $1.48$  \\ 
\midrule
$\left\langle ||f_x||^{2} \right\rangle$ & 1 & $1.00$ & $0.94$ & $0.20$ &  $9.1\cdot10^{-3}$ &  $5.1\cdot10^{-4}$  &  $9.9\cdot10^{-6}$ &     $1.1\cdot10^{-11}$ \\ 
$\left\langle ||f_y||^{2} \right\rangle$ & 1 & $1.00$ & $0.94$ & $0.15$ &  $5.9\cdot10^{-3}$ &  $4.1\cdot10^{-4}$  &  $8.7\cdot10^{-6}$ &     $1.0\cdot10^{-11}$ \\ 
\midrule
$\alpha^2 \left\langle ||f_x||^{2} \right\rangle$ & $0$ & $1.7\cdot10^{-8}$ & $1.6\cdot10^{-6}$ &  $3.5\cdot10^{-5}$ &  $1.5\cdot10^{-4}$ &  $8.6\cdot10^{-4}$  &  $1.7\cdot10^{-3}$ &     $1.9\cdot10^{-3}$ \\ 
$\alpha^2 \left\langle ||f_y||^{2} \right\rangle$ & $0$ & $2.9\cdot10^{-8}$ &  $2.7\cdot10^{-6}$ &  $4.3\cdot10^{-5}$ &  $1.7\cdot10^{-4}$ &  $1.2\cdot10^{-3}$  &  $2.6\cdot10^{-3}$  &  $2.9\cdot10^{-3}$ \\ 
\midrule
\end{tabular}
    \caption{Sensitivity of time-averaged total, internal, external errors (respectively $\langle {E} \rangle$, $\langle {E}_I \rangle$, and $\langle {E}_E \rangle$), and forcing magnitudes ($\langle||f_x||\rangle$ and $\langle||f_y||\rangle$) to the nudging coefficient $\alpha$ for the data-set of group VS with $\Delta s=0.125$ (see tables \ref{tab:datasetgroup}-\ref{tab:arrays}). In the fourth and fifth rows, the values of $\langle||f_x||\rangle$ and $\langle||f_y||\rangle$ are normalized by their values for standard URANS.
    }
    \label{tab:alpha}
\end{table}
In this appendix, the sensitivity of the reconstruction results with respect to the parameter $\alpha$ in the nudged URANS equations (\ref{eqn:nudged-urans}) is investigated. Time-averaged global errors $\langle {E} \rangle$, $\langle {E_I} \rangle$ and $\langle {E_E} \rangle$ (see (\ref{eqn:total-error}) and (\ref{eqn:internal-external-error})) are summarized in rows 2 to 4 of table \ref{tab:alpha} for values of the nudging coefficient that range from $\alpha=0$ (corresponding to standard URANS) to $\alpha=1000$. The considered measurement data-set belongs to group VS with $\Delta s=0.125$. Fifth and sixth rows in table \ref{tab:alpha} report the time average of the norm of the forcing vector
\begin{equation}
    \boldsymbol{f}=\mathcal{H}(\bar{\boldsymbol{u}}_{\alpha})-\boldsymbol{m},
\end{equation}
which thus allows to quantify the discrepancies between estimated and reference solution at the measurement locations, distinguishing between the streamwise and crossflow components. 

Table \ref{tab:alpha} confirms that the magnitude of the components of $\boldsymbol{f}$ (fifth and sixth rows) is reduced by more than three orders of magnitude for $\alpha \geq 0.1$ compared to standard URANS and, as expected, converges towards zero as $\alpha$ is further increased. On the other hand, the effective amplitude of the forcing that is introduced in the nudged URANS equations (\ref{eqn:nudged-urans}), which may be evaluated as $\alpha\langle||\boldsymbol{f}||\rangle$, converges towards constant values, as the errors with respect to the reference data seem to decrease proportionally to increasing values of the nudging coefficient (seventh and eighth rows in table \ref{tab:alpha}). As this effective amplitude but also the global errors in the second to fourth rows of table \ref{tab:alpha} show minor sensitivity with respect to $\alpha$ for $\alpha \ge 1$, it was decided to maintain $\alpha = 100$ for all runs in the present contribution. 

\bibliographystyle{jfm}  

\end{document}